\pdfoutput=1
\documentclass[aps,prd,twocolumn,superscriptaddress,preprintnumbers,floatfix,nofootinbib]{revtex4-2}

\usepackage{graphicx}
\usepackage{amsmath}
\usepackage[caption=false]{subfig}
\usepackage{siunitx}
\usepackage{placeins}
\usepackage{color}
\usepackage{standalone}
\usepackage{dcolumn}
\usepackage{tensor}
\usepackage{bm}
\usepackage{microtype}
\usepackage{etoolbox}
\usepackage{amssymb}
\usepackage{mathrsfs}
\usepackage{accents}
\usepackage[normalem]{ulem}
\usepackage[dvipsnames]{xcolor}
\usepackage[colorlinks,urlcolor=NavyBlue,citecolor=NavyBlue,linkcolor=NavyBlue,pdfusetitle]{hyperref}
\usepackage[all]{hypcap}
\usepackage[inline]{enumitem}
\usepackage[utf8]{inputenc}
\usepackage{csquotes}
\usepackage{array}
\usepackage{booktabs}
\usepackage{wasysym}
\usepackage[nolist,nohyperlinks]{acronym}
\usepackage[final]{changes}
\usepackage{subfig}

\newcommand{\beq}{\begin{equation}}
\newcommand{\eeq}{\end{equation}}

\newcommand*{\eq}[1]{Eq.~\eqref{eq:#1}}

\graphicspath{{./fig/}}

\newcommand{\dcc}{LIGO-P2200221}

\newcommand{\infd}{\mathrm{d}}

\begin{document}

\title{Parametrizing gravitational-wave polarizations}

\author{Maximiliano Isi}
\email[]{misi@flatironinstitute.org}
\affiliation{Center for Computational Astrophysics, Flatiron Institute, 162 5th Ave, New York, NY 10010}

\hypersetup{pdfauthor={Isi}}

\date{\today}

\begin{abstract}
We review the formalism underlying the modeling of gravitational wave (GW) polarizations, and the coordinate frames used to define them.
In the process, we clarify the notion of ``polarization angle'' and identify three conceptually distinct definitions.
We describe how those are related and how they arise in the practice of GW data analysis, explaining in detail the relevant conventions that have become the LIGO-Virgo standard.
Furthermore, we show that any GW signal can be expressed as a superposition of elliptical (i.e., fully-polarized) states, and examine the properties and possible parametrizations of such elementary states.
We discuss a variety of common parametrizations for fully-polarized modes, and compute Jacobians for the coordinate transformations relating them.
This allows us to examine the suitability of each parametrization for different applications, including unmodeled or semimodeled signal reconstructions.
We point out that analyses parametrized directly in terms of the plus and cross mode amplitudes will tend to implicitly favor high signal power, and to prefer linearly-polarized waves along a predefined direction; this makes them suboptimal for targeting face-on or face-off sources, which will tend to be circularly polarized.
We discuss alternative parametrizations, with applications extending to continuous waves, ringdown studies, and unmodeled analyses like \textsc{BayesWave}.
Code and additional material are made available in \url{https://github.com/maxisi/gwpols}.
\end{abstract}

\maketitle

\begin{acronym}
\acro{GR}{general relativity}
\acro{CBC}{compact-binary coalescence}
\end{acronym}

\section{Introduction}
\label{sec:intro}

Gravitational waves (GWs) come in two distinct polarization states, whose amplitude and phase evolution reflect the structure of \ac{GR} and the dynamics of the source.
As for electromagnetic waves, these states are only unambiguously defined up to rotations of the reference frame around the wave's direction of propagation.
When analyzing data from detectors like LIGO \cite{TheLIGOScientific:2014jea} and Virgo \cite{TheVirgo:2014hva}, it is natural to parametrize polarizations differently depending on the application.
For instance, searches for \acp{CBC} aim to relate the signal observed by different detectors to templates obtained from theory, and thus benefit from describing GW polarizations in the same frame as the predictions (e.g., \cite{Faye:2012we,Kidder:2007rt}).
On the other hand, unmodeled (or semimodeled) analyses aim to reconstruct GWs without relying on detailed input from theory, and must instead make an arbitrary choice in orienting the polarization frame \cite{Klimenko:2004qh,Klimenko:2005xv,Klimenko:2008fu,Cornish:2014kda,Cornish:2020dwh}.
Furthermore, lacking waveform templates, unmodeled analyses must also decide how to parametrize the GW polarization state and its time evolution in a way sufficiently flexible to capture a range of morphologies while parsimonious enough to remain computationally tractable.
Analyses that focus on recovering signal power without coherently modeling the phase evolution may use yet different conventions \cite{Romano:2016dpx}.

The abundance of polarization parametrizations and reference directions is visible in the literature as well as in the implementation of data analysis software.
Such variety can cause confusion, and hinder comparisons across analyses with different conventions, or even complicate the interpretation of individual analysis outputs.
As an example, this comes into play when parametrizing continuous GWs from galactic pulsars, and in relating such (projected) measurements to electromagnetic observations of the source orientation (e.g., \cite{Ng:2007te,Dupuis:2005xv,Isi:2017equ,Pitkin:2017qfy}).
They are relevant in parametrizing ringdown signals for black hole spectroscopy, where multiple factorizations are possible for the polarization amplitudes (e.g., \cite{Isi:2021iql,Carullo:2019flw,LIGOScientific:2020tif,LIGOScientific:2021sio}).
They are also important for understanding the implications of different treatments of polarization ellipticity in unmodeled analyses, e.g., with the \textsc{BayesWave} algorithm \cite{Cornish:2014kda,Cornish:2020dwh,Chatziioannou:2021mij}, and in comparing such results to modeled \ac{CBC} inference.

This paper provides a comprehensive exposition of the formalism underlying GW polarizations as it pertains practical applications.
The goal is twofold:
\emph{first}, pedagogical, in reviewing the relations between different polarization conventions, and in clarifying how these come to bear in real-world data analysis;
\emph{second}, technical, in explicitly working out the coordinate transformations that link different parametrizations, and providing ready-to-use expressions for the corresponding Jacobians---the mathematical factors that translate between posterior probability densities obtained under different parametrizations, which are required to exchange priors when carrying out Bayesian inference or similar applications.
In this work, the exposition is geared towards observers, or theorists interested in drawing connections to observation---as such, it strives for concreteness over abstraction, and, in particular, steers away from the rich formal connections between the treatment of GW polarizations and the mathematical structure of GR.

The review of GW polarizations begins in Sec.~\ref{sec:primer} with a derivation of signal decompositions into three different polarization bases: linear, circular and elliptical.
Having established the importance of elliptical (fully polarized) modes, Sec.~\ref{sec:ellip_modes} examines their key properties, outlines some of their uses, and sketches their connection to spin-weighted spherical harmonics.
Next, Sec.~\ref{sec:angles} carefully examines the different notions of ``polarization angle'' that arise for elliptical and nonelliptical signals, elucidating both their conceptual independence and their frequent interchangeability in practical applications.
Taking advantage of the mathematical formalism introduced in the preceding sections, Sec.~\ref{sec:jacobians} provides a census of different parametrizations of elliptical states, derives the Jacobians connecting them, and discusses their implications for parameter estimation.
Finally, Sec.~\ref{sec:nongr} briefly covers generalizations of these ideas to beyond-GR polarization states, and Sec.~\ref{sec:conclusion} concludes.
Code behind the figures and additional material are made available in an accompanying release \cite{release}.

\section{Polarization primer}
\label{sec:primer}

\subsection{Linear basis}
\label{sec:linear}

In GR, there exist two propagating gravitational degrees of freedom, corresponding to two independent GW polarizations (e.g., \cite{Thorne1983,Thorne:1987af,Poisson2014,BT}).
At any given time, their local effect can be encoded in a strain tensor $h_{ij}$ representing the transverse-traceless part of the metric perturbation, also known as the \emph{gravitational-wave field} \cite{Thorne1983,Thorne:1987af,Poisson2014,BT}.
In a Cartesian frame with $z$-axis along the direction of propagation, we can write this matrix as
\beq \label{eq:hij}
(h_{ij}) = \begin{pmatrix}
h_+ & h_\times  & 0 \\
h_\times  & - h_+ & 0  \\
0 & 0 & 0
\end{pmatrix} ,
\eeq
where the \replaced{instantaneous plus ($+$) and cross ($\times$) polarization amplitudes, $h_{+/\times}$, are given by some polarization functions $h_{+/\times}(t; \theta)$ depending}{plus ($+$) and cross ($\times$) polarization amplitudes, $h_{+/\times}$, depend} implicitly on the retarded time, $t - R/c$, in a way determined by the source dynamics and by the (luminosity) distance $R$ to the source, as well as on any other relevant source parameters, \added{$\theta$}, controlling the amplitude and phase of the wave, as dictated by Einstein's equations.

It can be useful to rewrite \eq{hij} as $h_{ij} = h_+ e^+_{ij} + h_\times e^\times_{ij}$, in terms of the $e^{+/\times}_{ij}$ polarization basis tensors given by
\begin{subequations} \label{eq:lin}
\begin{align}
e^+_{ij} &\equiv \hat{x}_i \hat{x}_j - \hat{y}_i \hat{y}_j \, ,\\
e^\times_{ij} &\equiv \hat{x}_i \hat{y}_j + \hat{y}_i \hat{x}_j\, ,
\end{align}
\end{subequations}
where $\hat{x}$ and $\hat{y}$ are arbitrary orthonormal vectors that, with $\hat{z}$, form a right-handed Cartesian basis; we will call this the \emph{wave frame}.
Since this frame is constructed to have $\hat{z}$ aligned with the wavevector $\vec{k}$ (i.e., $\hat{z} = \hat{k} \equiv \vec{k}/|k|$), the polarization tensors are implicit functions of the wave propagation direction $\hat{k}$, or, equivalently, the source sky location $\hat{n} = -\hat{k}$.
For a given $\hat{k}$, due to the orthonormality of $\hat{x}$ and $\hat{y}$, it is easy to check that these tensors are orthogonal such that $e^p_{ij} e^{p'ij}=2\delta^{pp'}$ for $p,p'$ in $\{+,\times\}$.

We will refer to plus and cross jointly as the \emph{linear} polarization basis.
Their physical interpretation is best illustrated by their instantaneous effect on a small, freely-falling ring of particles, as shown in Fig.~\ref{fig:rings}.
Other polarization bases can be constructed, as we will see below, but the linear polarizations are generally the most convenient for expressing measurements.

In the small-antenna (long-wavelength) limit, the signal induced by a passing GW on a given detector can be written as the dyadic projection
\beq \label{eq:h}
h(t) \equiv D^{ij} h_{ij} = F_+ h_+ + F_\times h_\times\, ,
\eeq
with antenna patterns $F_{+/\times} \equiv D^{ij} e^{+/\times}_{ij}$ defined in terms of a detector tensor $D_{ij}$ that encodes the geometry of the measurement.
\added{In the small-antenna limit, this tensor contains all relevant information about the detector's response to GWs \cite{Forward:1978zm,bluebook,Estabrook:1985id,Finn:2008np}; it represents the zero-frequency limit of the frequency-dependent transfer function of a stationary instrument \cite{Schilling:1997id,Rakhmanov2005a,Rakhmanov:2008is,Rakhmanov:2009zz,Essick:2017wyl}.}
For a differential-arm detector, like LIGO, with arms pointing along unit vectors $\hat{X}$ and $\hat{Y}$, this is just $D_{ij} = (\hat{X}_i \hat{X}_j - \hat{Y}_i \hat{Y}_j)/2$.%
\footnote{These expressions are valid in the local Lorentz frame of the detector, so we can raise and lower indices with the flat metric.}
In this limit, the antenna patterns are thus purely geometric factors that encode the relative orientations of the detector and wave frames, as defined by $\{\hat{X},\, \hat{Y},\, \hat{Z}\}$ and $\{\hat{x},\, \hat{y},\, \hat{z}\}$ respectively.

\begin{figure}
\includegraphics[width=0.4\columnwidth]{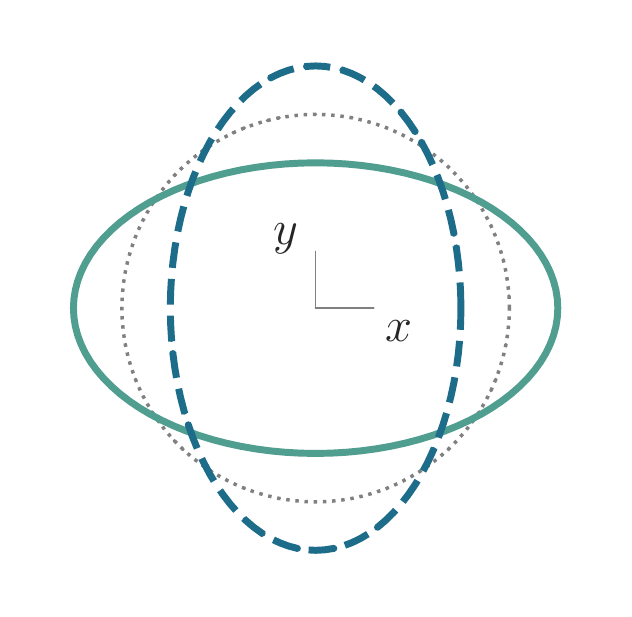}
\includegraphics[width=0.4\columnwidth]{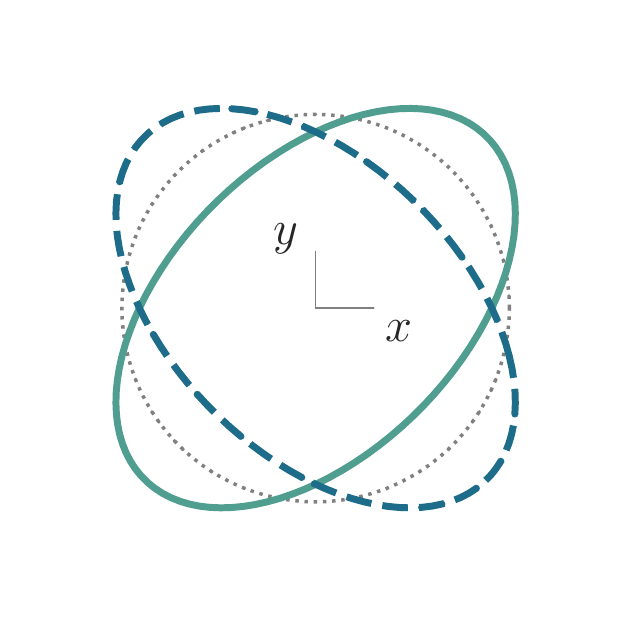}
\caption{Effect of plus (left) and cross (right) polarizations on a small, freely falling ring of particles. The wave propagates in the $z$ direction, perpendicular to the page. The effect is illustrated half a period apart (solid vs dashed); the unperturbed ring is also shown for reference (thin dotted line).
}
\label{fig:rings}
\end{figure}

After fixing the frame orientation, any plane GW may be expressed in terms of the Fourier components of its polarization \replaced{functions}{amplitudes} as
\begin{align}
\label{eq:planewave}
h_{ij}(t,\vec{x}) &= \frac{1}{2\pi}\int_{-\infty}^{+\infty} \tilde{h}_{ij}(\omega, \hat{k})\, e^{i\omega \left(\frac{\hat{k}\cdot\vec{x}}{c}-t\right)} \infd \omega \\
&= \frac{1}{2\pi} \sum_{p=+,\times} \int_{-\infty}^{+\infty} \tilde{h}_p(\omega)\, e^p_{ij}(\hat{k})\, e^{i\omega \left(\frac{\hat{k}\cdot\vec{x}}{c}-t\right)} \infd \omega \nonumber
\end{align}
where the sum is over linear polarization states ($+,\times$) defined in some wave frame attached to the propagation direction $\hat{k}$, and we obtained the second line using the fact that the $e^{+/\times}_{ij}$ are real valued.
\added[comment={FN}]{More broadly, the strain tensor at any point in spacetime may be expressed with full generality as a superposition of these planewaves
by integrating over all directions of propagation (e.g., \cite{Romano:2016dpx,Isi:2018miq}).}

Equation \eqref{eq:planewave} implicitly defines the complex-valued Fourier polarization functions $\tilde{h}_p(\omega)$ to correspond to the time-domain polarizations at the spatial origin, $h_p(t) \equiv h_p(t, \vec{x}=0)$, by
\beq \label{eq:ft}
\tilde{h}_p(\omega) \equiv \int_{-\infty}^{+\infty} h_p(t)\, e^{i\omega t} \infd t \, ,
\eeq
establishing our convention for the Fourier transform.

Since $h_{ij}$ is real valued, the Fourier strain \added{tensor} must  satisfy the complex-conjugate symmetry $\tilde{h}_{ij}(-\omega, \hat{k}) = \tilde{h}_{ij}^*(\omega,\hat{k})$, where the asterisk indicates complex conjugation.
For the linear polarizations, this directly reduces to
\beq \label{eq:sym_linear}
\tilde{h}_{+/\times}(\omega) = \tilde{h}_{+/\times}^*(-\omega)\, ,
\eeq
because the linear basis tensors are themselves real valued.
As usual, then, the positive and negative frequencies must be considered as inseparable contributions to a single Fourier mode.
The existence of this symmetry reveals a redundancy in the description that we can exploit to write Eq.~\eqref{eq:planewave} more concisely.

\begin{figure*}
\includegraphics[width=0.8\textwidth]{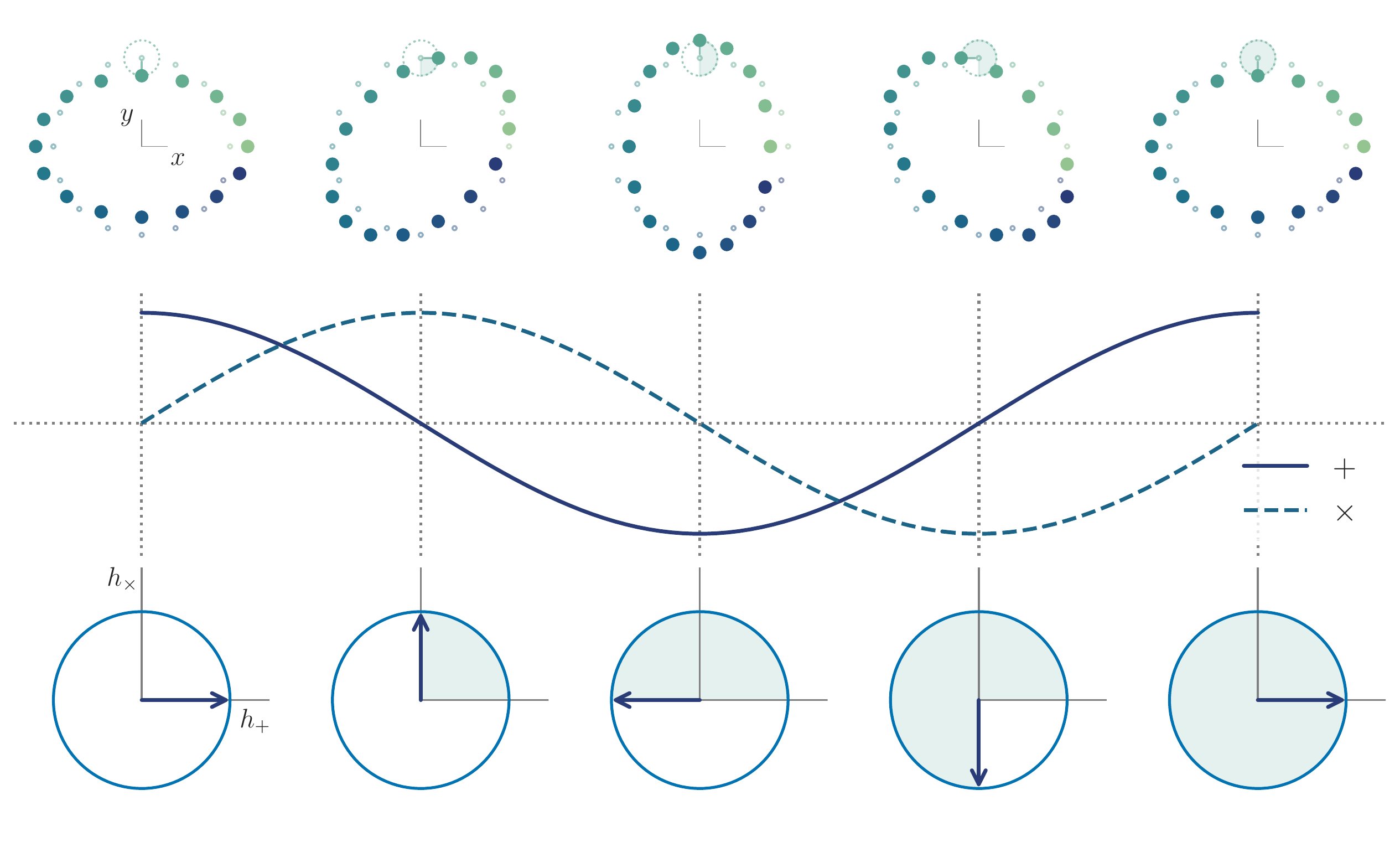}
\caption{A right-handed, circularly polarized GW as a function of time (left to right) over a period. \emph{Top:} as the wave propagates out of the page, it deforms a freely falling ring of particles (colored dots) into an ellipsoidal pattern, which rotates counterclockwise with time; each individual particle is pushed in a circle around its original location, i.e., the location it would have had in absence of the wave (small empty circles).
\emph{Middle:} amplitudes of the plus (solid) and cross (dashed) linear polarizations making up the wave as a function of time; the plus polarization is $\pi/2$ radians ahead of the cross polarization.
\emph{Bottom:} representation of the polarization state as a phasor vector in the $+$ and $\times$ space; the phasor rotates counterclockwise in a circle.
Reversing the direction of time by reading this diagram right-to-left gives the effect of a left-handed circularly polarized wave.
(See Ref.~\cite{release} for accompanying animations.)
}
\label{fig:pol_diagram_circ}
\end{figure*}

\subsection{Circular basis}

First, instead of the linear plus and cross polarizations above, we could equivalently work with the associated \emph{circular} right-handed (R) and left-handed (L) \replaced{polarizations}{modes}.
These are defined in the Fourier domain by the complex-valued basis tensors
\beq \label{eq:circ}
e^{R/L}_{ij} \equiv \frac{1}{\sqrt{2}} \left(e^+_{ij} \pm i e^\times_{ij} \right) ,
\eeq
with the plus (minus) sign corresponding to R (L).
These tensors are also orthogonal and normalized similarly to $e^{+/\times}_{ij}$ such that $(e^{p'ij})^* e^p_{ij} = 2 \delta^{pp'}$ for $p,p'$ in $\{R,L\}$.

\deleted[comment=$\downarrow$]{To understand the physical significance of the circular polarizations\dots}

The orthogonality and completeness of the tensors in Eq.~\eqref{eq:circ} mean that we can rewrite Eq.~\eqref{eq:planewave} in terms of the circular polarizations without loss of generality as the sum
\beq \label{eq:planewave_circ}
h_{ij}(t,\vec{x}) 
= \frac{1}{2\pi} \sum_{p=R,L} \int_{-\infty}^{+\infty} \tilde{h}_p(\omega)\, e^p_{ij}(\hat{k})\, e^{i\omega \left(\frac{\hat{k}\cdot\vec{x}}{c}-t\right)} \infd \omega \, ,
\eeq
where $\tilde{h}_{R/L}$ and $e^{R/L}_{ij}$ have replaced their $+/\times$ counterparts.
As is straightforward to show from Eq.~\eqref{eq:circ}, the Fourier amplitudes of the circular and linear polarizations are related by
\begin{align} \label{eq:circ_amps}
\tilde{h}_{R/L} = \frac{1}{\sqrt{2}} \left(\tilde{h}_+ \mp i\tilde{h}_\times \right) ,
\end{align}
with the minus (plus) sign for R (L).
Based on this, the complex-conjugate condition of Eq.~\eqref{eq:sym_linear} implies that
\beq \label{eq:sym_circular}
\tilde{h}_R(\omega) = \tilde{h}_L^*(-\omega) \, ,
\eeq
which again manifests the redundancy in Eq.~\eqref{eq:planewave_circ}, as in Eq.~\eqref{eq:planewave}.
It also reveals that R and L switch roles for $\omega \to - \omega$, \deleted{as we anticipated below Eq.~\eqref{eq:circ_example},} indicating that these states are invariant under parity-time reversals.

\added{To understand the physical significance of the circular
polarizations,
consider a purely R-polarized monochromatic mode $h^R_{ij}$ with positive frequency $f_0 > 0$, unit amplitude and zero phase offset at the spatial origin ($\vec{x} =0$).
Based on the above discussion, a positive-frequency right-handed mode is the same as a negative-frequency left-handed mode; therefore, in the Fourier domain $h^R$ must take the following form:}
\begin{equation}
  \tilde{h}^R_{ij}(f; f_0) = \frac{1}{2} \left[\delta(f - f_0) e^R_{ij} + \delta(f + f_0)\, e^L_{ij} \right]\, ,
\end{equation}
\added{in terms of the linear frequencies, $f = \omega/2\pi$ and $f_0 = \omega_0/2\pi$; in the time domain, this is}
\begin{align} \label{eq:circ_example}
  h_{ij}^R(t; f_0) &= \int_{-\infty}^{+\infty} \tilde{h}^R_{ij}(f)\, e^{-i2\pi f t}\, \infd f \nonumber\\
  &= \frac{1}{\sqrt{2}} (e^+_{ij} \cos \omega_0 t + e^\times_{ij} \sin \omega_0 t) \, ,
\end{align}
\added{using the definition from Eq.~\eqref{eq:circ}.}

\added{We can visualize the above result as follows.}
In the 2D Cartesian space defined by the linear polarization amplitudes, $\left(h_+, h_\times\right)$, $h^R$ defines a circle, around which the \emph{phasor} encoding the state of the wave rotates counterclockwise (for $\omega_0 > 0$).
This means that, at any given time, the wave will have a unit total amplitude (i.e., $h^2_+ + h^2_\times=1$) and the cross polarization will \emph{lag behind} the plus polarization by $\pi/2$ radians in phase.
Consequently, a purely R-polarized wave will deform a ring of freely-falling particles into an elliptical pattern that is seen to rotate counter-clockwise when looking towards the source (Fig.~\ref{fig:pol_diagram_circ}), i.e., it follows the right-hand rule relative to the direction of propagation (pointing away from the source).
The opposite will be true for purely L-polarized waves \added{with $\omega_0>0$}, which will result in a clockwise-rotating ellipse.
This assignment of the ``right'' and ``left'' labels is known as the ``source based'' handedness convention.

\subsection{Elliptical basis}
\label{sec:ellip}

Next, it is convenient to encode the two linear GW polarizations as quadratures of a single complex-valued scalar field, 
\beq
H(t) \equiv h_+ - i h_\times,
\eeq
in the time domain.
This complex number provides an alternative representation of the $\left(h_+, h_\times\right)$ phasor introduced in the previous section (see bottom panel of Fig.~\ref{fig:pol_diagram_circ}).
If this quantity, the \emph{complex strain}, is purely real (imaginary), then the wave is purely plus (cross) polarized.
In those same terms, a unit-amplitude circularly-polarized mode like the one in \eq{circ_example} can be expressed simply as $H(t)  = \exp(\mp i \omega t)/\sqrt{2}$, with the minus (plus) sign in the exponent corresponding to R (L) for $\omega > 0$.%
\footnote{The choice of sign in the definition of the complex strain as $h_+ - ih_\times$ matches the convention of the Fourier transform in Eq.~\eqref{eq:ft} in order to make it so that $\exp(-i|\omega| t)$ encodes a right-handed mode as defined in the source-based convention.}

Using this fact, an economic way of expressing the information in Eq.~\eqref{eq:planewave} for any given direction of propagation $\hat{k}$ is to write the time-domain complex strain at the spatial origin ($\vec{x}=0$) as a Fourier integral of the form
\begin{align} \label{eq:hcomp_fd}
H(t) = \frac{1}{2\pi} \int_{-\infty}^{+\infty} \tilde{H}(\omega)\, e^{-i \omega t} \,\infd \omega \, ,
\end{align}
where the complex-valued Fourier amplitudes are defined by $\tilde{H}(\omega) \equiv \int (h_+ - i h_\times) \exp(i\omega t)\, \infd t = \tilde{h}_+(\omega) - i \tilde{h}_\times(\omega)$, following our Fourier transform convention in Eq.~\eqref{eq:ft}.
Unlike in Eq.~\eqref{eq:planewave}, it is clear that these Fourier amplitudes will not generally satisfy the symmetry $\tilde{H}(-\omega) = \tilde{H}^*(\omega)$, since the quantity on the left hand side of Eq.~\eqref{eq:hcomp_fd} is not real-valued unless the wave is fully plus-polarized.

In fact, given the interpretation of $\exp(\pm i \omega t)$ discussed above, the positive (negative) frequency Fourier amplitudes in Eq.~\eqref{eq:hcomp_fd} must encode contributions from the R-polarized (L-polarized) portion of the waveform.
This becomes obvious if we note that, by Eq.~\eqref{eq:circ_amps} and the definition of $\tilde{H}$, it must be the case that $\tilde{H}(\omega) = \sqrt{2}\, \tilde{h}_R (\omega) = \sqrt{2} \tilde{h}_L^*(-\omega)$, the last equality being due to Eq.~\eqref{eq:sym_circular}.
We can leverage this to rewrite Eq.~\eqref{eq:hcomp_fd} as an integral restricted to positive frequencies,
\begin{equation} \label{eq:hcomp_fd_rl}
H(t) = \frac{1}{\sqrt{2\pi^2}} \int_{0}^{\infty} \left[ \tilde{h}_R(\omega)\, e^{-i \omega t} + \tilde{h}_L^*(\omega)\, e^{i \omega t}\right] \infd \omega \, .
\end{equation}
This expression carries the same information as Eq.~\eqref{eq:planewave} without any redundancies.

Equation \eqref{eq:hcomp_fd_rl} lends itself to a straightforward physical interpretation.
Any plane GW, with arbitrary time evolution and polarization state (including unpolarized states), can be expressed as a superposition of fully-polarized Fourier modes;
each such monochromatic mode of frequency $|\omega|$ is made up of two counterrotating circularly-polarized contributions (R and L, the two summands) that add up to a single elliptically polarized mode.
Such elliptical, or \emph{fully-polarized}, modes are thus of fundamental importance; we discuss their properties in detail below, beginning with modes of a definite frequency as they appear in \eq{hcomp_fd_rl}.

\section{Elliptical modes}
\label{sec:ellip_modes}

\begin{figure}
\includegraphics[width=0.65\columnwidth]{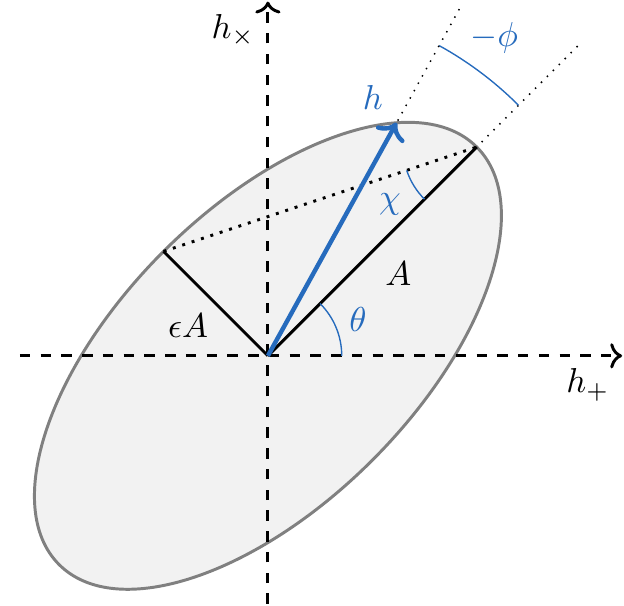}
\caption{Polarization ellipse. Following \eq{hcomp_ellip}, at any given time, the phasor (blue arrow) of an elliptically polarized signal lies on an ellipse with some maximum amplitude $A$ and ellipticity $\epsilon$, with semimajor axis tilted by an angle $\theta$ with respect to the plus-polarization axis (abscissa); a second angle, $\phi$, determines the initial location of the phasor within the ellipse.
The shape of the ellipse can also be parametrized in terms of the angle $\chi \equiv \arctan \epsilon$.
}
\label{fig:ellipse}
\end{figure}

\subsection{Monochromatic modes}
\label{sec:ellip:mono}

\subsubsection{Morphology}
\label{sec:ellip:mono:morph}

Elliptical GWs define an ellipse in the $\left(h_+, h_\times\right)$ phasor space (Fig.~\ref{fig:ellipse}).
We can see this explicitly for the Fourier modes in Eq.~\eqref{eq:hcomp_fd_rl} above by considering a monochromatic signal given by $\tilde{h}_{R/L}(\omega) = \pi\, \delta(\omega-\omega_0)\, C_{R/L} $,
isolating a single Fourier mode of frequency $\omega_0 >0$ and complex-valued amplitudes $\pi\, C_{R/L}$.
\added[comment=FN]{As in Eq.~\eqref{eq:circ_example}, this choice of prefactor can be motivated by noting that $\tilde{h}_{R/L}(\omega) = 2\pi \, \delta(\omega - \omega_0)\, C_{R/L}/2 = \delta(f - f_0)\, C_{R/L} / 2$ implies that $C_{R/L}/2$ are amplitude densities with respect to the frequency $f \equiv \omega/2\pi$; the additional factor of $1/2$ normalizes the signal power such that $h_+^2 + h_\times^2 = 1$ for $|C_R|=1, |C_L|=0$ or $|C_R|=0, |C_L|=1$.}

For such a signal, the result of the Fourier integral of $H(t) \equiv h_+ - i h_\times$ is then (relabeling $\omega_0 \to \omega$ after integration)
\begin{align} \label{eq:ellip_circ}
H(t) =\frac{1}{\sqrt{2}} \left( C_R\, e^{-i \omega t} + C^*_L\, e^{i\omega t}\right)\, ,
\end{align}
for complex amplitudes $C_{R/L} \equiv A_{R/L} \exp(i\phi_{R/L})$, where $A_{R/L}$ and $\phi_{R/L}$ are real valued.
Without loss of generality, the above expression can be refactored into%
\footnote{This is the same parametrization we defined in \cite{Isi:2021iql} up to a factor of $\sqrt{2}$ in the circular polarization amplitudes.}
\begin{align}
H(t) = \frac{1}{2}A\hspace{-2pt}\left[ \left(1+\epsilon\right) e^{-i (\omega t - \phi_R)} + \left(1-\epsilon\right) e^{i (\omega t - \phi_L)} \right]\hspace{-3pt} .
\end{align}
Here 
$A \equiv (A_R + A_L)/ \sqrt{2}$ is the peak amplitude of the mode, and $\epsilon = (A_R - A_L)/(A_R + A_L)$ is its ellipticity.
With some trigonometry, it is easy to show that this corresponds to linear polarization quadratures given by
\begin{subequations} \label{eq:hcomp_ellip}
\beq
h_+ = A \left[\cos \theta \cos(\omega t - \phi) - \epsilon \sin \theta \sin(\omega t - \phi)\right] ,
\eeq
\beq
h_\times = A \left[\sin \theta \cos(\omega t - \phi) + \epsilon \cos \theta \sin(\omega t - \phi)\right] ,
\eeq
\end{subequations}
with $\phi \equiv (\phi_L + \phi_R)/2$ and $\theta \equiv (\phi_L - \phi_R)/2$.%
\footnote{Since $\phi_{R/L}$ are $2\pi$-periodic, the most generic relation between them and $\{\theta$, $\phi\}$ is actually $\theta = [\phi_L - \phi_R  + 2\pi (k-j)]/2 \mod 2\pi$ and $\phi = [\phi_L + \phi_R  + 2\pi (k+j)]/2 \mod 2\pi$ for any integers $k,j$. \label{foot:angles}}
In the $\left(h_+,h_\times\right)$ plane, this defines an ellipse with semimajor axis $A$ and semiminor axis $\epsilon A$, oriented so as to subtend an angle $\theta$ between the semimajor axis and the $h_+$ axis, and with an initial location around the ellipse given by $-\phi$ (Fig.~\ref{fig:ellipse}).
The total power in this mode is given by the square of the \emph{intensity amplitude}, which we define as 
\begin{equation}
  \hat{A} \equiv \sqrt{A_R^2 + A_L^2} = A \sqrt{1 + \epsilon^2}\, .
\end{equation}

\begin{figure*}
\includegraphics[width=0.8\textwidth]{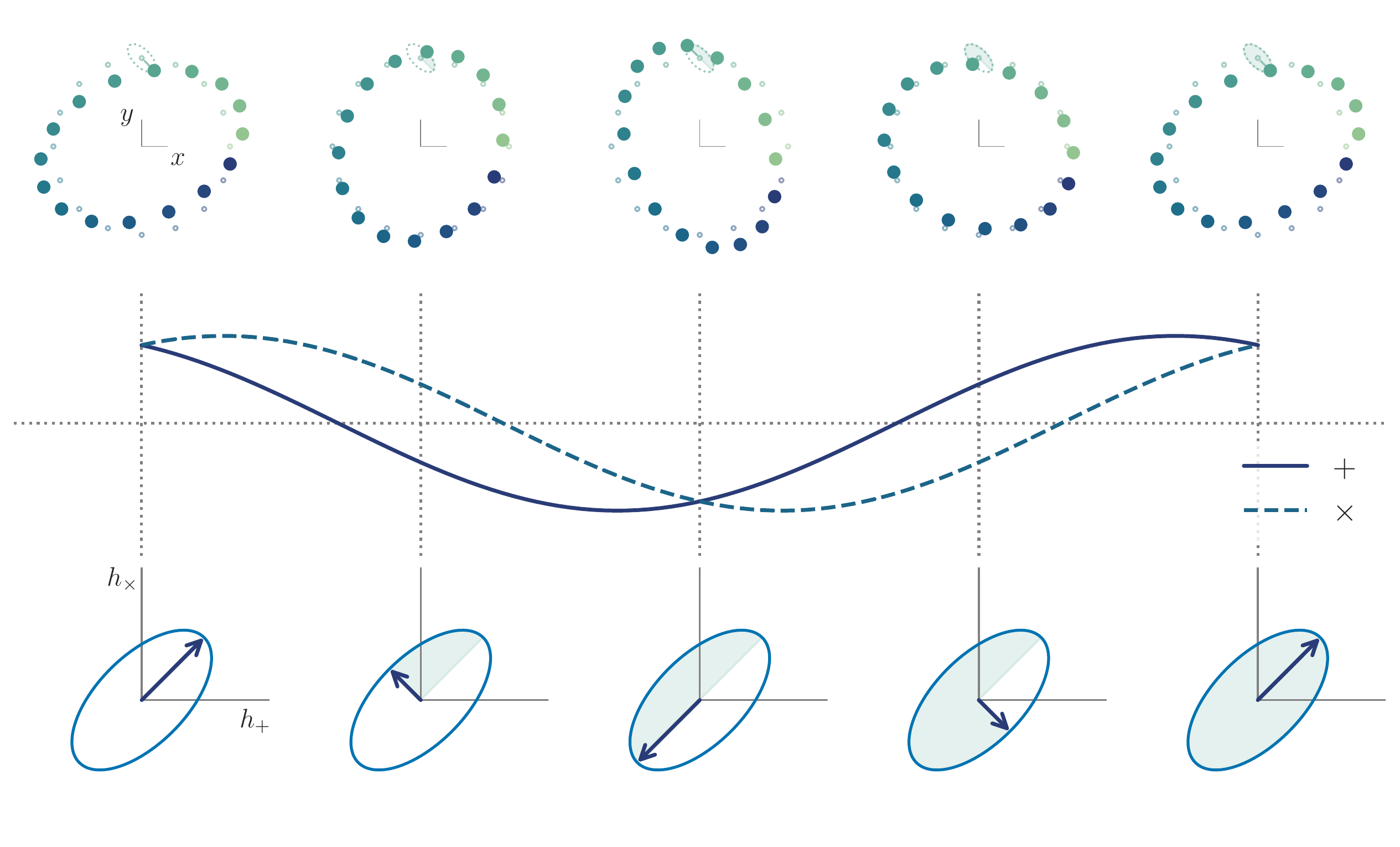}
\caption{An elliptically polarized GW as a function of time (left to right) over a period, described by Eq.~\eqref{eq:hcomp_ellip} with $\epsilon=1/2$, $\theta=\pi/4$, $\phi=0$ and arbitrary amplitude. \emph{Top:} as the wave propagates out of the page, it deforms a freely falling ring of particles (colored dots) into an ellipsoidal pattern, which in this case rotates counterclockwise with time, albeit nonrigidly; each individual particle is pushed in an ellipse with the same ellipticity as the wave itself, and oriented at an angle $\theta + \varphi$ from the $x$-axis, where $\varphi$ is the polar coordinate locating the particle around the ring.
\emph{Middle:} amplitudes of the plus (solid) and cross (dashed) linear polarizations making up the wave as a function of time.
\emph{Bottom:} representation of the polarization state as a phasor in the $+$ and $\times$ space; the phasor rotates counterclockwise as in Fig.~\ref{fig:ellipse}.
(See Ref.~\cite{release} for accompanying animations.)
}
\label{fig:pol_diagram_ellip}
\end{figure*}

Equation \eqref{eq:hcomp_ellip} encapsulates all possible morphologies of a monochromatic, fully polarized wave.
As special cases, $\epsilon = +1$ ($\epsilon = -1$) encodes an R (L) circularly-polarized wave, while $\epsilon =0$ encodes a $+$ ($\times$) linearly-polarized wave if $\theta = 0,\pi$ ($\theta = \pm \pi/2$);
an example in between, with $\epsilon=1/2$ and $\theta = \pi/2$, is illustrated in Fig.~\ref{fig:pol_diagram_ellip} (compare to Fig.~\ref{fig:pol_diagram_circ}, where $\epsilon=1$).
Each Fourier component in Eq.~\eqref{eq:hcomp_fd_rl} is a fully polarized mode of this kind, with ellipticity determined by the relative magnitudes of $\tilde{h}_{R/L}(\omega)$, and ellipse orientation determined by the difference in their Fourier phases, through $\theta = (\arg \tilde{h}_L - \arg \tilde{h}_R)/2$.
\added[comment=FN]{Since we use ``elliptical'' generically to also encompass circular and linear polarizations as special cases, ``elliptical'' and ``fully polarized'' are synonyms in this sense.}

\newcommand{\jonesbasis}{\vec{\mathfrak{e}}}

The domain for the parameters in \eq{hcomp_ellip} is $A \geq 0$ for the amplitude, $-1 \leq \epsilon \leq 1$ for the ellipticity, and $0 \leq \theta < 2\pi$ and $0 \leq \phi < 2\pi$ for the two phase angles (or, equivalently, $-\pi \leq \theta < \pi$ and $-\pi \leq \phi < \pi$).
However, allowing $\theta$ and $\phi$ to vary freely over this range results in a double covering of the waveform space; this is because the template is invariant under the addition or subtraction of $\pi$ to both $\theta$ and $\phi$, i.e., under the transformations $\{\theta, \phi\} \to \{\theta \pm \pi, \phi \pm \pi\}$, for any combination of plus and minus signs
The existence of this degeneracy is easy to infer from Fig.~\ref{fig:ellipse}, and can be traced back to the property discussed in footnote \ref{foot:angles} in relation to $\phi_{R/L}$.
Within the $[-\pi, \pi]$ branch cut, the $\{\theta, \phi\}$ space can therefore be restricted to a diamond bounded by the four diagonals satisfying $|\phi| = \pi \pm \theta$.
This comes into play in practice when translating between probability densities obtained under different parametrizations, as we do in Sec.~\ref{sec:jacobians} (see in particular Fig.~\ref{fig:jac_Aeps_Arl_angles}).

The requirement that $\theta$ extend all the way up to $2\pi$ (or $\pm \pi$) arises from our definition of the phase angle $\phi$ with respect to the semimajor axis of the ellipse (see Fig.~\ref{fig:ellipse}).
Fundamentally, however, $\theta$ need only be specified over half that range in order to determine the \emph{orientation} of the ellipse, disregarding the signal phase.
Indeed, if we instead chose to work in terms of a phase angle $\bar{\phi} \equiv \theta - \phi = -\phi_R$ measured counterclockwise from the $h_+$ axis (and thus decoupled from $\theta$), \eq{hcomp_ellip} would become
\begin{subequations} \label{eq:hcomp_ellip_2th}
\beq
h_+ = \frac{A}{2} \left[\left(1+\epsilon\right) \cos(\omega t + \bar{\phi}) + \left(1 - \epsilon\right) \cos(\omega t + \bar{\phi} - 2\theta) \right] ,
\eeq
\beq
h_\times = \frac{A}{2} \left[\left(1 + \epsilon\right) \sin(\omega t + \bar{\phi}) -\left(1-\epsilon\right) \sin(\omega t + \bar{\phi} - 2\theta) \right] ,
\eeq
\end{subequations}
where now $\theta$ only enters the template as $2\theta$, and so $0 \leq \theta < \pi$ (or $-\pi/2 \leq \theta < \pi/2$) spans the full space of waveforms, with the initial state set freely by $0 \leq \bar{\phi} < 2\pi$.

We can obtain another useful parametrization for fully polarized states by replacing the ellipticity parameter $\epsilon$ in \eq{hcomp_ellip} with an angle $\chi \equiv \arctan \epsilon$, which is also illustrated in Fig.~\ref{fig:ellipse}.
In terms of this quantity and the intensity amplitude $\hat{A}=A\sqrt{1+\epsilon^2}=A \sec\chi$, the elliptical mode of \eq{hcomp_ellip} becomes
\begin{subequations} \label{eq:hcomp_ellip_chi}
\beq
h_+ = \hat{A} \left[\cos\chi \cos \theta \cos(\omega t - \phi) - \sin\chi \sin \theta \sin(\omega t - \phi)\right] ,
\eeq
\beq
h_\times = \hat{A} \left[\cos\chi \sin \theta \cos(\omega t - \phi) + \sin\chi \cos \theta \sin(\omega t - \phi)\right] ,
\eeq
\end{subequations}
Now, $\chi = 0$ gives a linearly polarized state, while $\chi=\pm \pi/4$ gives a R/L circularly polarized state.
Its domain is given by $-\pi/4 \leq \chi \leq \pi/4$, as implied by $-1 \leq \epsilon \leq 1$.

\subsubsection{Mathematical framework}
\label{sec:math}

The mathematical treatment of polarized GW states is entirely analogous to the electromagnetic case.
To start, any of these states can be represented graphically by a series of phasor diagrams like the one in Fig.~\ref{fig:ellipse}, as in the bottom of Figs.~\ref{fig:pol_diagram_circ} and \ref{fig:pol_diagram_ellip}.
For monochromatic modes (i.e., of a definite frequency $\omega$), the same information can also be encoded algebraically in a complex valued \emph{Jones vector} $\vec{C}$ like 
\beq \label{eq:jones}
\begin{pmatrix}
h_+\\
h_\times
\end{pmatrix} \equiv
\Re \left[ \begin{pmatrix}
C_+\\
C_\times
\end{pmatrix} e^{-i\omega t}\right] \equiv
\Re \left[ \vec{C}\, e^{-i\omega t}\right] ,
\eeq
with $C_{+/\times} \equiv A_{+/\times} \exp(i\phi_{+/\times})$.
In that notation, $ \jonesbasis_+ \equiv \left(1, 0\right)$ encodes a unit-amplitude linearly polarized $+$ mode, and $\jonesbasis_\times \equiv \left(0,1\right)$ a $\times$ mode; meanwhile, the vectors $\jonesbasis_{R/L} \equiv \left(1,\pm i\right)/\sqrt{2}$ encode circular R/L modes, with the plus sign for R.
Thus, the generic signal in \eq{jones} can be equally conveyed by
\beq \label{eq:jones_bases}
\vec{C} = C_+ \, \jonesbasis_+ + C_\times \, \jonesbasis_\times = C_R \, \jonesbasis_R + C_L \, \jonesbasis_L\, ,
\eeq
with $C_{R/L} = (C_+ \mp i C_\times)/\sqrt{2}$ the same complex amplitudes as in \eq{ellip_circ}---although note that here $C_L$ appears without conjugation.
We will briefly make use of Jones vectors to facilitate coordinate transformations below.

Considering the parametrization in \eq{hcomp_ellip_chi}, we have two angles that fully define the shape of the polarization ellipse, $\chi$ and $\theta$.
If we interpret $-\pi/2 \leq 2\chi \leq \pi/2$ and $0 \leq 2\theta \leq 2\pi$ respectively as latitude and longitude coordinates, then the space of all unique polarization states can be arranged into a sphere such that linear polarization states of different orientations live on the equator ($\chi = 0$), and circular states live on the poles ($2\chi = \pm \pi/2$) \cite{poincare,goldstein}.
Any two antipodal states in this so-called \emph{Poincar\'e sphere} can function as a \replaced{basis for polarization states}{polarization basis}.
In this language, reexpressing Eq.~\eqref{eq:planewave} as Eq.~\eqref{eq:planewave_circ} amounted to effecting a Poincar\'{e} rotation of our basis vectors.
The polarization ellipse (Fig.~\ref{fig:ellipse}) can be recovered from the Poincar\'e sphere by a stereographic projection.

If we scale the radius of the Poincar\'{e} sphere to be the signal intensity $I \equiv \hat{A}^2$, then it can be defined in terms of Cartesian coordinates corresponding to the three other \emph{Stokes parameters} that characterize the distribution of power in the signal accross different polarization states \cite{Anile1974}.
For a fully polarized monochromatic mode, in addition to $I$ itself, these are given by
\begin{subequations} \label{eq:stokes}
\beq
Q \equiv |C_+|^2 - |C_\times|^2 = \hat{A}^2 \cos 2\chi \cos 2\theta \, , 
\eeq
\beq
U \equiv C_+ C_\times^* + C_+^* C_\times = \hat{A}^2 \sin2\chi \sin 2\theta \,  ,
\eeq
\beq
V \equiv |C_R|^2 - |C_L|^2 = \hat{A}^2 \sin 2\chi \, ,
\eeq 
\end{subequations}
for $C_+ = (C_R + C_L)/\sqrt{2}$ and $C_\times = i (C_R - C_L)/\sqrt{2}$.
As implied by the definitions above, $Q/I$ controls the (power) fraction of linear polarization, $U/I$ the orientation of the linear component, and $V/I$ the fraction of circular polarization.
The Poincar\'{e} sphere is then the sphere of radius $I$ centered on $\left(Q=0, U=0, V=0\right)$.

For a fully polarized state, the Stokes parameters (quantifying signal power) are equivalent to the polarization quantitites $\left\{A, \epsilon, \theta\right\}$ or $\{\hat{A}, \chi, \theta\}$ defining the ellipse in Fig.~\ref{fig:ellipse} (and quantifying signal amplitude).
Because they are defined in terms of power, Stokes parameters do not retain phasing information, but have the advantage of being easily generalizable to fully or partially unpolarized waves, which can be achieved by replacing the definition in Eq.~\eqref{eq:stokes} with corresponding two-point correlation functions (power spectra); in the fully-unpolarized case, $Q=U=V=0$ and there is no Poincar\'{e} sphere to speak of.
The Stokes parameters are thus especially useful when dealing with stochastic signals \cite{Romano:2016dpx,Conneely:2018wis,Seto:2008sr,Kato:2015bye}; since we will be dealing mainly with phase-coherent signals, we will not make further reference to Stokes parameters in what follows.

\subsection{Non-monochromatic modes}
\label{sec:ellip:gen}

We arrived at the expression for a fully-polarized, monochromatic GW in Eq.~\eqref{eq:hcomp_ellip} by way of the generic Fourier decomposition of a plane wave in Eq.~\eqref{eq:hcomp_fd_rl}, wherein elliptical modes appear naturally with a determinate frequency.
Yet, we may also speak of fully-polarized states even if the signal is not monochromatic.

The argument applies to any high-frequency coherent wave, i.e., any signal that can be written as a slow-varying amplitude modulating a fast phase.%
\footnote{In signal processing terminology, signals with such morphologies are often denoted ``chirps'' \citep[e.g.,][]{Candes:2006yv}; however, we avoid that nomenclature here to avoid confusion with compact binary chirps, to which this discussion applies but is not restricted.}
In that case, the polarization parameters $\{A, \epsilon, \theta\}$ can be defined instantaneously using the stationary phase approximation or similar procedures.
This way, any GW with a constant polarization state, i.e., whose polarization ellipse takes a fixed, determinate shape (but not necessarily scale), can be encapsulated by an expression of the form
\begin{subequations} \label{eq:ellip_gen}
\begin{equation} %
h_+ = \mathcal{A}(t) \left[\cos \Phi(t) \cos \theta - \epsilon \sin \Phi(t) \sin\theta \right] ,
\end{equation}
\begin{equation} %
h_\times = \mathcal{A}(t) \left[ \cos \Phi(t) \sin \theta + \epsilon \sin \Phi(t) \cos\theta \right] ,
\end{equation}
\end{subequations}
enhancing Eq.~\eqref{eq:hcomp_ellip} with a (slowly) time varying amplitude $\mathcal{A}(t)$ and a (quickly) time varying phase $\Phi(t)$, which need no longer grow linearly with time.
Following this expression, the aspect ratio and orientation of the polarization ellipse remains constant, while its size may increase or decrease according to $\mathcal{A}(t)$.
The initial state of the signal is defined by the initial amplitude $A = \mathcal{A}(t=0)$ and phase $\phi = \Phi(t=0)$.
\added[comment=FN]{The shape of the ellipse could also be made to vary adiabatically via $\epsilon$ and $\theta$ but that is seldomly done in real-world applications.}

Most conceivable signals are neither monochromatic nor fully polarized.
Nevertheless, a large variety of morphologies can be captured by a finite superposition of elliptically polarized modes, potentially with time-varying polarization parameters as above.
This should be apparent from the fact that an (uncountably) \emph{infinite} set of elliptical modes can describe \emph{any} GW signal, as we showed in \eq{hcomp_fd_rl}.
For many practical applications, it is advantageous to decompose signals into 
sums of fully-polarized modes in the shape of \eq{ellip_gen},
\begin{subequations} \label{eq:ellip_sum}
\begin{equation} \label{eq:ellip_sum_p}
h_+ = \sum \mathcal{A}_n(t) \hspace{-1pt} \left[\cos \Phi_n(t) \cos \theta_n - \epsilon_n \sin \Phi_n(t) \sin\theta_n \right] ,
\end{equation}
\begin{equation} \label{eq:ellip_sum_c}
h_\times = \sum \mathcal{A}_n(t) \hspace{-1pt} \left[ \cos \Phi_n(t) \sin \theta_n + \epsilon_n \sin \Phi_n(t) \cos\theta_n \right] ,
\end{equation}
\end{subequations}
with a sum over some number of modes indexed by $n$, with amplitudes and phases taking some prescribed functional form for each $n$.

The form of Eq.~\eqref{eq:ellip_sum} is flexible enough that it can be used in practice to model arbitrary signals in real detector data.
For example, that is the strategy taken by \textsc{BayesWave} \cite{Cornish:2014kda,Cornish:2020dwh}, which reconstructs generic GW signals by fitting a variable number of elliptically-polarized sine-Gaussians.%
\footnote{\textsc{BayesWave} can currently operate in two configurations: one which assumes the overall signal is elliptically polarized, and another which does not.}
It is also the case, in ringdown studies that fit the final portion of a compact binary signal as a superposition of elliptically polarized damped sinusoids \cite{Isi:2021iql}.

For such applications, each phasing function will usually correspond to some given frequency $\omega_n$ as in a Fourier expansion, so that $\Phi_n(t) = \omega_n t + \phi_n$; meanwhile, the $\mathcal{A}_n(t)$ functions encode amplitude envelopes evolving slowly over some timescale $\tau_n \equiv 1/\gamma_n$ (or, equivalently, with some quality factor $Q_n \equiv \omega_n \tau_n/2$).
For example, in the case of ringdown templates, $\mathcal{A}_n(t) = A_n \exp(-\gamma_n t)$ and $\Phi_n(t) = \omega_n t + \phi_n$, for some set of frequencies and damping rates to be inferred from the data together with polarization parameters $\{ A_n, \epsilon_n, \theta_n, \phi_n\}$.
Equations \eqref{eq:ellip_sum} can be equivalently written in the frequency domain, as done for the sine-Gaussian basis in \cite{Cornish:2014kda,Cornish:2020dwh}.

The elliptical decomposition of Eq.~\eqref{eq:ellip_sum} allows us to flexibly model a GW signal without assuming full independence of the two GW polarizations.
This is justified because, as argued in \cite{Chatziioannou:2021mij}, we expect both polarizations to be generated by the same physical processes, so that their spectral properties should not be totally independent.
Moreover, even if there was a choice of waveframe in which the two linear polarizations looked completely dissimilar, the polarizations will look spectrally similar to generic observers whose frame is randomly oriented (see the discussion of polarization mixing in Sec.~\ref{sec:angles} below).

Besides the modeling of generic signals, \eq{ellip_sum} serves as the exact representation of several classes of astrophysically-relevant signals.
The most salient example of this, as we will see below, is that of CBCs; in particular, a nonprecessing, quasicircular CBC dominated by the quadrupolar angular harmonic of the radiation can be described by a single, fully polarized component, as in \eq{ellip_gen}.
More generally, the signal from a precessing CBC is well represented by the superposition of five fully polarized modes \cite{Fairhurst:2019vut}.

\subsection{Relation to spherical harmonics}
\label{sec:harmonics}

When modeling specific sources (e.g., in a numerical-relativity simulation), it is common to decompose the outgoing strain in terms of spin-weighted spherical harmonics ${}_{-2} Y_{\ell m}$ in the frame of the source (e.g., \cite{Kidder:2007rt}), so that, for a detector infinitely far away, we can write
\begin{align} \label{eq:spherical}
H(t) = \sum_{\ell \geq 2} \sum_{-\ell \leq m \leq \ell} H_{\ell m}(t)\, {}_{-2}Y_{\ell m} (\iota, \varphi)\, ,
\end{align}
for a source seen with inclination $\iota$ and azimuthal angle $\varphi$, with intrinsic time-dependence encoded in the $H_{\ell m}$ functions as determined by Einstein's equations.
The decomposition into spherical harmonics presumes the choice of both (1) a polar frame defining $\iota$ and $\varphi$, and (2) an orientation of the waveframe vectors with respect to the direction of propagation to establish the meaning of $h_{+/\times}$ as in \eq{hij}.
In the LIGO-Virgo convention (which follows \cite{Blanchet:2008je,Faye:2012we}), the waveframe in \eq{spherical} is defined by $\hat{x} = -\hat{e}_\iota$ and $\hat{y} = - \hat{e}_\varphi$ \cite{LALSuite:source}, and the overall polar frame is centered on and comoving with the source, with an orientation respecting its symmetries (e.g., aligned with the orbital plane).

The different $H_{\ell m}$'s in \eq{spherical} are generated by the time evolution of specific current and mass moments of the source \cite{Thorne:1980ru}.
As such, their structure must inherit the symmetries of Einstein's equations, including parity.
In particular, for any source satisfying equatorial-reflection (planar) symmetry, like a nonprecessing inspiral, parity can be shown to imply that $H_{\ell -m} = (-1)^\ell H_{\ell m}^*$ \cite{Faye:2012we}, assuming that the coordinates in \eq{spherical} are oriented such that $\iota=\pi/2$ is the plane of symmetry.
Allowing for a generic (slow) amplitude and (fast) phase evolution by writing $H_{\ell m}(t) = \mathcal{A}_{\ell m}(t) \exp[-i \Phi_{\ell m}(t)]$, this symmetry reduces to $\mathcal{A}_{\ell -m}(t) = (-1)^\ell \mathcal{A}_{\ell m}(t)$ and $\Phi_{\ell m}(t) = - \Phi_{\ell -m}(t)$, where we have taken $\mathcal{A}$ and $\Phi$ to be real valued.
With that ansatz, \eq{spherical} can be rewritten with an explicit term for negative values of $m$ (and double counting $m=0$ modes) as
\begin{widetext}
\begin{subequations} \label{eq:spherical_modes}
\begin{align}
H(t) &= \sum_{\ell \geq 2} \sum_{0\leq m \leq \ell} \left[H_{\ell m}(t)\, {}_{-2}Y_{\ell m} (\iota, \varphi) + H_{\ell -m}(t)\, {}_{-2}Y_{\ell -m} (\iota, \varphi) \right] \\
&= \sum_{\ell \geq 2} \sum_{0\leq m \leq \ell} \left[\mathcal{A}_{\ell m}(t)\, e^{-i\Phi_{\ell m} (t)} {}_{-2}Y_{\ell m}(\iota, \varphi) +  \mathcal{A}_{\ell m}(t)\, e^{i\Phi_{\ell m} (t)} {}_{-2}Y_{\ell m}^*(\pi-\iota, \varphi) \right] \\
&= \sum_{\ell \geq 2} \sum_{0\leq m \leq \ell} \left[\mathcal{C}_{\ell m}(t)\, e^{-i\Phi_{\ell m} (t)}  +  \mathcal{C}_{\ell -m}(t)\, e^{i\Phi_{\ell m} (t)} \right]\, ,
\end{align}
\end{subequations}
\end{widetext}
for some overall complex-valued amplitudes $\mathcal{C}_{\ell \pm m}$, which absorb the angular dependence of the spherical harmonics and any potential (slow) time variation in $\mathcal{A}_{\ell m}$.
In the second line above, we took advantage of the identity relating spherical harmonics for different signs of $m$, ${}_{-2} Y_{\ell -m}(\iota,\varphi) = (-1)^{\ell} {}_{-2} Y_{\ell m}^*(\pi-\iota,\varphi)$ \cite{goldberg:1967}.

The summand in the last line of \eq{spherical_modes} takes the form of \eq{ellip_circ}, and its interpretation is the same for any fixed observation direction: each $(\ell$, $|m|)$ angular harmonic contributes a single, elliptically polarized mode to the waveform, composed of right- and left-handed pieces corresponding to the $m>0$ and $m<0$ modes respectively.
Thus the overall strain for such a source must be a superposition of purely polarized modes, with adiabatically evolving amplitudes as in \eq{ellip_sum}.

The amplitude and ellipticity of each mode are determined by a combination of the intrinsic amplitudes $\mathcal{H}_{\ell \pm m}$, and the viewing angle $(\iota, \varphi)$---the latter through the ${}_{-2} Y_{\ell m}$ factors.
The intensity of the mode will vary with time following $\mathcal{A}_{\ell m}(t)$; meanwhile, its ellipticity, as observed from a given $\iota$ and $\varphi$, will be fixed by the relative amplitudes of the $\pm|m|$ spherical harmonics,
\begin{align}
\epsilon_{\ell|m|}(\iota) = \frac{\left|{}_{-2} Y_{\ell m}(\iota,\varphi)\right| - \left|{}_{-2} Y_{\ell -m}(\iota,\varphi)\right|}{\left|{}_{-2} Y_{\ell m}(\iota,\varphi)\right| + \left|{}_{-2} Y_{\ell -m}(\iota,\varphi)\right|} ,
\end{align}
which is exclusively a function of the inclination $\iota$, because $\varphi$ only affects the phase (not the magnitude) of the spin-weighted spherical-harmonic factors, with ${}_{-2} Y_{\ell m}(\iota,\varphi) = {}_{-2} Y_{\ell m}(\iota) \exp(i m \varphi)$ factoring out the $\varphi$ dependence.

\begin{figure}
\includegraphics[width=\columnwidth]{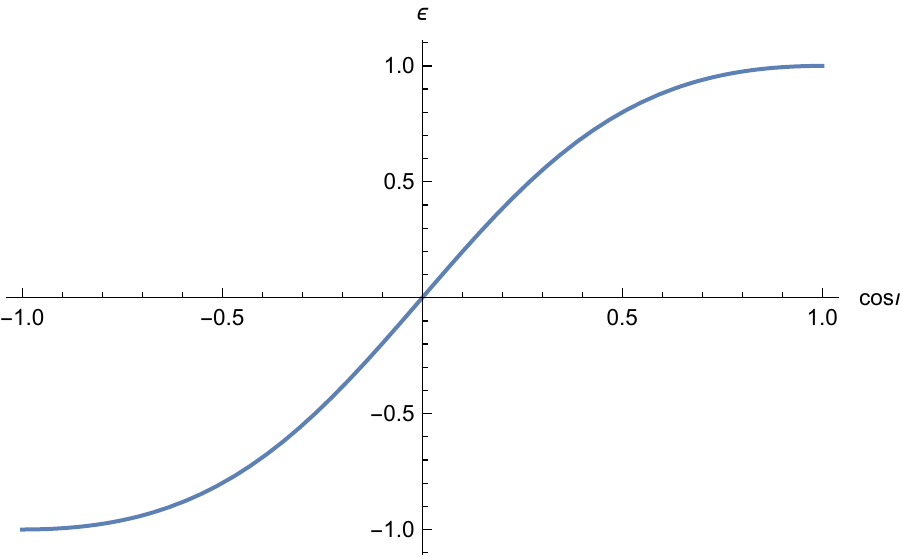}
\caption{Ellipticity ($\epsilon$, ordinate) as a function of the cosine of the inclination ($\cos\iota$, abscissa) for the $\ell = |m| = 2$ GW strain from a nonprecessing compact binary inspiral, Eq.~\eqref{eq:ellip_cosi}. The signal from a face-on (face-off) binary has ellipticity $\epsilon = +1$ ($\epsilon=-1$), meaning it has a right-handed (left-handed) circular polarization; an edge-on source has a linear polarization.}
\label{fig:ellip_cosi}
\end{figure}

The complex strain $H_{\ell|m|}(t)$ for a given elliptical $(\ell, |m|)$ mode, as given by the summand in \eq{spherical_modes}, can be further rewritten as
\begin{equation}
H_{\ell|m|}(t) = \mathcal{A}_{\ell m}(t) \left[ Y^+_{\ell m} \cos \Phi_{\ell m}'(t) - 
i Y^\times_{\ell m} \sin \Phi_{\ell m}'(t) \right],
\end{equation}
where we have defined $\Phi_{\ell m}'(t) \equiv \Phi_{\ell m}(t) - m \varphi$, and
\begin{equation}
Y_{\ell m}^{+/\times}(\iota) \equiv {}_{-2} Y_{\ell m}(\iota) \pm {}_{-2} Y_{\ell m}(\pi-\iota) \, ,
\end{equation}
with the plus (minus) sign for $+$ ($\times$), and noting that, after factoring out the $\varphi$ dependence, the ${}_{-2} Y_{\ell m}(\iota)$ quantities are real valued.
For the special case of the dominant $\ell=|m|=2$ mode, the strain $H_{\ell|m|} = h_+ - i h_\times$ thus reduces to
\begin{subequations} \label{eq:nonprecessing}
\beq
h_+ = \frac{1}{2} \sqrt{\frac{5}{4\pi}} \mathcal{A}_{22}(t) \left(1 + \cos^2\iota\right) \cos \Phi'_{22}(t) \, , 
\eeq
\beq
h_\times = \sqrt{\frac{5}{4\pi}} \mathcal{A}_{22}(t)  \cos\iota \sin \Phi'_{22}(t) \, ,
\eeq
\end{subequations}
as can be checked by computing explicit expressions for ${}_{-2} Y_{22}(\iota)$.
This is exactly of the form of \eq{ellip_gen}, with amplitude $\mathcal{A} = \sqrt{5/16\pi}\,\mathcal{A}_{22}\left(1+\cos^2\iota\right)$, ellipticity
\beq \label{eq:ellip_cosi}
\epsilon = \frac{2 \cos\iota}{1+\cos^2\iota}\, ,
\eeq
which we illustrate in Fig.~\ref{fig:ellip_cosi}, and $\theta = 0$.
The fact that $\theta = 0$ is a consequence of our special choice of coordinate frame in \eq{spherical}, which we constructed to reflect the symmetries of the planar source so that the equator is the plane of symmetry (we return to this point in Sec.~\ref{sec:position}).

The above results, Eqs.~(\ref{eq:spherical_modes}--\ref{eq:ellip_cosi}), hold only for sources with equatorial-reflection symmetry.
The GWs for more generic, precessing, sources will not generally be given by the superposition of fully polarized modes with constant ellipticity \cite{Buonanno:2002fy,Schmidt:2010it,Schmidt:2012rh,Ochsner:2012dj,Boyle:2011gg,Blanchet:2013haa,Lousto:2018dgd}.
However, some of such signals may be decomposed into elliptical modes with a slowly-evolving ellipticity; that is the case, for example, for the early stages of precessing compact binary inspirals, whose signal can be well approximated by \eq{nonprecessing} with a slowly varying inclination.

In some cases, nonplanar sources can also give rise to superpositions of fully polarized modes.
For example, this is the case for black-hole ringdown signals \cite{Vishveshwara:1970cc, Press:1971wr, Teukolsky:1973ha, Chandrasekhar:1975zza}, which can be written as a harmonic expansion similar to \eq{spherical},
\beq \label{eq:spheroidal}
H(t) = \sum_{\ell \geq 2} \sum_{-\ell \leq m \leq \ell} \sum_{n\geq 0} C_{\ell m n} e^{-i\tilde{\omega}_{\ell m n} t} {}_{-2}S_{\ell m n} (\iota, \varphi) ,
\eeq
for complex frequencies $\tilde{\omega}_{\ell m n} \equiv \omega_{\ell m n} - i/\tau_{\ell m n}$ indexed by the usual angular numbers $\ell$ and $m$, as well as an overtone number $n$, which orders modes of a given $(\ell, m)$ by decreasing damping time; the angular dependence is encoded in the spin-weighted spheroidal harmonics, ${}_{-2}S_{\ell m n}$ \cite{Teukolsky:1973ha,Press:1973zz,Leaver:1985ax,Berti:2005gp,Cook:2014cta}, which have replaced the spherical harmonics in \eq{spherical}.
Parity in this decomposition implies $\tilde{\omega}_{\ell m n} = -\tilde{\omega}_{\ell-m n}^*$; it can thus be shown that, for fixed $\iota$ and $\varphi$, \eq{spheroidal}, is equivalent to
\beq \label{eq:ringdown}
H(t) = \sum \left( C_{\ell m n}' e^{-i\omega_{\ell m n} t} + C_{\ell -m n}' e^{i\omega_{\ell m n} t} \right) e^{-t/\tau_{\ell m n}} ,
\eeq
where the $m$ sum is now restricted to nonnegative values, $0 \leq m \leq \ell$, and $C'_{\ell \pm m n}$ are redefined amplitudes absorbing angular factors.
Comparing to \eq{ellip_circ}, it is evident from \eq{ringdown} that the ringdown strain is made up from elliptically polarized components, with exponentially decaying amplitudes.
If the ringdown excitations had equatorial symmetry, then the initial amplitudes in \eq{spheroidal} would satisfy $C_{\ell -m n }= (-1)^{\ell} C^*_{\ell m}$, and the ellipticity of the observed modes would only be a function of the observing direction.
(See Sec.~IIA and Appendix B of \cite{Isi:2021iql} for an extended discussion.)

\section{Polarization angles}
\label{sec:angles}

\subsection{Wave-frame and the angle $\psi$}
\label{sec:pol}

\begin{figure}
  \includegraphics[width=\columnwidth]{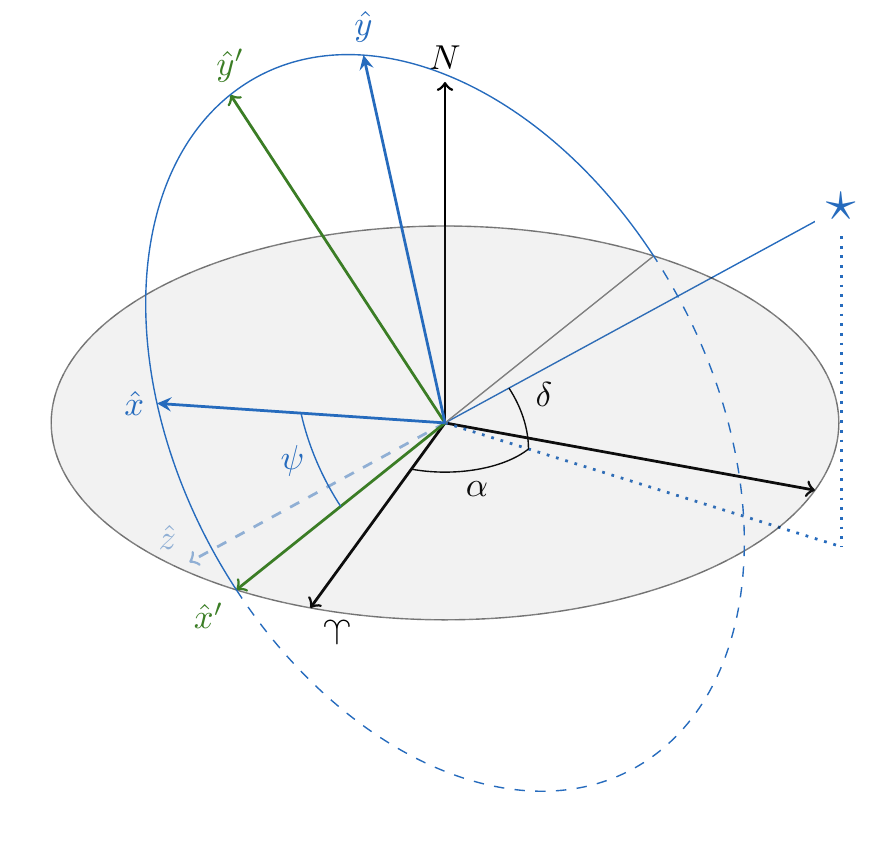}
  \caption{Standard construction of the polarization waveframe used in LIGO-Virgo data analysis \cite{LALSuite:wave,Anderson:T010110}. Earth sits at the origin of the equatorial coordinate system defined by the celestial north ($N$) and the vernal equinox ($\vernal$). The sky location of a source ($\star$) is encoded in its right ascension ($\alpha$) and declination ($\delta$), so that the wave propagation direction is $\hat{k}=\hat{z}=-(\cos\delta \cos\alpha, \cos\delta\sin\alpha, \sin\delta)$. The canonical polarization frame is defined by $\hat{x}'=(\sin\alpha, -\cos\alpha,0)$ as being the intersection between the plane of the sky (blue circle) and the celestial equator (gray circle), and $\hat{y}'=(-\sin\delta \cos\alpha, -\sin\delta\sin\alpha, \cos\delta)$ the projection of $N$ onto the plane of the sky, completing the right-handed basis with $\hat{z}$.
  In terms of these vectors, an arbitrary polarization frame rotated by some angle $\psi$ around $\hat{z}$ is given by $\hat{x}=\cos\psi\,\hat{x}' + \sin\psi\,\hat{y}'$ and $\hat{y}=-\sin\psi\,\hat{x}' + \cos\psi\,\hat{y}'$. Dashed trace indicates elements below the equator.}
  \label{fig:diagram_waveframe}
  \end{figure}

Equation \eqref{eq:hij} presumes a specific choice of frame orientation that defines the basis in which the $h_{ij}$ components are written and, therefore, the physical meaning of $h_{+}$ and $h_\times$.
Although \eq{hij} requires that $\hat{z}$ be parallel to the (spatial) wave vector $\vec{k}$, there is no a priori restriction on the orientation of the $x$ and $y$ axes within the plane perpendicular to $\vec{k}$.
This freedom is usually encapsulated in the choice of an arbitrary \emph{polarization angle} $\psi$, defined with respect to some convenient reference direction.
For instance, in the LIGO-Virgo convention, this angle is defined with respect to celestial coordinates such that $\psi=0$ means that the waveframe $\hat{x}$ is parallel to the celestial equator due west, and $\psi$ is measured following the right hand rule around $\hat{z}$ \cite{LALSuite:wave,Anderson:T010110}; we illustrate this in Fig.~\ref{fig:diagram_waveframe}.

With some trigonometry, it is straightforward to show that a \emph{clockwise}%
\footnote{This \emph{passive} clockwise rotation of the waveframe corresponds to an \emph{active} counterclockwise rotation of the polarization state.}
rotation of $\hat{x}$ and $\hat{y}$ by some angle $\Delta \psi$ around $\hat{z}$ leaves the form of \eq{hij} unchanged after redefining
\begin{subequations} \label{eq:htransf}
\beq
h_+ \rightarrow h_+' = h_+ \cos 2\Delta \psi - h_\times \sin 2\Delta\psi \, ,
\eeq
\beq
h_\times \rightarrow h_\times' = h_\times \cos 2\Delta \psi + h_+ \sin 2\Delta\psi \, .
\eeq
\end{subequations}
This contravariant transformation gives the polarization amplitudes that would be measured by an observer in the rotated (primed) frame, as a function of the amplitudes in the original frame.
The $2\Delta\psi$ dependence in \eq{htransf} reveals the fact that $h_+$ and $h_\times$ are nothing but the two components of a tensor field with spin-weight $|s|=2$, and the two polarizations are only defined up to an arbitrary choice of $\psi$.

Under clockwise rotations of the wave frame the antenna patterns of \eq{h} transform through an expression complementary to \eq{htransf},
\begin{subequations} \label{eq:Ftransf}
\beq
F_+ \rightarrow F_+' = F_+ \cos 2\Delta \psi - F_\times \sin 2\Delta\psi \, ,
\eeq
\beq
F_\times \rightarrow F_\times' = F_\times \cos 2\Delta \psi + F_+ \sin 2\Delta\psi \, ,
\eeq
\end{subequations}
ensuring that the observable $h(t)$ in \eq{h} is independent of the arbitrary angle $\psi$.
More generally, any scalar like $D^{ij} e^{p}_{ij}$ will necessarily be frame invariant.%
\footnote{This extends to gauge transformations: the spacetime tensors $D_{ab}$ and $h_{ab}$ are gauge dependent, but their inner product is not.}

Unlike the linear modes of Eq.~\eqref{eq:lin}, the tensors of Eq.~\eqref{eq:circ} do not mix under rotations around the direction of propagation:
the circular polarizations are eigenstates of the helicity operator with weight $\pm 2$, corresponding to the two helicities of a spin-2 massless particle (see, e.g., \cite{Hinterbichler2011}).
The equivalent transformation to Eq.~\eqref{eq:htransf} is
\begin{subequations} \label{eq:htransf_circ}
\begin{align}
h_R &\rightarrow h_R' = h_R \exp(- i2  \Delta \psi) \, ,\\
h_L &\rightarrow h_L' = h_L \exp(+ i2  \Delta \psi)\, ,
\end{align}
\end{subequations}
meaning that a rotation around $\hat{z}$ is equivalent to a simple change in the overall phase of the circular polarization components.
As such, a change in $\psi$ can be absorbed by a redefinition of the Fourier phases in \eq{hcomp_fd_rl}, multiplying the integral through by $\exp(-i2\Delta\psi)$.

Equations~(\ref{eq:htransf}--\ref{eq:htransf_circ}) allow us to transform predictions for the strain $h_{ij} = h_+ e^{+}_{ij} + h_{\times} e^{\times}_{ij}$ in some waveframe $\{\hat{x}, \hat{y}, \hat{z}\}$ to a different one $\{\hat{x}', \hat{y}', \hat{z}'\}$, rotated clockwise around $\hat{k}=\hat{z}=\hat{z}'$ by $\Delta \psi$ (simply labeled $\psi$, if the primed frame corresponds to the reference frame that defines $\psi=0$).
In real-world data analysis applications, however, we simply write the unprimed basis vectors in the primed basis and evaluate \eq{h} through numerical dot products using \eq{lin}.
To do this, we express the components of $e^{+/\times}_{ij}$ and $D_{ij}$ in a common basis suitably aligned with the reference waveframe---for ground-based detectors, where we take $\hat{x}'$ to be parallel to the celestial equator, these are equatorial celestial coordinates (Fig.~\ref{fig:diagram_waveframe}).
Knowing how the $\{\hat{x}', \hat{y}'\}$ vectors are expressed in such coordinates, we can construct $h_{ij}'$ by noting that $\hat{x} = \cos\psi\, \hat{x}' + \sin\psi\, \hat{y}'$ and $\hat{y}= - \sin\psi\, \hat{x}' + \cos\psi\, \hat{y}'$.
It is then straightforward to write the signal at any given detector in terms of the polarization amplitudes $h_{+/\times}$ computed in the original frame.

The definition of the angle $\psi$ (as in Fig.~\ref{fig:diagram_waveframe}) is not intrinsically related to any feature of the signal: it simply chooses an absolute reference direction that defines an arbitrary frame in which to prescribe the \replaced{$h_+(t)$ and $h_\times(t)$}{$h_+$ and $h_\times$} polarization functions in \eq{hij}, or equivalently, the frame in which to measure the phases of the circularly polarized Fourier components.
Nevertheless, even though any choice of assignment for $\psi$ is formally valid, specific signal morphologies may make some choices more convenient than others.

\subsection{Elliptical waves and the angle $\theta$}
\label{sec:theta}

\begin{figure}[tbp]
  \includegraphics[width=0.65\columnwidth]{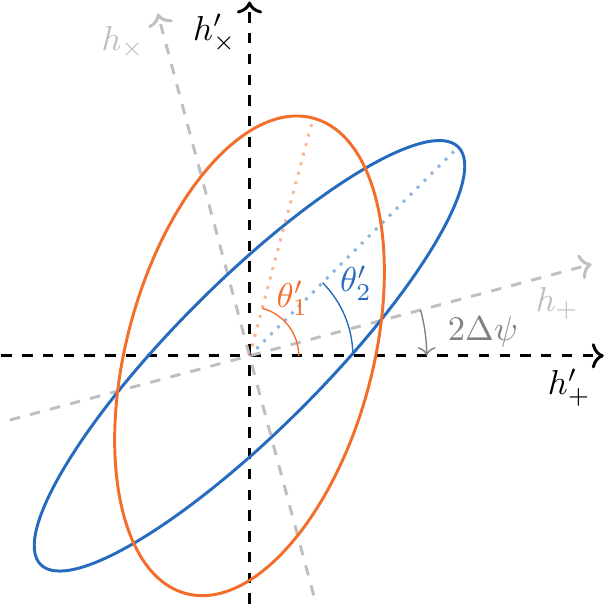}
  \caption{A counterclockwise rotation of the physical waveframe of Fig.~\ref{fig:diagram_waveframe} by an angle $\Delta\psi$ manifests as a rotation by $2\Delta\psi$ in the $(h_+,h_\times)$ polarization space, so that elliptical modes with original orientation angles $\theta_n$ are seen to have orientations $\theta_n'=\theta_n + 2\Delta\psi$ in the new (primed) frame.
  The diagram illustrates two such modes (blue and orange) with ellipse orientations $\theta_{1/2}$ (not labeled) with respect to $h_+$, and  $\theta_{1/2}'$ with respect to $h_+'$.
  }
  \label{fig:diagram_ellipse_extra}
  \end{figure}

Another notion of ``polarization angle'' arises naturally in the description of elliptically polarized signals.
The expression for an elliptical wave in \eq{ellip_gen} presumes some specific choice of $\psi$ that defines the meaning of plus vs cross by orienting $\hat{x}$, as explained in the previous section.
The expression simplifies if we choose that angle such that the plus and cross axes are aligned with the principal components of the ellipse, i.e., constructing the polarization frame to ensure that $\theta = 0$ (see Fig.~\ref{fig:ellipse}).

With such a choice of wave frame (equivalently, choice of $\psi$), \eq{ellip_gen} becomes just
\begin{subequations} \label{eq:ellip_frame}
\beq
h_+ = \mathcal{A}(t) \cos \Phi(t) \, ,
\eeq
\beq
h_\times = \epsilon \mathcal{A}(t) \sin \Phi(t)\, ,
\eeq
\end{subequations}
and we may simply read off the ellipticity $\epsilon$ as the ratio of the $\times$ to $+$ amplitudes.
Crucially, an elliptical wave will generally not take the form of Eq.~\eqref{eq:ellip_frame} unless $\hat{x}$ is chosen appropriately; only circularly polarized signals ($\epsilon=\pm1$) will take  this simplified form irrespective of the wave frame orientation (again showing that these are eigenstates of the helicity operator).

Given the above, when working with a single elliptically-polarized wave, Eq.~\eqref{eq:ellip_frame} defines a privileged orientation of the wave frame, unique up to rotations by $\pi/2$ around $\hat{z}$.
If we adopt $\theta =0$ as a convention (or, equivalently, $\theta=\pi$), then we \emph{define} our wave frame to lie along the principal axes of the polarization ellipse and, thus, the polarization angle $\psi$ becomes synonymous with the polarization ellipse orientation by construction.
However, the two angles $\theta$ and $\psi$ are conceptually distinct; in particular, $\theta$ is defined only for elliptically polarized waves, whereas $\psi$ is always defined.

As for any GW, the detector output for an elliptically polarized wave will be given by \eq{h}.
In this case, however, \eq{Ftransf} implies that $\psi$ and $\theta$ are degenerate, as detailed in Appendix A of \cite{Isi:2017equ}.
Concretely, for a fixed sky location (i.e., propagation direction), rotating the waveframe \emph{counterclockwise} around $\hat{z}$ results in a change from $\psi \to \psi' = \psi - \Delta\psi$ in the antenna patterns, which can be absorbed by a change in $\theta$.
This is because the expression for the strain at a given detector,
\beq
h = F_+(\psi - \Delta \psi)\, h_+ + F_\times(\psi - \Delta \psi)\, h_\times \, ,
\eeq
can be expanded by means of \eq{Ftransf} to read
\begin{align}
h = &\left[ F_+(\psi) \cos 2\Delta\psi + F_\times(\psi) \sin 2\Delta\psi \right] h_+\, + \nonumber \\
 &\left[F_\times(\psi) \cos 2\Delta\psi - F_+(\psi)\sin 2\Delta\psi\right] h_\times \, .
\end{align}
Plugging in the expressions for an elliptical wave in \eq{ellip_gen} and taking advantage of trigonometric identities, this can be rearranged into
\begin{widetext}
\begin{align} \label{eq:theta_psi}
h = & \mathcal{A}(t) \left[\cos \Phi(t) \cos(\theta + 2\Delta\psi) -  \epsilon \sin \Phi(t)\sin(\theta + 2\Delta\psi) \right] F_+(\psi) +\nonumber\\
&\mathcal{A}(t) \left[\cos \Phi(t) \sin(\theta + 2\Delta\psi) + \epsilon \sin \Phi(t) \cos(\theta + 2\Delta\psi) \right] F_\times(\psi), 
\end{align}
\end{widetext}
which is the same result we would have obtained by replacing $\theta \to \theta' = \theta + 2 \Delta\psi$ in \eq{ellip_gen};
for a signal made up of multiple fully-polarized components, as in \eq{ellip_sum}, the waveframe orientation affects all ellipse orientations in the same way, i.e., $\theta_n \to \theta_n'=\theta_n + \Delta\psi$ (Fig.~\ref{fig:diagram_ellipse_extra}).
We could have equivalently (and more quickly) derived this by noting that $\theta$ is related to the phases of the circularly-polarized components of the signal by $\theta = \left(\phi_L - \phi_R\right)/2$, as in \eq{hcomp_ellip}; the transformation rule for $\theta$ then follows from \eq{htransf_circ}, which implies $\phi_{R/L} \to \phi'_{R/L} \mp 2\Delta\psi$, with the negative (positive) sign for R (L).

This relation between $\theta$ and $\psi$ implies that elliptical-wave analyses that allow $\theta$ to vary freely should avoid degeneracies by fixing $\psi$ to an arbitrary a priori value.
Choosing this fiducial value to be $\psi=0$, the template at a given detector would be constructed as
\begin{equation}
h = F_+(\psi=0)\, h_+ + F_\times(\psi=0)\,  h_\times \, ,
\end{equation}
for $h_{+/\times}$, as in Eq.~\eqref{eq:ellip_sum}, functions of $\{\theta_i, \epsilon_i\}$ and whatever other parameters are needed to evaluate the amplitude and phasing functions $\{\mathcal{A}_i(t), \Phi_i(t)\}$ (or their frequency-domain analogs).
The antenna patterns $F_{+/\times}$ are evaluated for some sky location and arrival time, which can be allowed to vary so as to measure them from the observed data.
On the other hand, the angle $\psi$ is fixed; allowing it to vary would amount to shifting all $\theta_i$ values by $2\psi$, per Eq.~\eqref{eq:theta_psi} and Fig.~\ref{fig:diagram_ellipse_extra}.
Fixing $\psi$ to some fiducial value was the approach taken in \cite{Isi:2017equ,Chatziioannou:2021mij,Isi:2021iql}.

\subsection{Compact binaries and the angles $\Psi$ and $\Omega$}
\label{sec:position}

\begin{figure}
\includegraphics[width=\columnwidth]{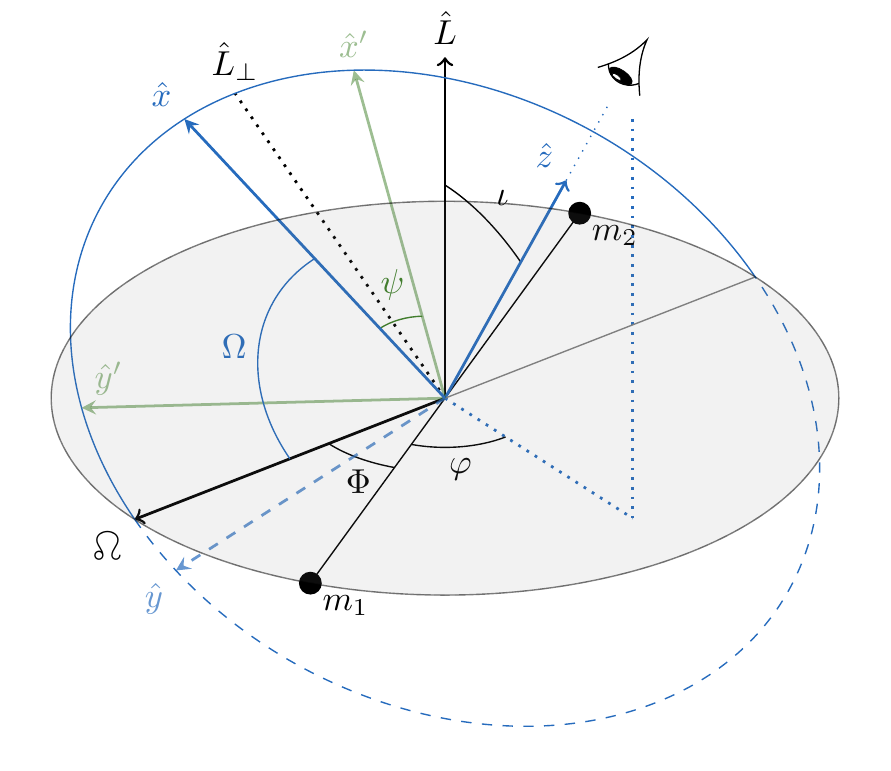}
\caption{Standard construction of the source frame used in LIGO-Virgo data analysis of compact binaries \cite{LALSuite:source}. Two inspiraling objects ($m_1$ and $m_2$) define an orbital plane (gray disk) perpendicular to the \replaced{direction of orbital angular momentum,}{orbital angular momentum direction} $\hat{L}$. At a given time (say, when the observed GW signal reaches 20 Hz) observers on Earth will be oriented with an inclination $\iota$ relative to $\hat{L}$ and an azimuthal angle $\varphi$ with respect to the line from $m_2$ to $m_1$ (related to the reference orbital phase $\Phi=\pi/2-\varphi$).
The intersection between the orbital plane and the plane of the sky (blue circle) defines the line of nodes, with the \emph{ascending node} ($\ascnode$) the point where the orbiting objects cross the sky-plane into the side of the observer.
The polarization vectors $\{\hat{x}',\hat{y}'\}$ and $\{\hat{x},\hat{y}\}$ of Fig.~\ref{fig:diagram_waveframe} lie in the plane of the sky, separated by an angle $\psi$ (clockwise around $\hat{z}$ from $\hat{x}$ to $\hat{x}'$).
The preferred vectors used to predict waveforms as in Sec.~\ref{sec:harmonics}, $\{\hat{x},\hat{y}\}$, define an angle $\Omega$, the longitude of ascending nodes, separating $\ascnode$ from $\hat{x}$ (the origin of longitude).
The LIGO-Virgo convention sets $\Omega=\pi/2$, so that $\hat{y}$ is the ascending node and $\hat{x}$ lies along the projection of $\hat{L}$ onto the plane of the sky ($\hat{L}_\perp$).
\added{While $h_{+/\times}$ are defined relative to $\{\hat{x},\hat{y}\}$, the antenna patterns are computed using $\{\hat{x}',\hat{y}'\}$ (see Fig.~\ref{fig:diagram_skyview}).}
}
\label{fig:diagram_sourceframe}
\end{figure}

When modeling GW waveforms from specific systems, it is useful to tie the polarization frame to the geometry of the source.
This is advantageous because, in order to write out explicit expressions for $h_+$ and $h_\times$, we must make \emph{some} definite choice of frame orientation, and doing so in a way that respects the symmetries of the source (if any) can lead to simplified expressions.
That was the case in going from Eq.~\eqref{eq:ellip_gen} to Eq.~\eqref{eq:ellip_frame} above: if we know \emph{a priori} that the waves from a given source will always be elliptically polarized, then it makes sense to anchor our wave frame to some feature of the source orientation that will ensure alignment with the principal directions of the polarization ellipse (i.e., $\theta=0$).

For a nonprecessing compact binary, as we saw in Sec.~\ref{sec:harmonics}, it is natural to orient our coordinates so as to respect the planar symmetry of the source.
With that standard choice, we find that the linear polarizations take the simple form of \eq{nonprecessing}, which matches the expression for an elliptical mode with $\theta=0$ as in \eq{ellip_frame}.
This again reveals that our choice of coordinates was a good one in modeling that source: because this wave-frame orientation preserves the symmetries of the binary, it also happens to be aligned with the principal directions of the polarization ellipse.
When making predictions for the signal we may always choose this frame to simplify calculations.

Of course, the frame that is most convenient for source modeling need not be the best frame to describe measurements.
In order to compare predictions to measurements, we need to understand how the frame in which the $h_{+/\times}$ polarizations were predicted is oriented with respect to the detectors.
The frame in \eq{nonprecessing}, which we here denote with unprimed symbols $\{\hat{x}, \hat{y}, \hat{z}\}$, was constructed such that the GW direction of propagation, $\hat{z}=\hat{k}$, is purely radial, with the remaining basis elements purely polar or azimuthal.
Although different definitions may be found in the literature (e.g., \cite{Faye:2012we,Kidder:2007rt}), the LIGO-Virgo convention is to choose $\hat{y}$ such that it points towards the ascending node, i.e., parallel to the \emph{line of nodes} defined by the intersection of the orbital plane with the plane of the sky \cite{LALSuite:source}; $\hat{x}$ completes the triad (Fig.~\ref{fig:diagram_sourceframe} with $\Omega = \pi/2$).
In this convention, then, $\hat{x} = -\hat{e}_\iota$, $\hat{y} = -\hat{e}_\varphi$ and $\hat{z}=\hat{e}_r$, where $\left(r, \iota, \varphi\right)$ are the spherical coordinates associated with the spherical-harmonic frame in \eq{spherical_modes}.

\begin{figure}
\includegraphics[width=0.8\columnwidth]{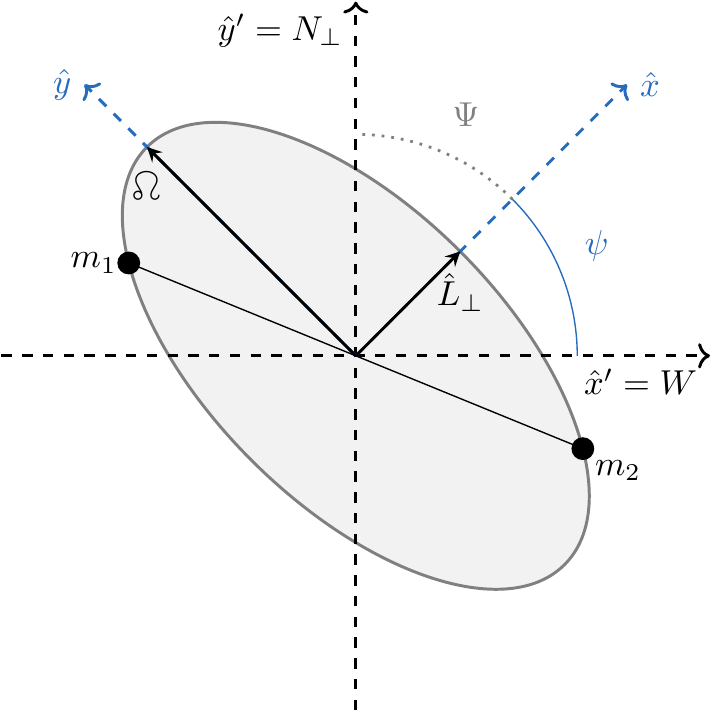}
\caption{View of Fig.~\ref{fig:diagram_sourceframe} as seen by the observer, with the LIGO-Virgo convention that $\Omega=\pi/2$. The masses rotate counterclockwise in the plane of the sky, spanned by the local west ($W$) and the in-sky projection of the celestial north ($N_\perp$).
}
\label{fig:diagram_skyview}
\end{figure}

Having specified $h_+$ and $h_\times$ in that standard, source-based frame, all we need to do to predict the signal at a given detector is to evaluate \eq{h}.
As described in Sec.~\ref{sec:pol}, this is done in practice by expressing $\{\hat{x}, \hat{y}\}$ in terms of canonical reference vectors $\{\hat{x}',\hat{y}'\}$, which are themselves tied to an Earth-centered celestial coordinate system (Fig.~\ref{fig:diagram_waveframe}).
By convention, we specify the relative orientation between the two frames through the angle $\psi$ defined clockwise from $\hat{x}$ to $\hat{x}'$ around $\hat{z}$, where $\hat{x}'$ is the intersection of the celestial equator with the plane of the sky \cite{LALSuite:wave,Anderson:T010110}.
Knowing that, for CBCs, $\hat{y}$ was constructed to lie along the line of nodes, then $\psi = 0$ must mean that the ascending node points towards the projected celestial north ($\hat{y}'=\hat{y}=\ascnode$), and that the projection of the orbital angular momentum onto the plane of the sky is parallel to the horizon due west ($\hat{x}'=\hat{x}=\hat{L}_\perp$); we illustrate this in Fig.~\ref{fig:diagram_skyview} from the point of view of the observer.

In this convention, $\psi$ is identical to the complement of the \emph{position angle} of the source's orbital angular momentum $\Psi$, defined to be the angle between the projected orbital angular momentum and the celestial north in the plane of the sky (i.e., the angle between $\hat{L}_\perp$ and $\hat{y}'$ in Fig.~\ref{fig:diagram_sourceframe}, shown explicitly in Fig.~\ref{fig:diagram_skyview} for $\Omega=\pi/2$).
More generally, $\Psi = \pi - \psi - \Omega$ in terms of the \emph{longitude of the ascending node} $\Omega$ with $\hat{x}$ as the origin of longitude.
LIGO and Virgo always fix $\Omega = \pi/2$ \cite{LALSuite:source}, tying the primed polarization frame, in which $h_{+/\times}$ are predicted, to the source geometry.
Thus, when LIGO-Virgo report measurements of the polarization angle in CBCs, the quantity reported is the in-plane sky angle of the orbital ascending node relative due north.

In fact, with these conventions for a nonprecessing binary, the three angles $\psi$, $\Psi$ and $\theta$ can all be subsumed by a single parameter (usually written $\psi$) simultaneously encoding the orientation of the polarization basis, the alignment of the source in the sky, and the principal axes of the GW polarization ellipse.
We can then think of this angle as a property of the source to be measured from our data, rather than an arbitrary parameter orienting our frame.
Although this equivocation vastly simplifies analyses, it is helpful to keep in mind that the three angles are conceptually distinct: $\psi$ can always be defined, but $\theta$ only exists for fully polarized waves, and $\Psi$ is an orbital element, not defined for arbitrary sources (say, a stochastic source, or a supernova).

If the component spins are not (anti)aligned with the orbital angular momentum, the spins and the orbital plane will both precess.
As a consequence, the system will not be reflection symmetric and the GW signal will not be elliptically polarized overall (see Sec.~\ref{sec:harmonics}).
Nonetheless, it is still conventional to tie $\hat{y}$ to the source as in Fig.~\ref{fig:diagram_sourceframe}, referring to the line of nodes as oriented at some specific point in the binary evolution (e.g., when the detected GW signal reaches 20 Hz, or at a mass-invariant reference point \cite{Varma:2021csh,Mould:2021xst}).
In that case, yet another coordinate frame is used to specify the component spins at the reference time, as specified in \cite{LALSuite:spins} and illustrated in App.~\ref{app:spins}.

In summary, we can identify three conceptually distinct Cartesian frames: a wave frame that determines the principal directions along which we \emph{define} the effect of a plus vs cross wave; for an elliptical wave, an intrinsic polarization frame, encoding the principal directions of the polarization ellipse; and a source frame, aligned with the symmetries of the source, or otherwise anchored to some defining feature of it; all of these can be specified in some astronomical frame, like ecliptic celestial coordinates.
For nonprecessing binaries, which are highly symmetric, we can define the source frame to make it always align with the polarization frame.

In unmodeled analyses, as those discussed in Sec.~\ref{sec:ellip:gen}, it is not possible or useful to explicitly tie the polarization frame to properties of the source, since these analyses are not tailored to any specific source to begin with, or they purposely disregard source orientation information for the sake of generality.
In that case, the model for $h_+$ and $h_\times$ can be defined in any arbitrary wave frame.
A common choice is to simply set $\psi = 0$ in the standard coordinates described above, i.e., with $\hat{y}$ pointing towards the celestial north (Fig.~\ref{fig:diagram_sourceframe}).
Having done so, all information regarding polarization orientation will be encoded in the $\theta$ parameter of Fig.~\ref{fig:ellipse}, with one value per elliptical mode in the decomposition of \eq{ellip_sum}.
Varying both $\psi$ and $\theta$ simultaneously is ill-advised in that context, since the two parameters will be fully degenerate (see end of Sec.~\ref{sec:ellip}, including Fig.~\ref{fig:diagram_ellipse_extra}).

\section{Coordinate transformations}
\label{sec:jacobians}

In the previous sections, we have introduced different parametrizations of elliptical (i.e., fully polarized) waves, including Eqs.~\eqref{eq:ellip_circ}, \eqref{eq:hcomp_ellip} and \eqref{eq:hcomp_ellip_chi}.
Their use varies depending on the specific application, according to convenience and convention.
Understanding the relation between the different parametrizations becomes especially important when implementing and interpreting measurements, since the choice of parametrization often influences the prior \added{(implicitly or explicitly)} specified in \replaced{Bayesian analyses}{the analysis}.
\replaced{Probability densities obtained under different parametrizations, including posteriors from a Bayesian measurement,}{Measurements obtained from different parametrizations} can be related via a Jacobian.%
\footnote{\protect\added{Many quantities used in frequentist analyses, like the maximum likelihood estimator, are invariant under reparametrizations and thus have no need for Jacobians; however, note that this is not true of analyses that maximize a \emph{marginal} likelihood that has been averaged over nuisance parameters.}}

We may also want to switch parametrizations for technical reasons.
Although conceptually insightful, the manifestly-elliptical parameterization in terms of $\{A, \epsilon, \theta, \phi\}$ of \eq{hcomp_ellip} contains multiple degeneracies that make it less than ideal for sampling purposes.
For instance, the angles $\theta$ and $\phi$ become totally degenerate when $\epsilon = \pm 1$.
To circumvent this, we may switch to a more suitable parametrization in the sampling process, and then translate the result back into $\{A, \epsilon, \theta, \phi\}$ for interpretation.
In that case, we can still specify a prior in terms of the elliptical quantities by again making use of a Jacobian.

\begin{figure}
\includegraphics[width=0.49\columnwidth]{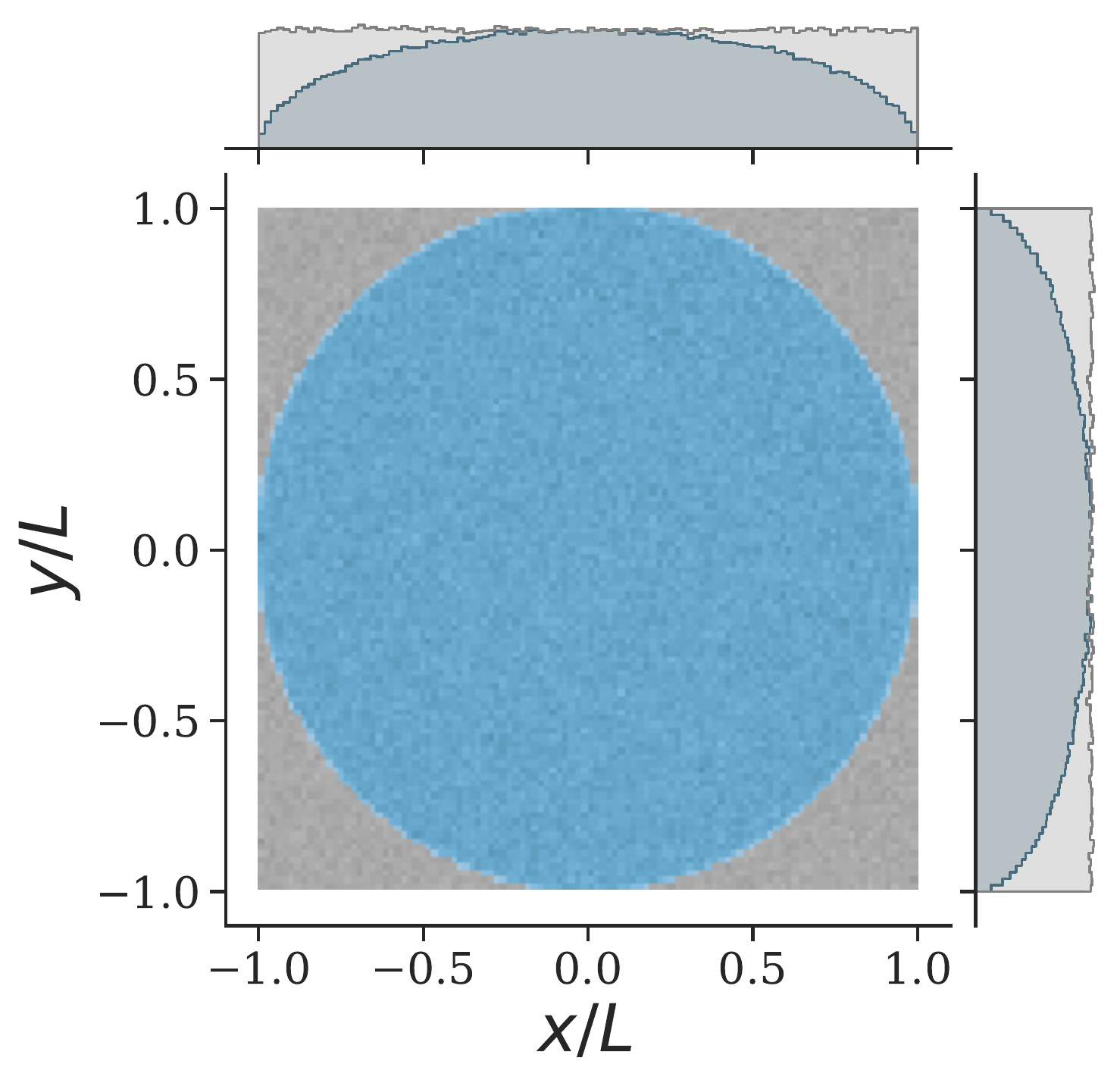}
\includegraphics[width=0.49\columnwidth]{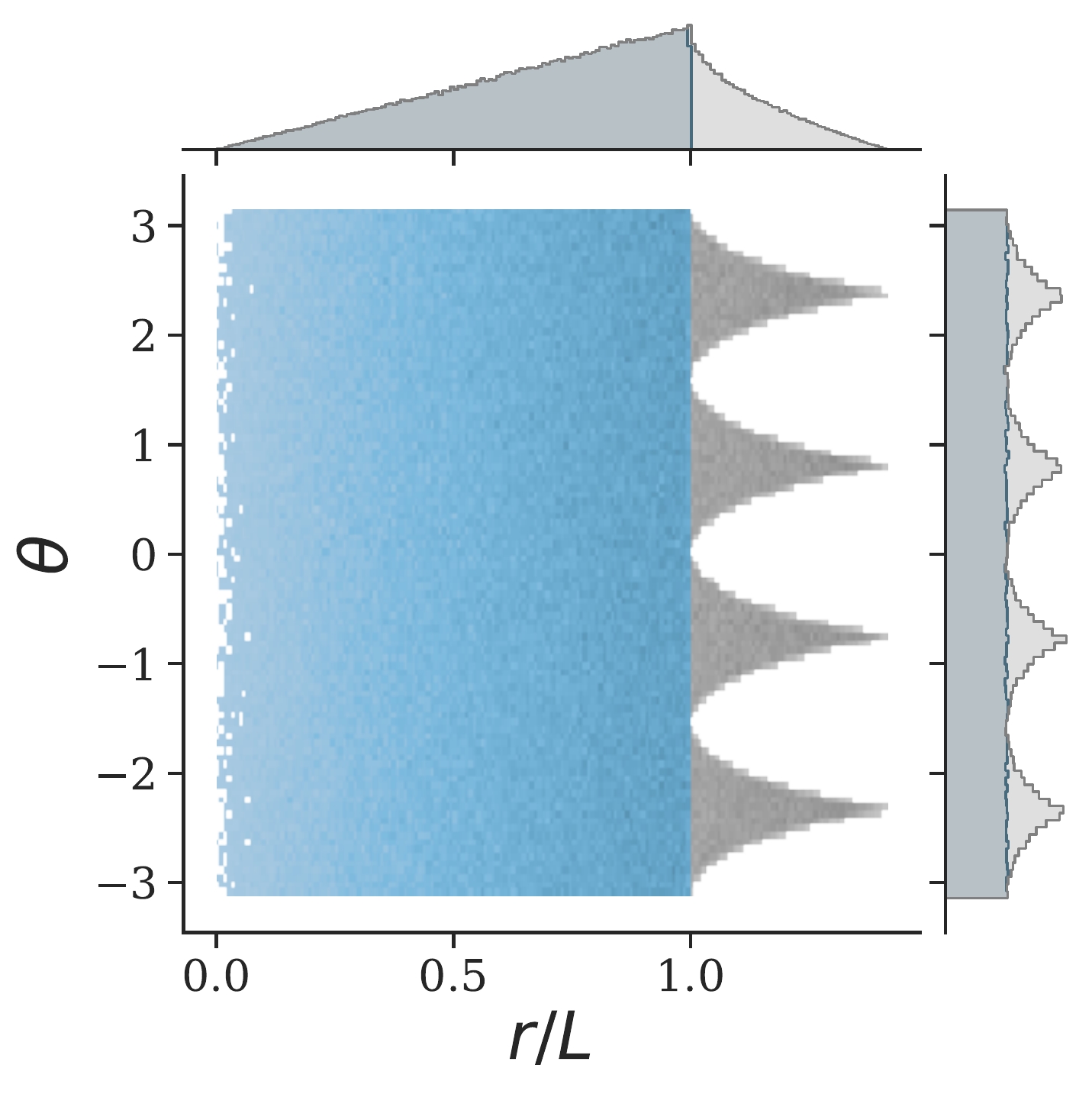}
\caption{Example of coordinate transformations for probability distributions.
Sampling uniformly in some Cartesian coordinates $-L \leq x \leq L$ and $-L \leq y \leq L$ (gray left), where $L$ is some arbitrary scale, imposes a nonuniform distribution in the corresponding polar coordinates $r = \sqrt{x^2 + y^2}$ and $\theta = \arctan(y/x)$ (gray right); we obtain a more restricted distribution if we limit the sampling to a disk $x^2 + y^2 < L^2$ instead of a square (blue). 
Up to $r < L$ the density is uniform in $\theta$ and proportional to $r$ (top right), while for $r>L$ there are spikes of probability at $\theta = \pm \pi/4,\pm 3\pi/4$ (right side), corresponding to the corners of the square on the left.
We can recover a uniform distribution in $(r, \theta)$ by applying a Jacobian $\propto 1/r$ per \eq{jac}, and explicitly restricting to $r< L$ (blue), thus cutting out the corners of the left-hand square.}
\label{fig:jac_example}
\end{figure}

If we parametrize our analysis in terms of some alternative set of parameters $\vec{\xi}$, we can impose some prior distribution defined in the space of elliptical quantities, $p({A, \epsilon, \theta, \phi})$, by choosing a corresponding prior $p(\vec{\xi})$ for the $\vec{\xi}$ quantities such that
\begin{equation} \label{eq:jac}
p \left( \vec{\xi} \right) = p \left( A, \epsilon, \theta, \phi \right) \left| \frac{\partial \{A, \epsilon, \theta, \phi\}}{\partial \vec{\xi}} \right| ,
\end{equation}
where the last factor $J \equiv | \partial (A, \epsilon, \theta, \phi)/\partial \vec{\xi} |$ is the determinant of the Jacobian matrix.
Applying the Jacobian without any further reweighting \replaced{effects}{yields} a flat prior on the  $\{A, \epsilon, \theta, \phi\}$ quantities over the region covered by the original prior.
As with any coordinate transformation, the integration limits must be adjusted to ensure that they correspond to the targeted region in the $\{A, \epsilon, \theta, \phi\}$ space---for example, sampling uniformly in the two Cartesian quadratures $(x, y)$, we can effect a uniform prior on the polar quantities $(0 < r=\sqrt{x^2+y^2} \leq L, \theta= \arctan y/x)$ by applying a Jacobian $\propto 1/r$ and explicitly enforcing $r \leq L$ (Fig.~\ref{fig:jac_example}).

In this section, we will consider four different parametrizations of an elliptical wave, and present the Jacobians relating them to the $\{A, \epsilon, \theta, \phi\}$ parametrization.
We will focus on a single elliptical component as a standin for any individual term in the sum of Eq.~\eqref{eq:ellip_sum}, so that the results are trivially generalizable to decompositions of GWs with arbitrary polarizations, as would be used by \textsc{BayesWave} or other generic analyses.
We assume the amplitude could potentially subsume any (slow) time dependence endowed by $\mathcal{A}(t)$ in Eq.~\eqref{eq:ellip_gen}, e.g., the amplitude parameters below could correspond to a reference amplitude $A=\mathcal{A}(t=0)$.

\subsection{Amplitude and ellipticity}
\label{sec:jac:Achi}

\begin{figure}
\includegraphics[width=0.9\columnwidth]{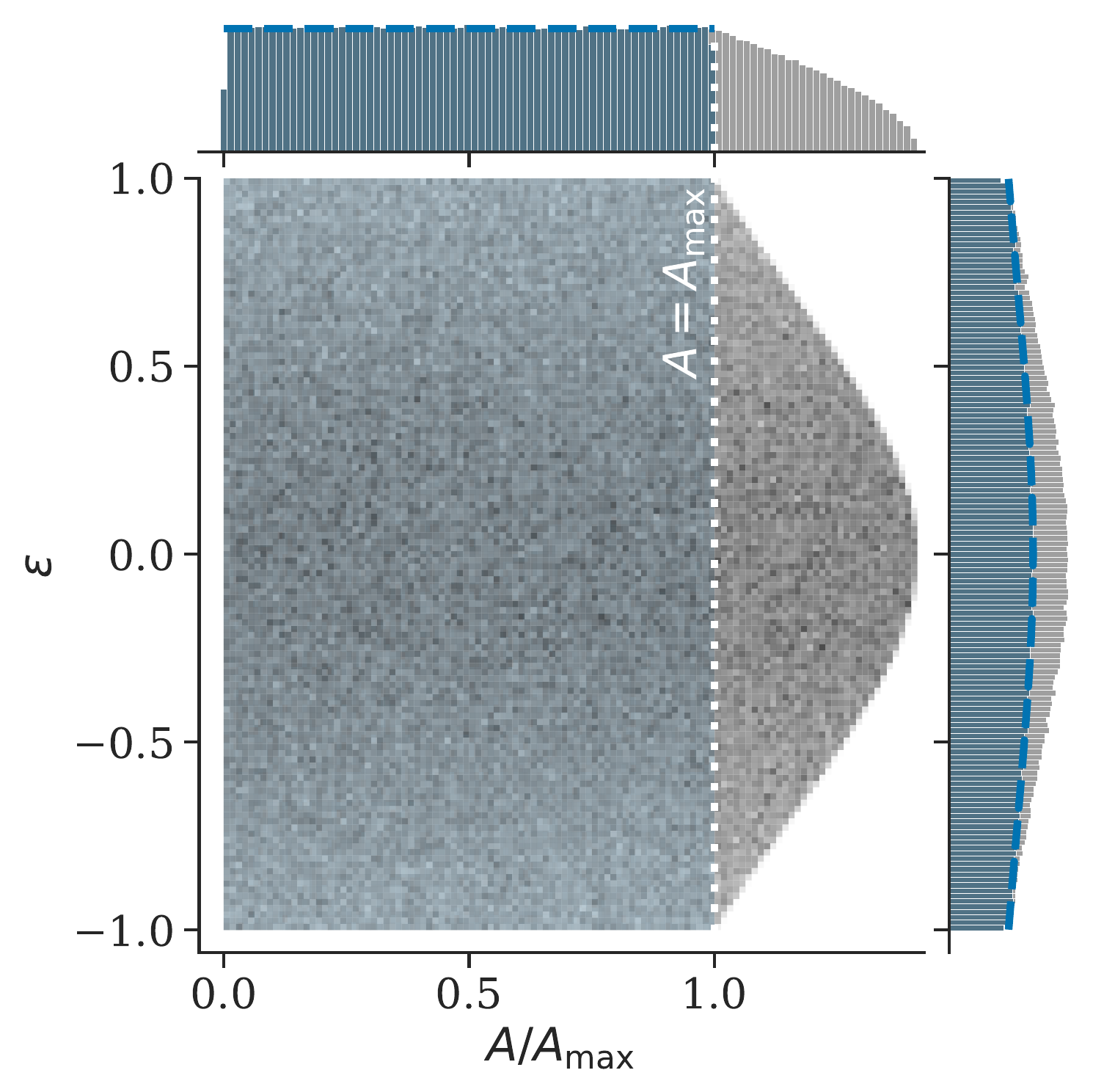}
\caption{Distribution imposed on $\{A,\epsilon\}$ by applying a flat prior on $\{\hat{A},\chi\}$ over the ranges $0 < \hat{A} \leq \hat{A}_{\max}$ and $-\pi/4 \leq \chi \leq \pi/4$.
Since $\hat{A} = A \sqrt{1+\epsilon^2}$ and $\max(\epsilon)= 1$, we should make sure to bound $A \leq  A_{\max} \equiv \hat{A}_{\max} /\sqrt{2}$ (dotted line), as in the example of Fig.~\ref{fig:jac_example}.
With that constraint (blue), the $A,\epsilon$ distribution is inversely proportional to the Jacobian in \eq{jac_Aeps_Achi}: a uniform prior on $\hat{A}$ and $\chi$ is uniform in $A$ but slightly favors linear polarizations ($\epsilon = 0$) over circular ones ($\epsilon = \pm 1$), with probability density $\propto 1/J_0$ (dashed curves). 
The marginal histograms are normalized by bin count.
}
\label{fig:jac_Aeps_Achi}
\end{figure}

In Sec.~\ref{sec:ellip:mono}, we presented two equivalent parametrizations of the $h_{+/\times}$ components of an elliptical wave, Eqs.~\eqref{eq:hcomp_ellip} and \eqref{eq:hcomp_ellip_chi}, illustrated in Fig.~\ref{fig:ellipse}.
Equation ~\eqref{eq:hcomp_ellip} parametrizes the signal strength via the maximum amplitude achieved by the wave, $A$ (the semimajor axis in Fig.~\ref{fig:ellipse}), and the shape of the polarization ellipse via the ellipticity, $\epsilon$ (the ratio between the semiminor and semimajor axes);
meanwhile, Eq.~\eqref{eq:hcomp_ellip_chi} parametrizes the strength via the intensity amplitude $\hat{A}$, which is the square-root of the signal intensity $I$, and the shape of the ellipse through the angle $\chi$.
The two parametrizations are straigthforwardly related by
\begin{equation} \label{eq:Aellip_Ahatchi}
\begin{cases}
\hat{A} = A \sqrt{1 + \epsilon^2} \\
\chi = \arctan \epsilon 
\end{cases} 
\end{equation}
and the inverse transformation
\begin{equation} \label{eq:Ahatchi_Aellip}
\begin{cases}
A = \hat{A} \cos \chi \\
\epsilon = \tan \chi \\
\end{cases} ,
\end{equation}
with no change to the angles $\theta$ and $\phi$.
The Jacobian relating these two transformations is simply
\begin{equation} \label{eq:jac_Aeps_Achi}
J_0 \equiv \left| \frac{\partial(A,\epsilon,\theta,\phi)}{\partial(\hat{A}, \chi, \theta, \phi)}\right| =  \sec \chi = \sqrt{1 + \epsilon^2} \, ,
\end{equation}
or, equivalently $J_0 = \hat{A}/A$.
This Jacobian, illustrated in Fig.~\ref{fig:jac_Aeps_Achi}, indicates that a uniform prior in the $\{\hat{A}, \chi\}$ quantities implicitly favors linear polarizations ($\epsilon = 0$) over circular ones, although the preference is mild.

\subsection{Circular components}
\label{sec:jac:Arl}

A monochromatic elliptical wave, Eq.~\eqref{eq:hcomp_ellip}, can be specified in terms of the circular polarization basis elements as in \eq{ellip_circ},
where the $C_{R/L} \equiv A_{R/L} \exp(i\phi_{R/L})$ quantities control the amplitude and phase of the right and left circularly-polarized components of the signal.
The representation in terms of such circular-mode amplitudes and phases is equivalent to Eq.~\eqref{eq:hcomp_ellip} if we impose
\begin{equation} \label{eq:Cphi_to_Aellip}
\begin{cases}
A = \frac{1}{\sqrt{2}}\left(A_R + A_L\right) \\
\epsilon = (A_R - A_L)/(A_R + A_L) \\ 
\theta = \frac{1}{2}(\phi_L - \phi_R)\\
\phi = \frac{1}{2}(\phi_L + \phi_R)\\
\end{cases} 
\end{equation}
(although see footnote \ref{foot:angles} for the angles).
Equivalently, the inverse transformation is 
\begin{equation} \label{eq:Aellip_to_Cphi}
\begin{cases}
A_R = \frac{1}{\sqrt{2}} A \left(1 + \epsilon\right) \\
A_L = \frac{1}{\sqrt{2}} A \left(1 - \epsilon\right) \\
\phi_R = \phi - \theta \\ 
\phi_L = \phi + \theta \\ 
\end{cases} .
\end{equation}
These expressions are particularly simple: amplitude parameters $\{ A_R, A_L\}$ transform directly into amplitude parameters $\{A, \epsilon\}$, irrespective of phasing angles.
This is a consequence of the fact that the circular polarizations are defined to be invariant under rotations around the direction of propagation, up to an overall phase as shown in \eq{htransf_circ}.

\begin{figure}
\includegraphics[width=0.9\columnwidth]{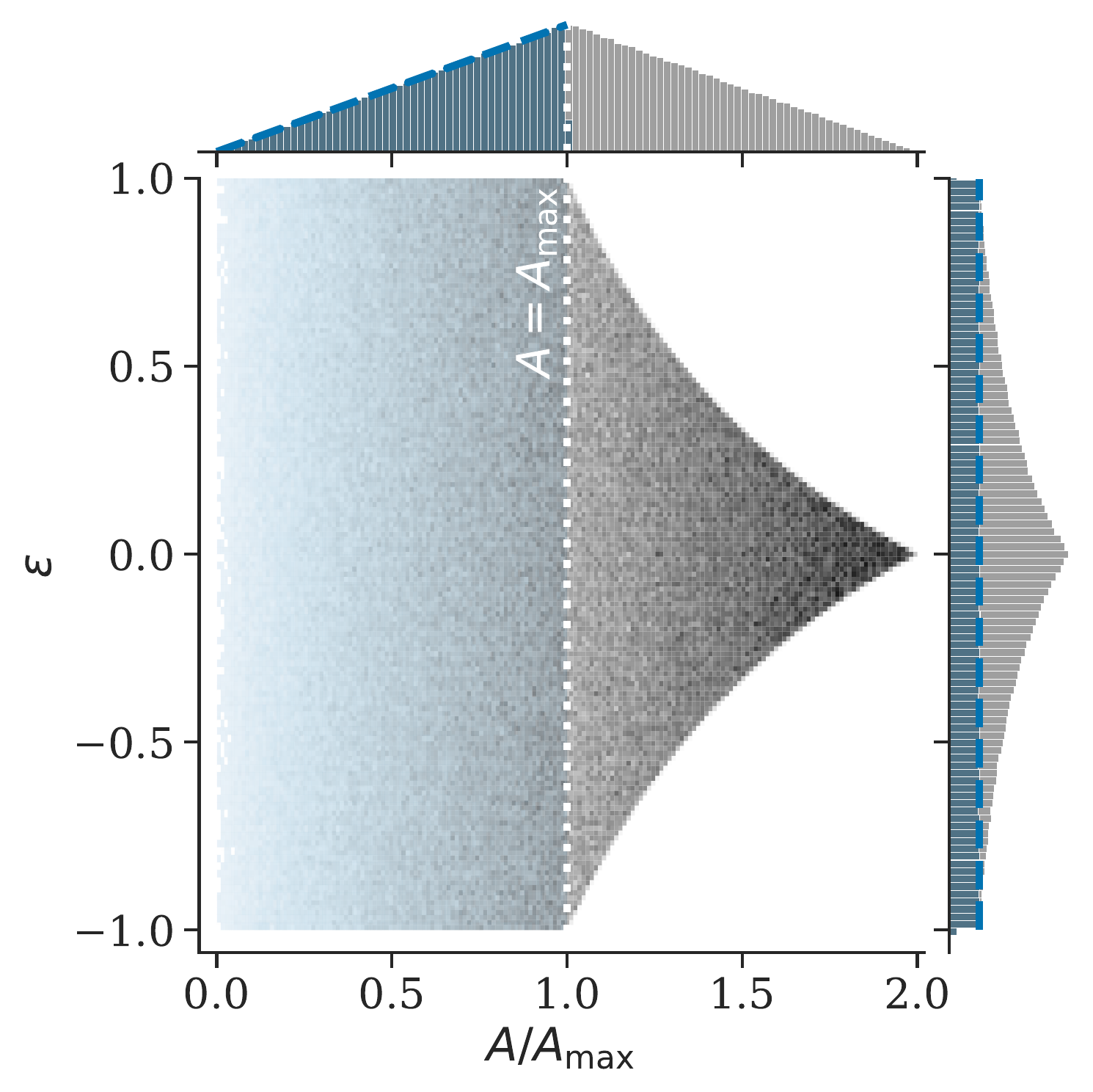}
\caption{Distribution imposed on $\{A,\epsilon\}$ by applying a flat prior on $\{A_R,A_L\}$ over the ranges $0 < A_{R/L} \leq A_{R/L,\max}$.
Since $A = (A_R + A_L)/\sqrt{2}$, we must constrain $A \leq A_{\max} \equiv A_{R/L,\max}/\sqrt{2}$ (dotted line), as in the example of Fig.~\ref{fig:jac_example};
failing to do so would result in a triangular distribution for $A$, i.e., upward sloping for $A < A_{\max}$ and downward sloping for $A > A_{\max}$ (top panel), and a nonuniform distribution on $\epsilon$ (right panel).
Restricting to $A < A_{\max}$ (blue), the density is proportional to $A$ and flat in $\epsilon$, following $\propto 1/J_1$ for the Jacobian in \eq{jac_Aeps_Arl} (dashed curves).
The marginal histograms are normalized by bin count.
}
\label{fig:jac_Aeps_Arl}
\end{figure}

The above transformations imply a Jacobian
\begin{equation} \label{eq:jac_Aeps_Arl}
J_1 \equiv \left| \frac{\partial(A,\epsilon,\theta,\phi)}{\partial(A_R, A_L, \phi_R, \phi_L)}\right| \propto \frac{1}{A_R + A_L}\, ,
\end{equation}
with a proportionality constant of $1/\sqrt{2}$, which is ignored in most applications as it can be absorbed by an overall normalization; based on \eq{Cphi_to_Aellip}, this is also proportional to $1/A$.
Therefore, a prior uniform in $A_R$ and $A_L$ results in a triangular prior in the overall amplitude of the mode defined by $A$ (Fig.~\ref{fig:jac_Aeps_Arl}).

\begin{figure}
\includegraphics[width=0.9\columnwidth]{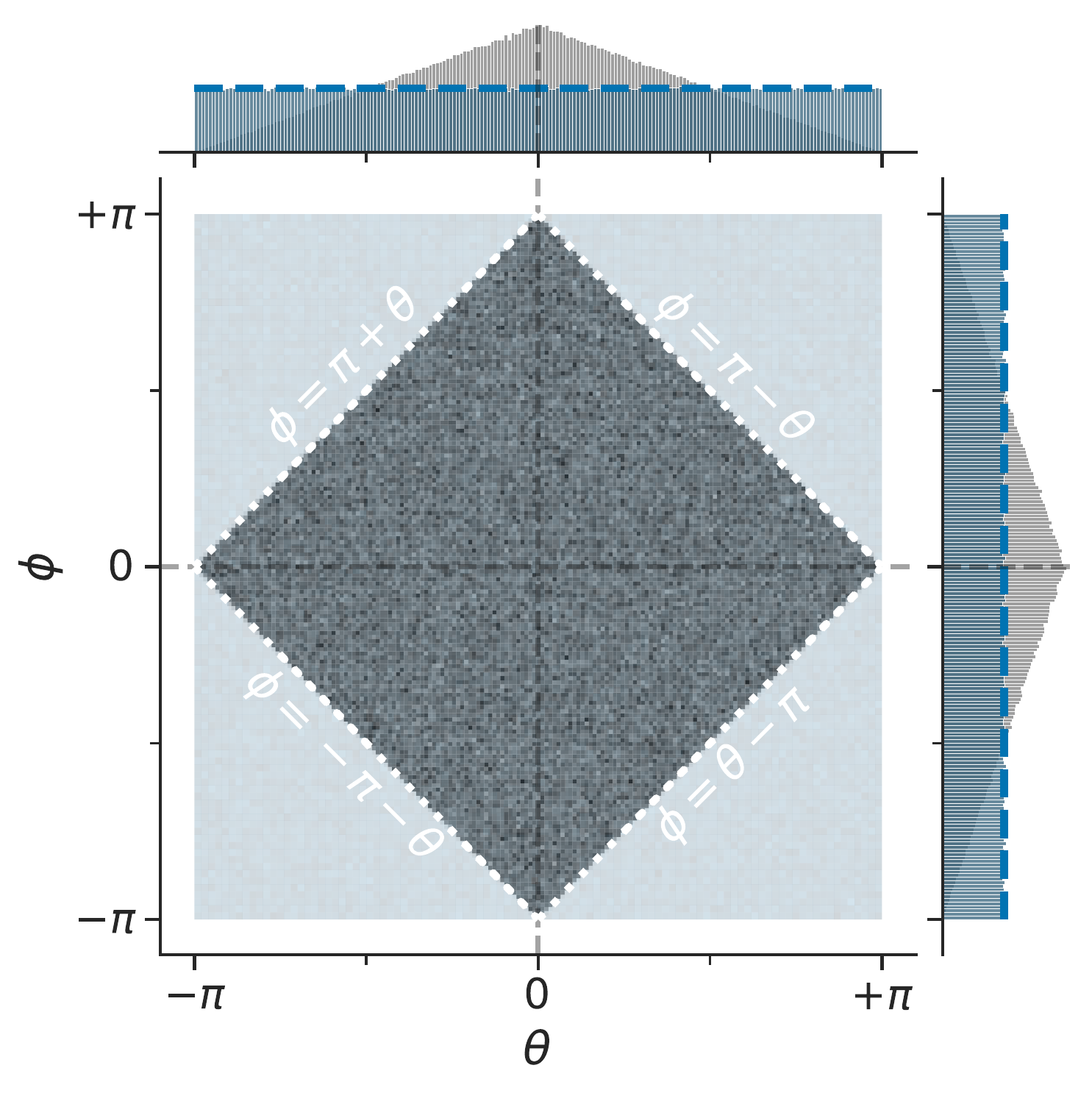}
\caption{Distribution imposed on $\{\theta,\phi\}$ by applying a flat prior on $\{\phi_R,\phi_L\}$ over $-\pi/2 < \phi_{R/L} \leq \pi/2$, through the simple transformation of \eq{Aellip_to_Cphi} (gray) or the more generic version in footnote \ref{foot:angles} (blue).
The latter spans the full domain by populating the corners outside the diagonals defined by $|\phi| = \pi \pm \theta$ (dotted lines), restoring uniformity in the marginals (blue dashed) at the expense of double covering the waveform space.
In practice, the samples from the blue distribution can be obtained by randomly adding $\pm \pi$ to both $\theta$ and $\phi$ for  half of the samples drawn from the gray distribution.
The marginal histograms are normalized by bin count.
}
\label{fig:jac_Aeps_Arl_angles}
\end{figure}

Equation \eqref{eq:jac_Aeps_Arl} implies that an analysis that samples uniformly in $A_R$ and $A_L$ within some range $0 \leq A_{R,L} \leq A_{R/L,\mathrm{max}}$ actually favors large overall mode amplitudes $A$, with a triangular distribution that vanishes at $A=0$ and $A=\sqrt{2}\,A_{R/L,\mathrm{max}}$, and peaks at $A = A_{R/L,\mathrm{max}}/\sqrt{2} \equiv A_{\max}$ (top panel of Fig.~\ref{fig:jac_Aeps_Arl}).
Without enforcing the $A \leq A_{\max}$ constraint, the ellipticity distribution will no longer be uniform, instead favoring linear polarizations (right panel of Fig.~\ref{fig:jac_Aeps_Arl}).
This was the case, e.g., for one of the ringdown analyses in \cite{LIGOScientific:2020tif}, which sampled uniformly in amplitude coefficients equivalent to $A_{R/L}$ up to an overall scaling.

The absence of angles in \eq{jac_Aeps_Arl} indicates that a uniform distribution in $\phi_{R/L}$ is also uniform in terms of $\theta$ and $\phi$.
However, this feature can be obfuscated by the fact that the relation between the two sets of angles is not strictly bijective due to their $2\pi$-periodicities (see footnote \ref{foot:angles}).
When applied as written, \eq{Aellip_to_Cphi} transforms a uniform distribution over $-\pi/2 < \phi_{R/L} < \pi/2$ into a uniform distribution over a $\pi/4$-rotated square domain in the $\{\theta, \phi\}$-space, as implied by the discussion in Sec.~\ref{sec:ellip:mono:morph} and illustrated in Fig.~\ref{fig:jac_Aeps_Arl_angles}; the corresponding marginals appear to favor $\theta=0$ and $\phi = 0$ (gray in Fig.~\ref{fig:jac_Aeps_Arl_angles}).
The uniformity over the full range of angles can again be made manifest by applying the more generic transformation of footnote \ref{foot:angles} (blue in Fig.~\ref{fig:jac_Aeps_Arl_angles}), at the expense of restoring the double-covering of the waveform space described in Sec.~\ref{sec:ellip:mono:morph}.

We can also relate the circular amplitudes to the alternative parametrization of \eq{hcomp_ellip_chi}.
The straightforward relation is given by the transformations
\begin{equation} \label{eq:Arl_to_Ahatchi}
\begin{cases}
\hat{A} = \sqrt{A_R^2 + A_L^2} \\
\chi = \arctan\left( \frac{A_R - A_L}{A_R + A_L}\right)
\end{cases} ,
\end{equation}
and
\begin{equation} \label{eq:Ahatchi_to_Arl}
\begin{cases}
A_R = \frac{1}{\sqrt{2}} \hat{A} \left(\cos\chi + \sin \chi\right) \\
A_L = \frac{1}{\sqrt{2}} \hat{A} \left(\cos\chi - \sin \chi\right)
\end{cases} ,
\end{equation}
while the remaining angles are related as in Eqs.~\eqref{eq:Cphi_to_Aellip} and \eqref{eq:Aellip_to_Cphi}.
Accordingly, the Jacobian that takes us from the circular parametrization to one flat in $\{\hat{A},\chi\}$ can be shown to be $J \propto \hat{A}^{-1}$.
Thus, as expected from the composition of Eqs.~\eqref{eq:jac_Aeps_Achi} and \eqref{eq:jac_Aeps_Arl}, a prior uniform in $A_{R/L}$ will also favor large intensity amplitudes with probability $\propto \hat{A}$, when restricted to the appropriate range; it will also be uniform in $\chi$.

\subsection{Linear components}
\label{sec:jac:Apc}

Rather than using the circular basis, we could instead work with the linear polarization modes as the fundamental quantity, parametrizing them directly as%
\footnote{Or, equivalently, using $\sin$ for $h_\times$ instead of $\cos$, to resemble \eq{nonprecessing}; this amounts to a redefinition of $\phi_\times$.}
\begin{subequations} \label{eq:Aphi}
\begin{equation}
h_+ = A_+ \cos (\omega t - \phi_+)\, ,
\end{equation}
\begin{equation}
h_\times = A_\times \cos (\omega t - \phi_\times) \, ,
\end{equation}
\end{subequations}
where $A_{+/\times}$ and $\phi_{+\times}$ are initial amplitudes and phases for each polarization, as elsewhere in the text.
Structurally, this mimics the parametrization adopted by \textsc{BayesWave} for each wavelet \cite{Cornish:2020dwh}.

\newcommand{\xp}{x_{+}}
\newcommand{\xc}{x_{\times}}
\newcommand{\xpc}{x_{+/\times}}
\newcommand{\yp}{y_{+}}
\newcommand{\yc}{y_{\times}}
\newcommand{\ypc}{y_{+/\times}}

\newcommand{\xr}{x_{R}}
\newcommand{\xl}{x_{L}}
\newcommand{\xrl}{x_{R/L}}
\newcommand{\yr}{y_{R}}
\newcommand{\yl}{y_{L}}
\newcommand{\yrl}{y_{R/L}}

\begin{figure}
\includegraphics[width=\columnwidth]{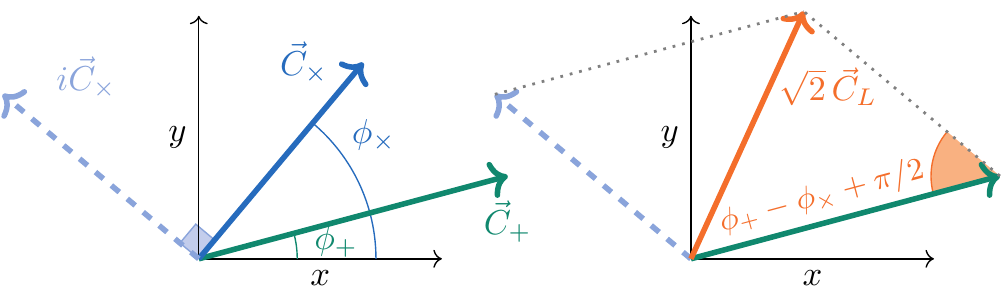}
\caption{Geometric derivation of Eq.~\eqref{eq:Aphi_to_Cphi} for the left-handed polarization case.
\emph{Left:} Following Eq.~\eqref{eq:Aphi}, the plus and cross Jones-vecor amplitudes, $\vec{C}_{+/\times} = A_{+/\times} \exp(i \phi_{+/\times})$, subtend angles $\phi_{+/\times}$ relative to the real axis ($x$, abscissa) and thus have Cartesian coordinates $\left(\xpc, \ypc \right)$, in terms of the quadratures defined in Eq.~\eqref{eq:xy}; the vector $i\vec{C}_\times$ is orthogonal to $\vec{C}_\times$, so that it subtends an angle $\pi/2 + \phi_\times-\phi_+$ with respect to $\vec{C}_+$.
\emph{Right:} the left-handed Jones amplitude, $\vec{C}_L = (\vec{C}_+ + i \vec{C}_\times)/\sqrt{2}$, has components $\left(\xp - \yc, \yp + \xc\right)/\sqrt{2}$ and, therefore, $\phi_R = \mathrm{atan2}( \yp + \xc, \xp - \yc)$; since the acute angle between $\vec{C}_+$ and $i\vec{C}_\times$ is $\pi - (\pi /2 + \phi_\times - \phi_+) = \phi_+ - \phi_\times + \pi/2$, the law of cosines implies Eq.~\eqref{eq:Aphi_to_Cphi} for $A_L$. (The right-handed case is analogous).
}
\label{fig:diagram_apac}
\end{figure}

Equation \eqref{eq:Aphi} still represents an elliptically polarized mode.
To relate this parametrization to that in Eq.~\eqref{eq:hcomp_ellip}, it is convenient to first map Eq.~\eqref{eq:Aphi} into the circular-basis parameters of the previous section.
We can do this geometrically by considering the respective Jones vectors (Sec.~\ref{sec:math}), from which we get $C_{R/L} = (C_+ \mp i C_\times)/\sqrt{2}$, for $C_{+/\times} \equiv A_{+/\times} \exp(i\phi_{+/\times})$ as in \eq{jones_bases}.
As illustrated in Fig.~\ref{fig:diagram_apac}, trigonometry then implies that
\begin{equation} \label{eq:Aphi_to_Cphi}
\begin{cases}
A_R^2 = \frac{1}{2}\left[A_+^2 + A_\times^2 + 2 A_+ A_\times \sin(\phi_\times - \phi_+)\right] \\
A_L^2 = \frac{1}{2}\left[A_+^2 + A_\times^2 - 2 A_+ A_\times \sin(\phi_\times - \phi_+)\right] \\
\phi_R = \mathrm{atan2}\left(\yp -\xc, \xp + \yc \right)\\
\phi_L = \mathrm{atan2}\left(\yp + \xc, \xp - \yc \right) 
\end{cases} ,
\end{equation}
where, to simplify the notation, we have defined the cosine and sine quadratures
\begin{subequations} \label{eq:xy}
\begin{equation}
\xpc \equiv A_{+/\times} \cos \phi_{+/\times} \, ,
\end{equation}
\begin{equation}
\ypc \equiv A_{+/\times} \sin \phi_{+/\times} \, .
\end{equation}
\end{subequations}
Together with Eq.~\eqref{eq:Cphi_to_Aellip}, this allows us to compute $\{A, \epsilon, \theta, \phi\}$ as a function of $\{A_+, A_\times, \phi_+, \phi_\times\}$.
This transformation is clearly less straightforward than those for the circular components in the previous section, with amplitude and phase parameters mixing into each other.
This is because this coordinate transformation encodes the frame rotation that would bring an arbitrarily-oriented elliptical wave into the simple form of Eq.~\eqref{eq:Aphi}, which is nothing but the special frame we identified in Eq.~\eqref{eq:ellip_frame}.

The overall Jacobian relating $\{A, \epsilon, \theta, \phi\}$ to $\{A_+, A_\times, \phi_+, \phi_\times\}$ is quite simple, however, when expressed in terms of the former set of parameters,
\begin{subequations} \label{eq:jac_Aphi}
\begin{align}
J_2 &\equiv \left| \frac{\partial(A, \epsilon, \theta, \phi)}{\partial(A_+, A_\times, \phi_+, \phi_\times)}\right| \nonumber \\
&= 2 A_+ A_\times \left[ \sqrt{A_+^4 + A_\times^4 + 2 A_+^2 A_\times^2 \cos 2(\phi_\times - \phi_+)} \right. \nonumber \\
& \times \left( \sqrt{A_+^2 + A_\times^2 -2 A_+ A_\times \sin(\phi_\times-\phi_+)} \right. \nonumber \\
&\left.\left. +  \sqrt{A_+^2 + A_\times^2 +2 A_+ A_\times \sin(\phi_\times-\phi_+)}\right)\right]^{-1}\\
&= \frac{1}{2 A} \sqrt{\left(\frac{1 + \epsilon^2}{1 - \epsilon^2}\right)^2 - \cos^2 2\theta} \, .
\end{align}
\end{subequations}
The Jacobian factorizes into a piece for the size of the ellipse ($1/A$), and a less trivial piece for its shape and orientation (function of $\epsilon$ and $\theta$).
The $J_2 \propto 1/A$ dependence implies that an analysis with uniform priors in the linear polarization amplitudes will implicitly favor high overall signal power, as was the case for the circular amplitudes in Fig.~\ref{fig:jac_Aeps_Arl}.
Additionally, the dependence on the ellipse's shape implies that the Jacobian diverges to positive infinity for $\epsilon = \pm 1$, meaning that circular polarizations will be disfavored in this scenario.

\begin{figure}
  \includegraphics[width=\columnwidth]{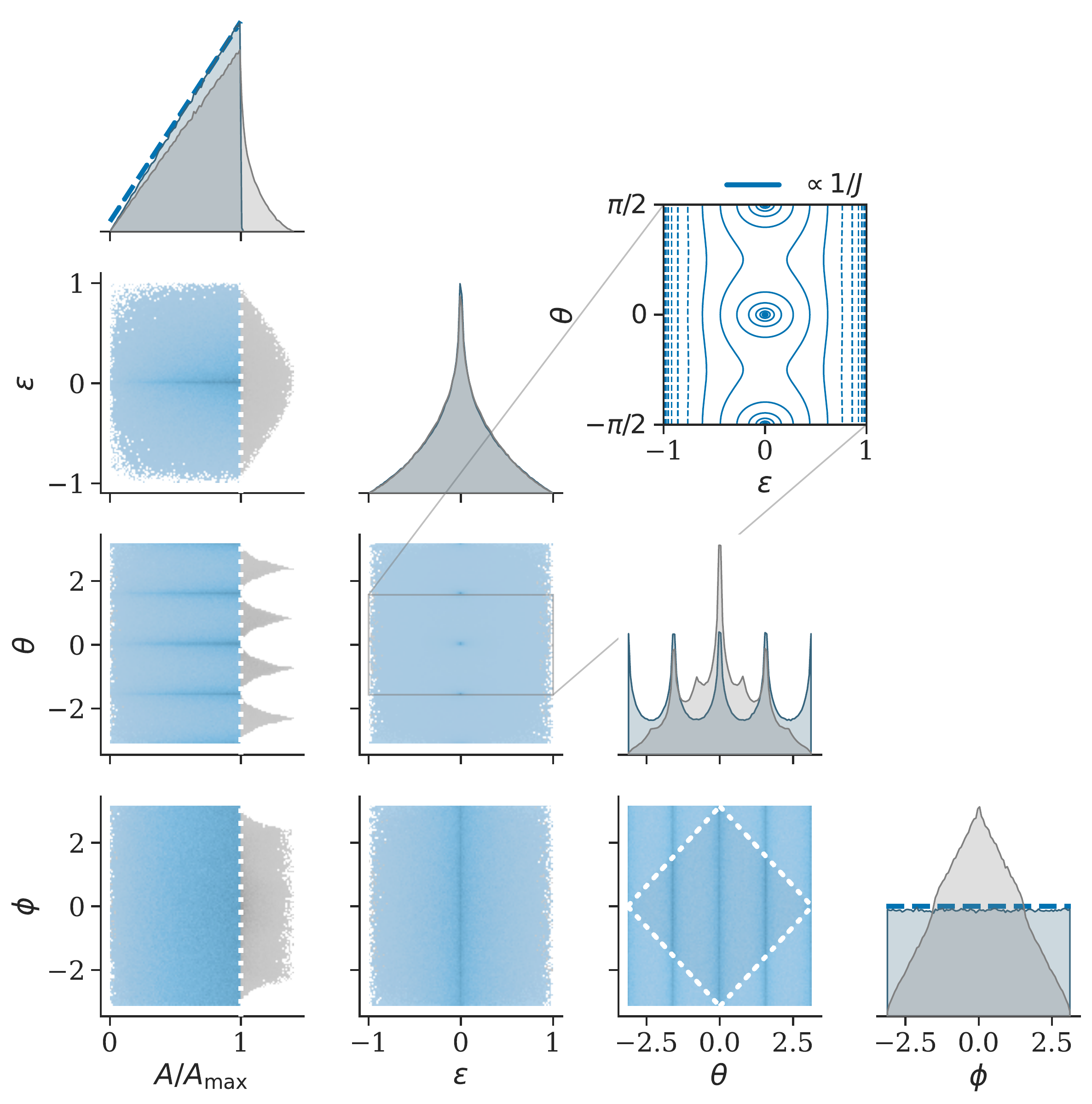}
  \caption{Distribution imposed on $\{A,\epsilon, \theta, \phi\}$ by applying a flat prior on $\{A_+,A_\times, \phi_+, \phi_\times\}$, as defined in \eq{Aphi}, over the ranges $0 < A_{+/\times} \leq A_{\max}$ and $0 < \phi_{+/\times} < 2\pi$.
  The blue distribution restricts parameters to the targeted range $0 < A < A_{\max}$, as in the example of Fig.~\ref{fig:jac_example}.
  The 4D density is inversely proportional to the $J_2$ Jacobian in \eq{jac_Aphi}, which is represented by blue lines (over the marginals for $A$ and $\phi$, and in the inset for $\epsilon$ and $\theta$); in particular, the inset shows logarithmically-spaced contours of the $(\epsilon,\theta)$-dependent piece of $1/J_2$.
  The difference between the blue and gray distributions for $\theta$ and $\phi$ is explained in the same way as in Fig.~\ref{fig:jac_Aeps_Arl_angles}.
  The distribution represented here heavily favors linear polarizations ($\epsilon = 0$) along the $+/\times$ directions defined by \eq{Aphi}.
  }
  \label{fig:jac_Aphi}
  \end{figure}

Both those features are visible in Fig.~\ref{fig:jac_Aphi}, which shows the distribution imposed on all of our canonical parameters, $\{A,\epsilon, \theta, \phi\}$, by drawing uniformly in $0 < A_{+/\times} < A_{\max}$ and $0 < \phi_{+/\times} < 2\pi$, for some arbitrary scale $A_{\rm max}$.
The distribution increases proportionally with $A$ up to $A_{\max}$ and peaks strongly at $\epsilon = 0$, sharply favoring linear polarizations.
In fact, the $\theta$ dependence of \eq{jac_Aphi} implies that pure $+$ or $\times$ polarizations ($\theta=0$ or $\theta = \pm\pi/2$, respectively) will be favored over any other orientation, i.e., pure linear polarizations aligned with the frame used to define $+$ and $\times$ in \eq{Aphi}.

The sharpness of the $\epsilon$ and $\theta$ features in Fig.~\ref{fig:jac_Aphi} suggests that fully correcting for the Jacobian in \eq{jac_Aphi} will be challenging in sampling applications.
Therefore, the parametrization of \eq{Aphi} is likely nonperformant if the goal is to obtain results under a uniform prior in $\{A,\epsilon\}$---we found this to be the case in practice in the context of \cite{Chatziioannou:2021mij}.
The parameterization is otherwise also likely undesirable if there is no known orientation of the polarization frame to favor in writing down \eq{Aphi}, i.e., in the language of Sec.~\ref{sec:angles}, if there is no a priori preferred polarization angle $\psi$.

\subsection{Linear polarization quadratures}
\label{sec:jac:Axy}

In the previous section we introduced the linear polarization quadratures $\xpc \equiv A_{+/\times} \cos \phi_{+/\times}$ and $\ypc \equiv A_{+/\times} \sin \phi_{+/\times}$, \eq{xy}, which are the Cartesian components (real and imaginary) corresponding to the complex-valued Jones amplitudes that encode the polarization state of the signal (see Fig.~\ref{fig:diagram_apac} and Sec.~\ref{sec:math}).
In \eq{Aphi_to_Cphi} we used these quantities to conveniently express the relation between the phases of the linear components of \eq{Aphi} and those of their circular counterparts, \eq{ellip_circ}, but their usefulness extends more widely.
Notably, the quadratures are usually more suitable for sampling applications, since working with periodic phases like $\phi_{+/\times}$ can be problematic for stochastic algorithms like Markov chain Monte Carlo (MCMC) \cite{Hogg:2017akh}.
Together with Gaussian priors, they can also make some problems analytically integrable \cite{Hogg:2020jwh}.

The usefulness of the linear-polarization quadratures stems from the fact that, unlike phase parameters like $\phi_{+/\times}$, they enter the waveform linearly.
Concretely, the expression for an elliptical monochromatic mode in terms of these quantities is
\begin{subequations} \label{eq:Axy}
\begin{equation}
h_+ = \xp \cos \omega t + \yp \sin \omega t \, 
\end{equation}
\begin{equation}
h_\times = \xc \cos \omega t + \yc \sin \omega t \, .
\end{equation}
\end{subequations}
The relation of $\xpc$ and $\ypc$ to the linear polarizations of \eq{Aphi} is given directly by the definition in \eq{xy}; such relation implies a transformation into the circular-polarization parameters given by
\begin{equation} \label{eq:Cphi_to_Axy}
\begin{cases}
A_R^2 = \frac{1}{2}\left[\left(\yp - \xc\right)^2 + \left(\xp + \yc\right)^2\right] \\
A_L^2 = \frac{1}{2}\left[\left(\yp + \xc\right)^2 + \left(\xp - \yc\right)^2\right] \\
\phi_R = \mathrm{atan2}\left(\yp -\xc, \xp + \yc \right)\\
\phi_L = \mathrm{atan2}\left(\yp + \xc, \xp - \yc \right) 
\end{cases} ,
\end{equation}
where last two lines are the same as in \eq{Aphi_to_Cphi}.
The inverse transformation is, as one might expect from Fig.~\ref{fig:diagram_apac},
\begin{equation} \label{eq:Axy_to_Arl}
\begin{cases}
\xp = \frac{1}{\sqrt{2}} \left( \xr + \xl \right)\\
\yp = \frac{1}{\sqrt{2}} \left( \yl - \yr \right)\\
\xc = \frac{1}{\sqrt{2}} \left( \yr + \yl \right)\\
\yc = \frac{1}{\sqrt{2}} \left( \xr - \xl \right)\\
\end{cases} ,
\end{equation}
for circular-polarization quadratures defined as $\xrl \equiv A_{R/L}\cos\phi_{R/L}$ and $\yrl \equiv A_{R/L} \sin \phi_{R/L}$.

From Eqs.~\eqref{eq:Cphi_to_Aellip} and \eqref{eq:Axy_to_Arl}, we can then derive a relation between $\{\xpc, \ypc\}$ and the canonical parameters $\{A,\epsilon,\theta,\phi\}$.
The inverse transformation, from  $\{A,\epsilon,\theta,\phi\}$ into $\{\xpc, \ypc\}$ is easier to express succinctly and is given by
\begin{equation} \label{eq:Axy_to_Aeps}
\begin{cases}
\xp = A \left(\cos\theta \cos\phi + \epsilon \sin\theta \sin\phi\right)\\
\yp = A \left(\cos\theta \sin\phi - \epsilon \sin\theta \cos\phi\right)\\
\xc = A \left(\sin\theta \cos\phi - \epsilon \cos\theta \sin\phi\right)\\
\yc = A \left(\sin\theta \sin\phi + \epsilon \cos\theta \cos\phi\right)
\end{cases} ,
\end{equation}
as is straightforward to check based on \eq{Axy} and \eq{hcomp_ellip} by basic trigonometry.

The corresponding Jacobian is remarkably simple when expressed in terms of the ellipse amplitude and shape,
\begin{subequations} \label{eq:jac_Aeps_Axy}
\begin{align}
J_3 &\equiv 2 \left\{\sqrt{\left(\yp -\xc \right)^2 +\left(\xp +\yc\right)^2}
\right. \nonumber \\
&\times \sqrt{\left(\yp + \xc\right)^2 + \left(\xp - \yc\right)^2}  \nonumber\\
&\times \left[ \sqrt{\left(\yp -\xc \right)^2 +\left(\xp +\yc\right)^2} \right.  \nonumber\\
&+ \left.\left. \sqrt{\left(\yp + \xc\right)^2 + \left(\xp - \yc\right)^2} \right]\right\}^{-1} \\
&= \frac{1}{A^{3} \left(1 - \epsilon^2\right)}  .
\end{align}
\end{subequations}
This Jacobian again factorizes into a piece for the size of the polarization ellipse and another for its shape, but without a dependence on the ellipse orientation.
The scale-dependent factor ($1/A^3$) indicates that a flat prior on $\{\xpc,\ypc\}$ will strongly favor large signal amplitudes.
Like in \eq{jac_Aphi}, this $J_3$ Jacobian diverges for $\epsilon = \pm 1$ which means that circular polarizations will be disfavored, albeit less strongly than $J_2$.
The lack of dependence of $J_3$ on the orientation ellipse $\theta$ indicates that no specific polarization frame is preferred by this prior, reflecting the isotropy built into the definition of $\xpc, \ypc$.

\begin{figure}
  \includegraphics[width=\columnwidth]{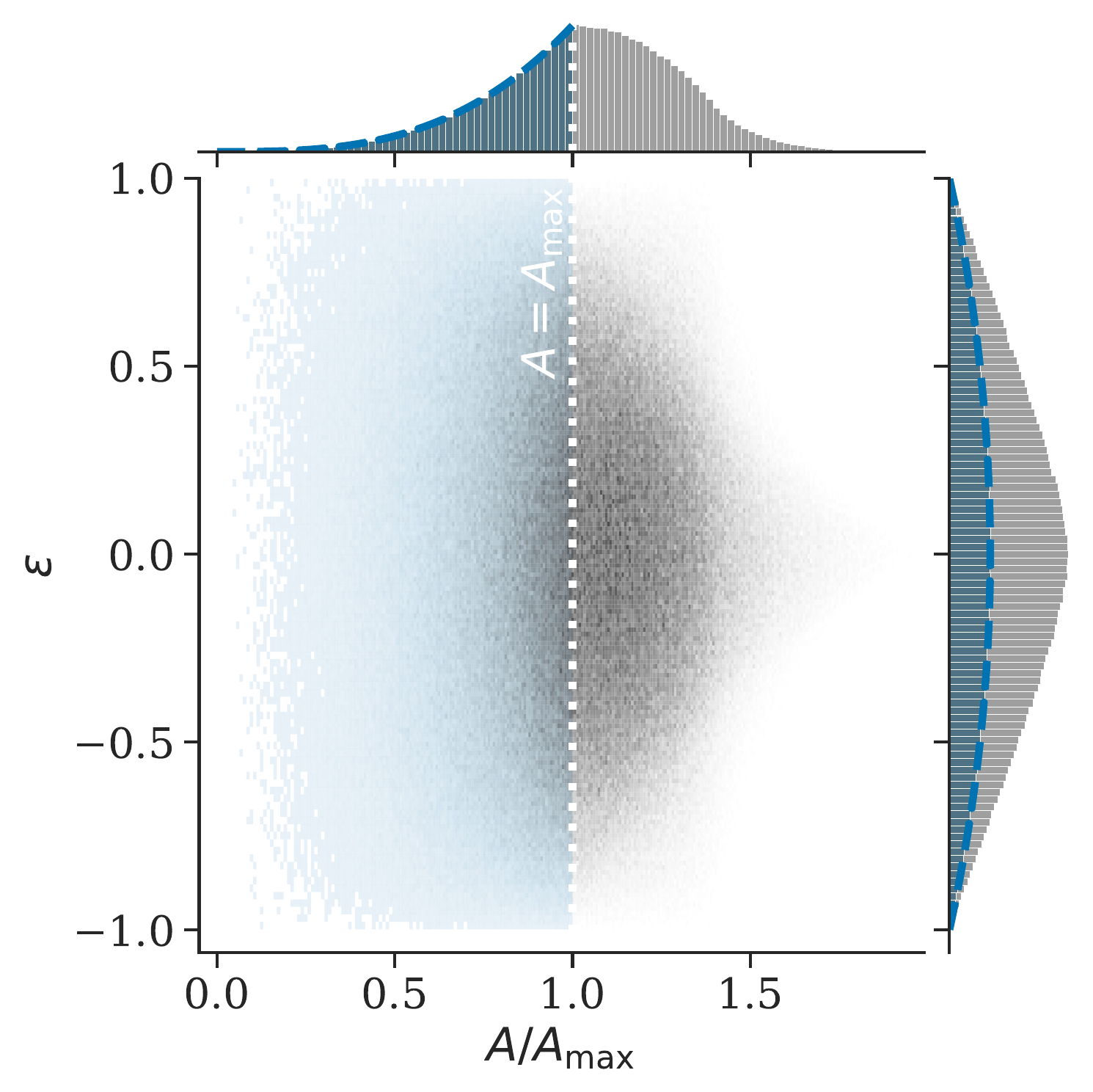}
  \caption{Distribution imposed on $\{A,\epsilon\}$ by applying a flat prior on $\{\xp, \yp, \xc, \yc\}$, as defined in \eq{xy}, over the ranges $- A_{\max} < \xpc,\ypc \leq A_{\max}$.
  The blue distribution restricts the parameters to the targeted region, $A < A_{\max}$, as explained in the example of Fig.~\ref{fig:jac_example}.
  Over that range, the probability density is inversely proportional to the Jacobian in \eq{jac_Aeps_Axy}: favoring large amplitudes and linear polarizations (dashed blue curves).
  As in Fig.~\ref{fig:jac_Aphi}, the corresponding distribution over the angles $\theta$ and $\phi$ is isotropic when restricted to the appropriate range of $-\pi/2/ \leq \theta,\phi \leq \pi/2$ (not shown).
  }
  \label{fig:jac_Axy}
  \end{figure}

The features described above are visible in the distribution imposed on $\{A,\epsilon,\theta,\phi\}$ by drawing uniformly on $\{\xpc, \ypc\}$, as shown in Fig.~\ref{fig:jac_Axy} for $A$ and $\epsilon$.
Over the targeted region ($A < A_{\max}$) the distribution steeply favors high signal amplitudes; it also favors linear polarizations ($\epsilon = 0$), although less sharply than in Fig.~\ref{fig:jac_Aphi}.
Enforcing $A < A_{\max}$, no specific value of $\theta, \phi$ is preferred; however, a similar structure to that in Fig.~\ref{fig:jac_Aphi} would appear unless explicitly mitigated, as explained in that case.
The constraint on the amplitude, $A < A_{\rm max}$, is crucial to guarantee isotropy in the ellipse orientation: without it, the corners of the squares defined by $-A_{\max} < \xpc,\ypc < A_{\max}$ would result in special directions of high probability, just as in the example of Fig.~\ref{fig:jac_example}.
The same result could be obtained by applying an intrinsically isotropic prior in the $\{\xpc, \ypc\}$ space, e.g., uncorrelated Gaussians.

\subsection{Inclination of a planar source}
\label{sec:jac:cosi}

There are several applications for which it is desirable to make a connection between the ellipticity of a signal and the corresponding inclination angle of a source via \eq{ellip_cosi}.
This is because even unmodeled \added{signal} analyses, like \textsc{BayesWave}, often target sources that are dominated by the quadrupolar harmonic of the radiation ($\ell=|m|=2$), and that can be presumed to respect the planar symmetry that gave rise to that equation (see Sec.~\ref{sec:harmonics}).

In that case, a physically meaningful prior for the shape of the polarization ellipse is usually one that is uniform in $\cos\iota$, corresponding to an isotropic prior on the source orientation.
Such a prior is necessarily nonuniform in $\epsilon$, as can be inferred from the relation between the two quantities, illustrated in Fig.~\ref{fig:ellip_cosi}: uniform draws in $\cos\iota$ will necessarily favor circular polarizations over linear ones, since the $\epsilon$-vs-$\cos\iota$ curve flattens at the edges as $\cos\iota\to\pm 1$ and $\epsilon \to \pm 1$.
Conversely, a prior uniform in $\epsilon$ will necessarily disfavor face on ($\cos\iota=+1$) or face off ($\cos\iota=-1$) sources.

\begin{figure}
\includegraphics[width=\columnwidth]{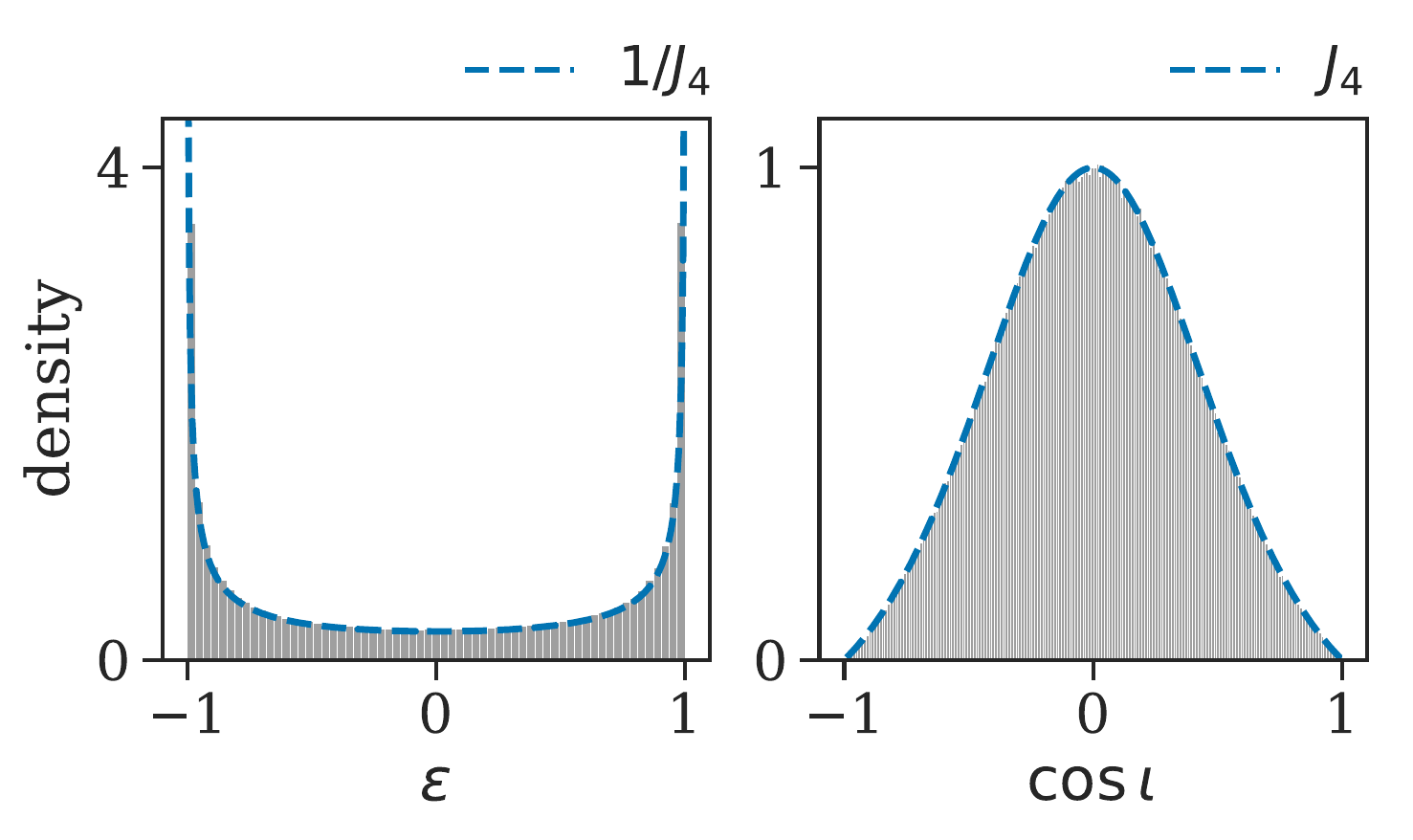}
\caption{\emph{Left:} Probability density imposed on $\epsilon$ by applying a flat prior in $\cos\iota$, which is inversely proportional to the Jacobian $J_4$ in \eq{jac_eps_cosi}; this diverges at $\epsilon = \pm 1$.
\emph{Right:} Distribution imposed on $\cos\iota$ by a flat prior in $\epsilon$, which is directly proportional to \eq{jac_eps_cosi}; this vanishes at $\cos\iota = \pm 1$.
}
\label{fig:jac_cosi}
\end{figure}

Indeed, the Jacobian $J_4 \equiv \left|\partial\epsilon/\partial \cos\iota\right|$, transforming from $\cos\iota$ to $\epsilon$, is
\begin{equation} \label{eq:jac_eps_cosi}
J_4 =1 - \epsilon^2 + \sqrt{1-\epsilon^2} 
\propto \frac{1 - \cos^2\iota}{\left(1+\cos^2\iota\right)^2} \, ,
\end{equation}
which vanishes for $\cos\iota = \pm 1$ (or, equivalently, $\epsilon = \pm 1$) and peaks at $\cos\iota = 0$ ($\epsilon= 0$).
This indicates that a distribution uniform in $\cos\iota$ will place infinite weight on $\epsilon = \pm 1$, while a distribution uniform in $\epsilon$ will place no weight on $\cos\iota = \pm 1$ (left and right panels in Fig.~\ref{fig:jac_cosi}, respectively).

The divergences of the Jacobian above complicate transformations from one prior to the other, and suggest their implementation in sampling applications is likely nonperformant---in other words, if the goal is to apply a uniform prior in $\cos\iota$, then we should sample in that quantity directly, not in $\epsilon$.

This issue becomes more pronounced if, rather than being uniform in $\epsilon$, the original prior itself disfavored $\epsilon = \pm 1$ in the first place.
This was the case for the parametrization in terms of $\hat{A}$ and $\chi$ in Sec.~\ref{sec:jac:Achi}, the linear polarization amplitudes in Sec.~\ref{sec:jac:Apc} and the linear polarization quadratures in Sec.~\ref{sec:jac:Axy}.
Of all these, the problem is most severe for the linear polarization amplitudes, since that parametrization places heavy weight (formally infinite) on $\epsilon = 0$ (Fig.~\ref{fig:jac_Aphi}).
Unfortunately, this was the parametrization used in \cite{Chatziioannou:2021mij}, which likely explains the difficulty in recovering the sky location of the circularly-polarized ($\cos\iota=1$) signal simulated in Fig.~11 of that work.

\section{Nontensor polarizations}
\label{sec:nongr}

\newcommand{\xsym}{\ensuremath{x}}
\newcommand{\ysym}{\ensuremath{y}}
\newcommand{\bsym}{\ensuremath{b}}
\newcommand{\lsym}{\ensuremath{l}}
\newcommand{\hx}{h_{\xsym}}
\newcommand{\hy}{h_{\ysym}}
\newcommand{\hb}{h_{\bsym}}
\newcommand{\hlon}{h_{\lsym}}

Metric theories beyond GR may allow for up to six independent polarizations, including the two tensor $+$ and $\times$ \replaced{polarizations}{modes} expected in GR \cite{Eardley:1973br,Eardley:1973zuo}.
\added{The presence of additional nontensor polarizations is a generic feature of many extensions of GR, and their detection would represent a smoking gun for new physics \cite{Thorne:1973zz,Will:2014kxa,Chatziioannou:2012rf}.
Some constraints on their existence have been placed by LIGO-Virgo \cite{LIGOScientific:2017ycc,LIGOScientific:2017ous,LIGOScientific:2018czr,LIGOScientific:2018dkp,LIGOScientific:2019fpa,LIGOScientific:2020tif}, and are expected to improve significantly with future observations \cite[e.g.,][]{Chatziioannou:2021mij,Cornish:2017oic}. 
Data analysis methods targeting nontensor GWs require extending the formalism of GW polarizations beyond the plus and cross polarizations treated above \cite{Isi:2015cva,Isi:2017equ,Callister:2017ocg,Isi:2017fbj,Chatziioannou:2021mij,Romano:2016dpx}.}

\replaced{Fortunately, the generalization to nontensor polarizations is straightforward}{The discussion above generalizes easily to include those additional modes}, starting with an enhanced version of the strain tensor in Eq.~\eqref{eq:hij},
\beq \label{eq:hij_ngr}
(h_{ij}) = \begin{pmatrix}
\hb + h_+ & h_\times  & \hx  \\
h_\times  & \hb - h_+ & \hy  \\
\hx    & \hy    & \hlon
\end{pmatrix} ,
\eeq
where, in addition to plus and cross, also appear the vector-$x$ (\xsym) and vector-$y$ \replaced{polarizations}{modes} (\ysym), as well as the scalar breathing (\bsym) and longitudinal (\lsym) \replaced{polarization}{mode}.%
\footnote{There are other possible normalizations in use in the literature, e.g., $\hlon \to \sqrt{2} \hlon$.}
Equivalently, as above, we can write this as a weighted sum over generalized polarization tensors,
\beq
h_{ij} = \sum_p h_p\, e^p_{ij} \, ,
\eeq
for $p$ in $\{+,\times, \xsym, \ysym, \bsym, \lsym\}$, and polarization tensors $e^p_{ij}$ defined implicitly by comparison with Eq.~\eqref{eq:hij_ngr}.
Generally, the $h_p$ are functions of time, as for plus and cross above.
With similar assumptions as in the GR case, the detector output can be written as a sum over polarizations weighted by antenna patterns,
\beq \label{eq:h_ngr}
h(t) = \sum_p F_p(\alpha, \delta; \psi)\, h_p(t)\, ,
\eeq
with $F_p \equiv D^{ij} e^p_{ij}$ as before. 
The physical effect of the non-GR polarizations is encoded in the antenna patterns, and is illustrated in, e.g., Fig.~1 of \cite{Isi:2017equ}.

\begin{figure}
  \subfloat[Breathing scalar]{\includegraphics[trim={2cm 2cm 2cm 2cm},width=0.19\columnwidth]{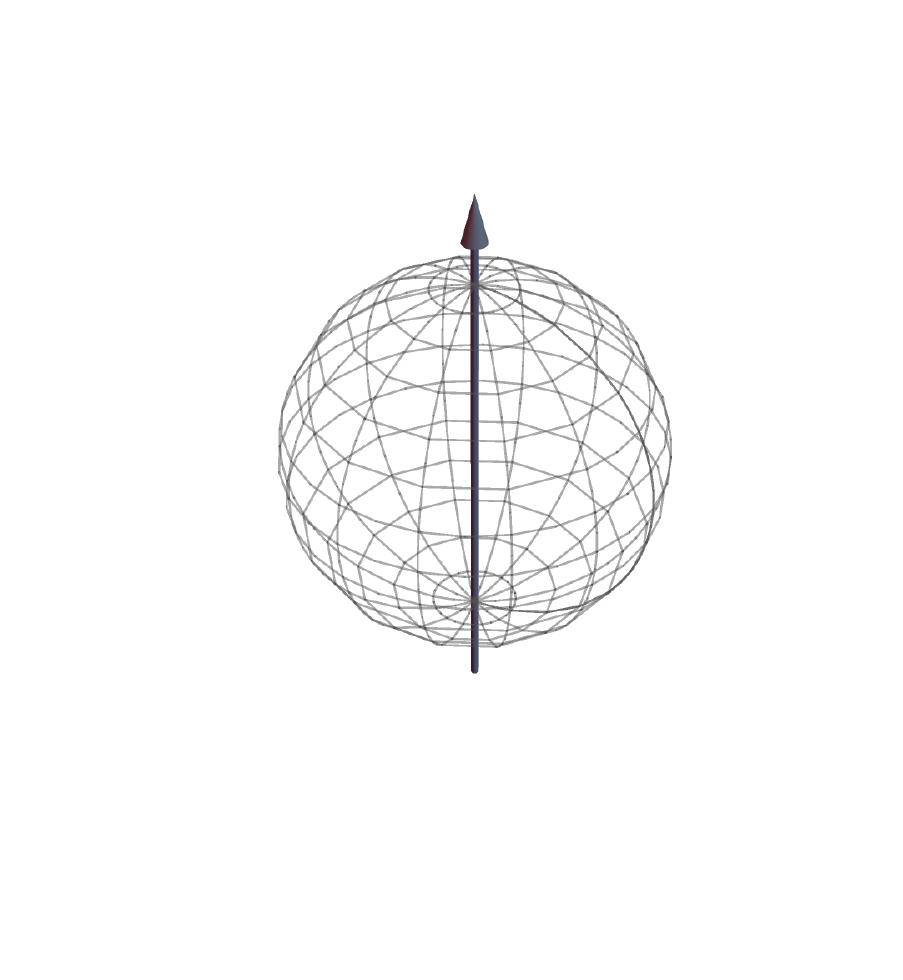}
\includegraphics[trim={2cm 2cm 2cm 2cm},width=0.19\columnwidth]{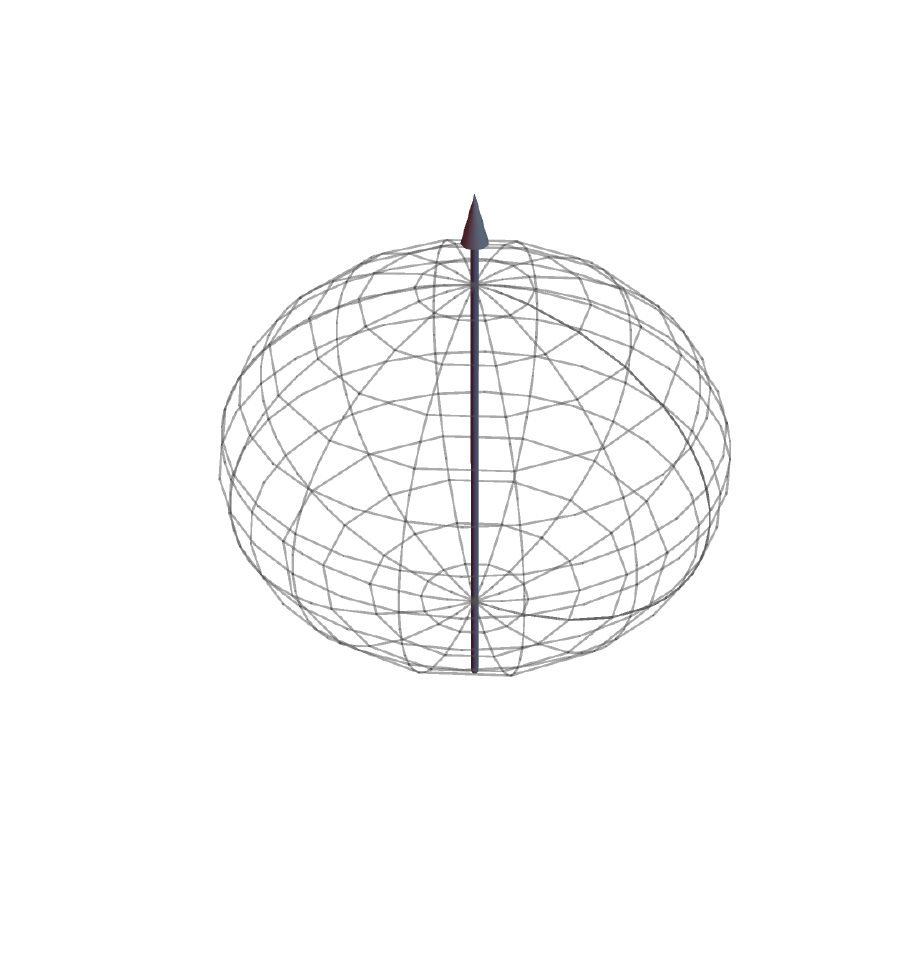}
\includegraphics[trim={2cm 2cm 2cm 2cm},width=0.19\columnwidth]{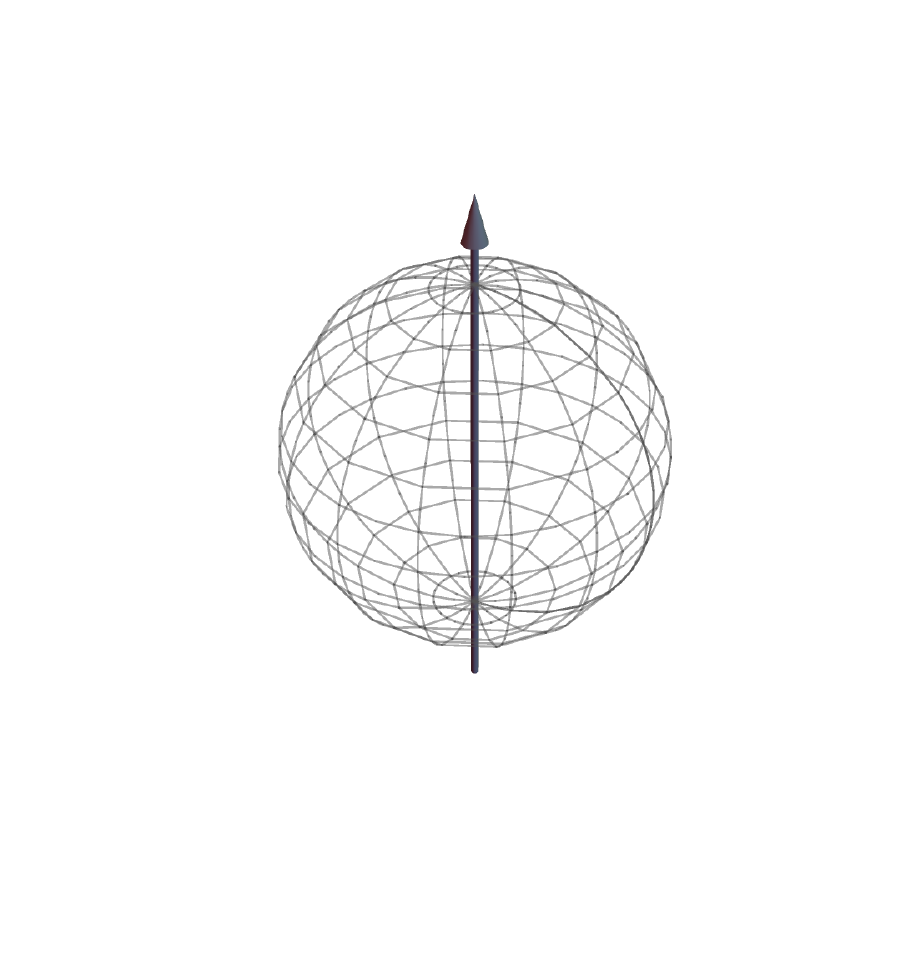}
\includegraphics[trim={2cm 2cm 2cm 2cm},width=0.19\columnwidth]{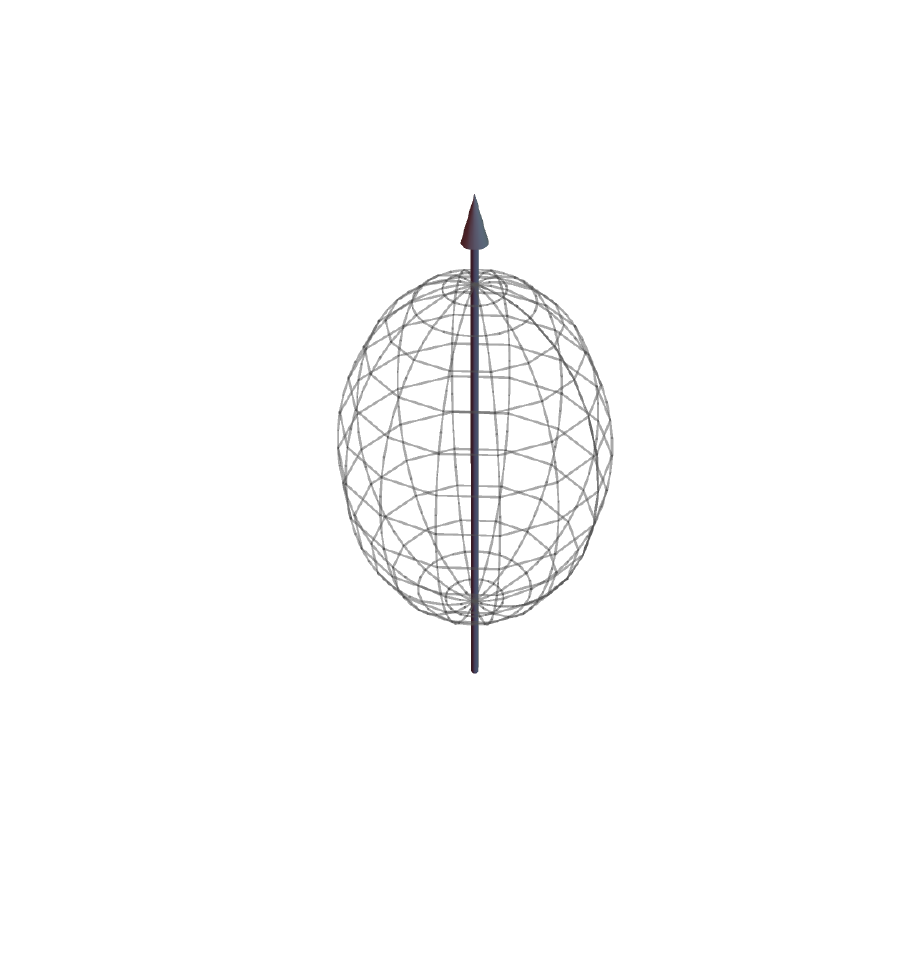}
\includegraphics[trim={2cm 2cm 2cm 2cm},width=0.19\columnwidth]{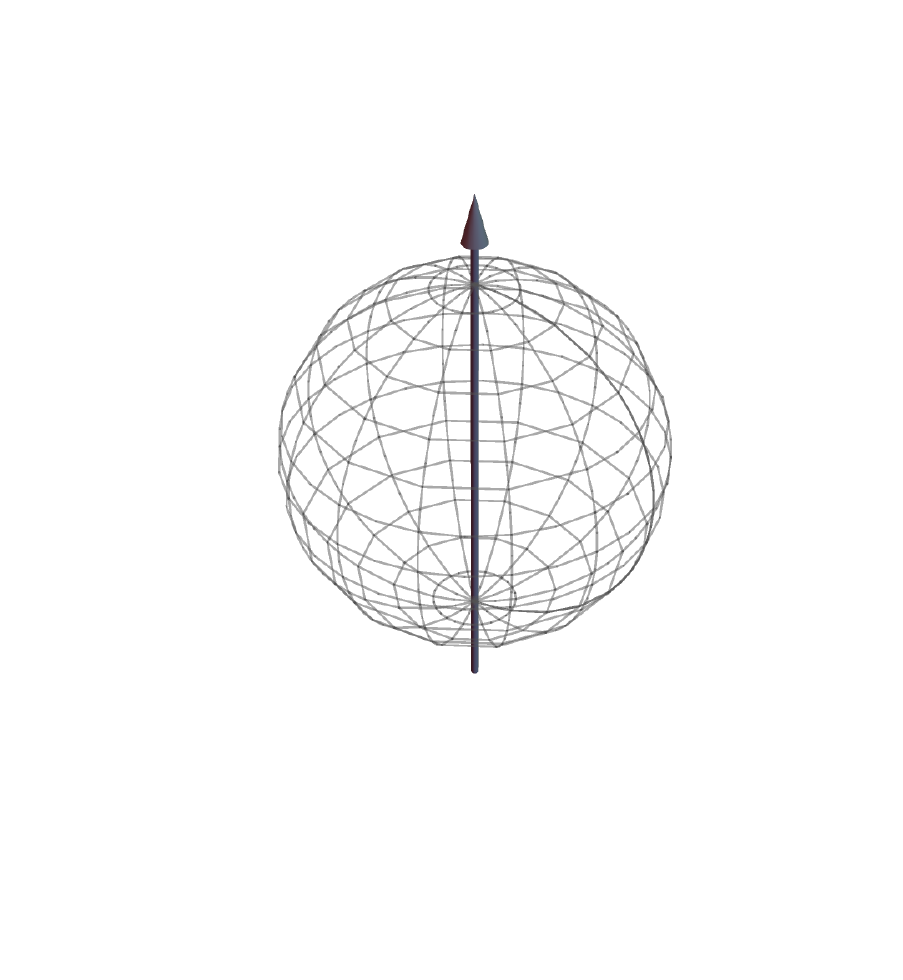}}\\
\subfloat[Longitudinal scalar]{\includegraphics[trim={2cm 2cm 2cm 2cm},width=0.19\columnwidth]{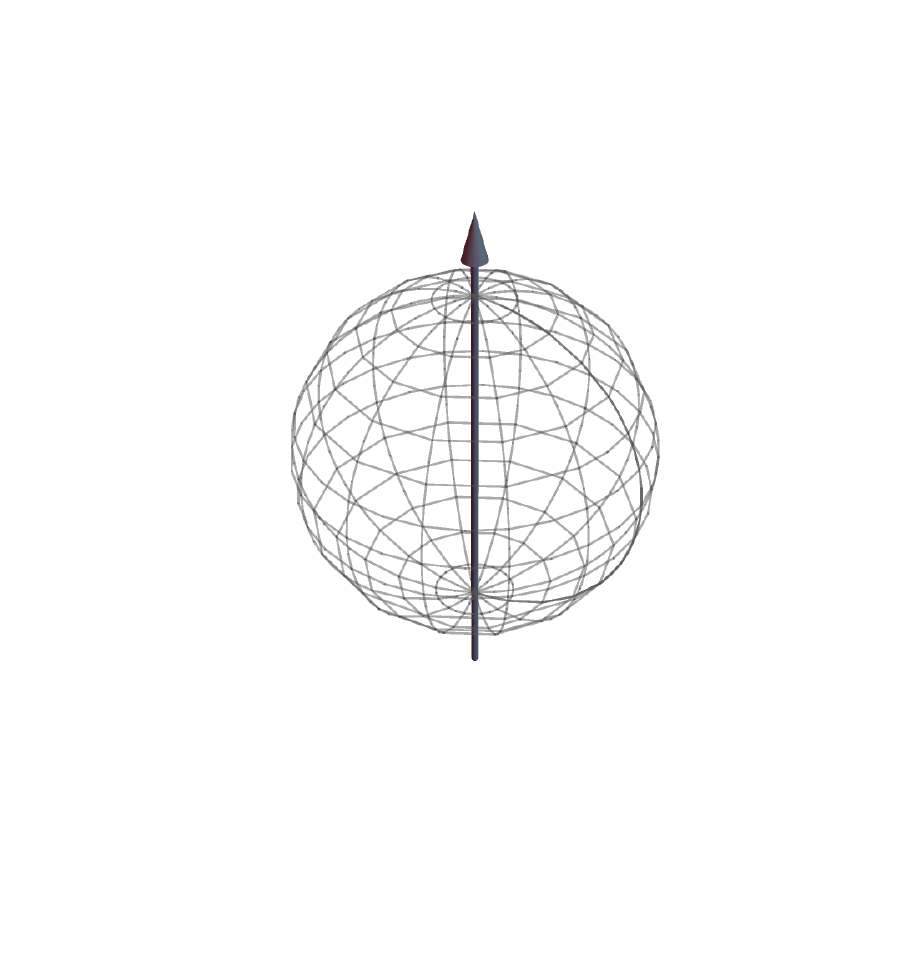}
\includegraphics[trim={2cm 2cm 2cm 2cm},width=0.19\columnwidth]{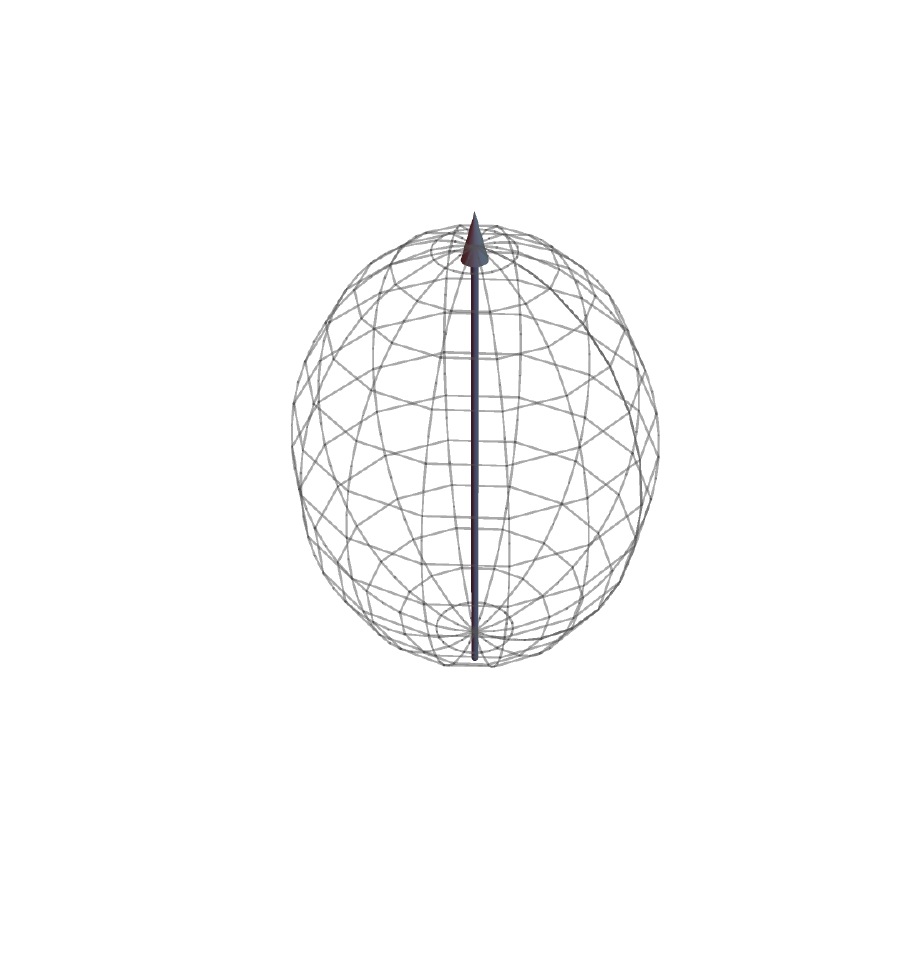}
\includegraphics[trim={2cm 2cm 2cm 2cm},width=0.19\columnwidth]{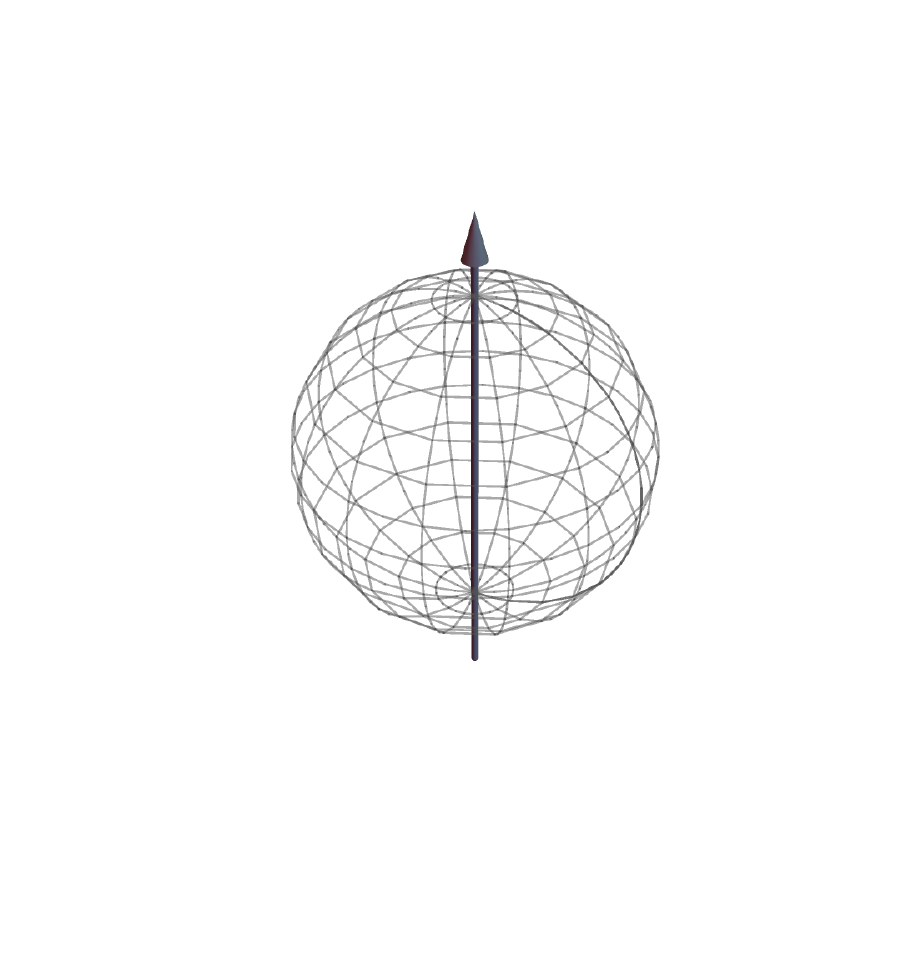}
\includegraphics[trim={2cm 2cm 2cm 2cm},width=0.19\columnwidth]{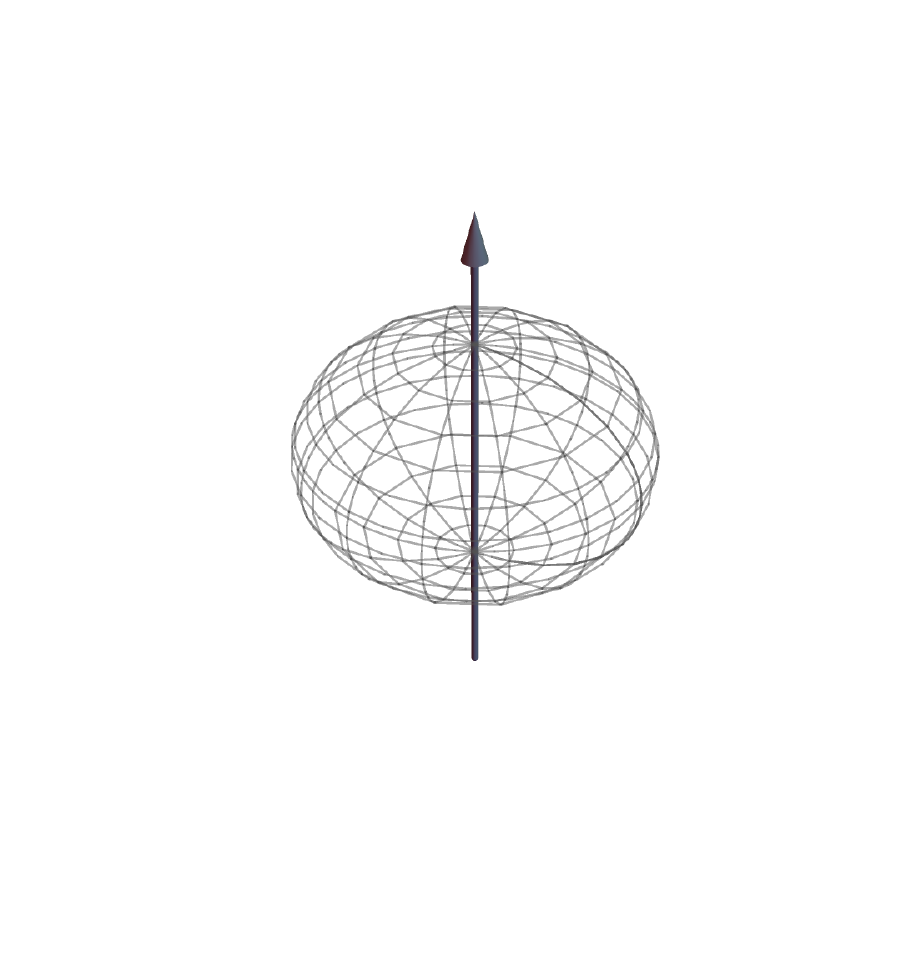}
\includegraphics[trim={2cm 2cm 2cm 2cm},width=0.19\columnwidth]{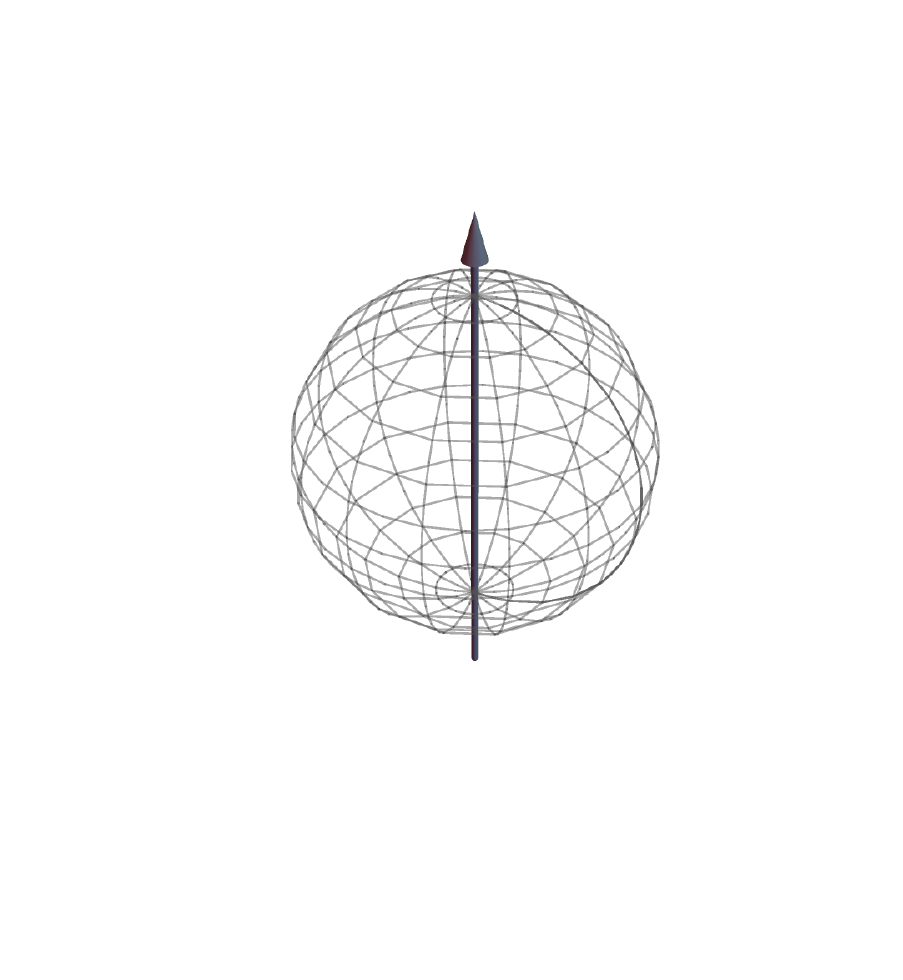}}\\
\subfloat[Traceless scalar]{\includegraphics[trim={2cm 2cm 2cm 2cm},width=0.19\columnwidth]{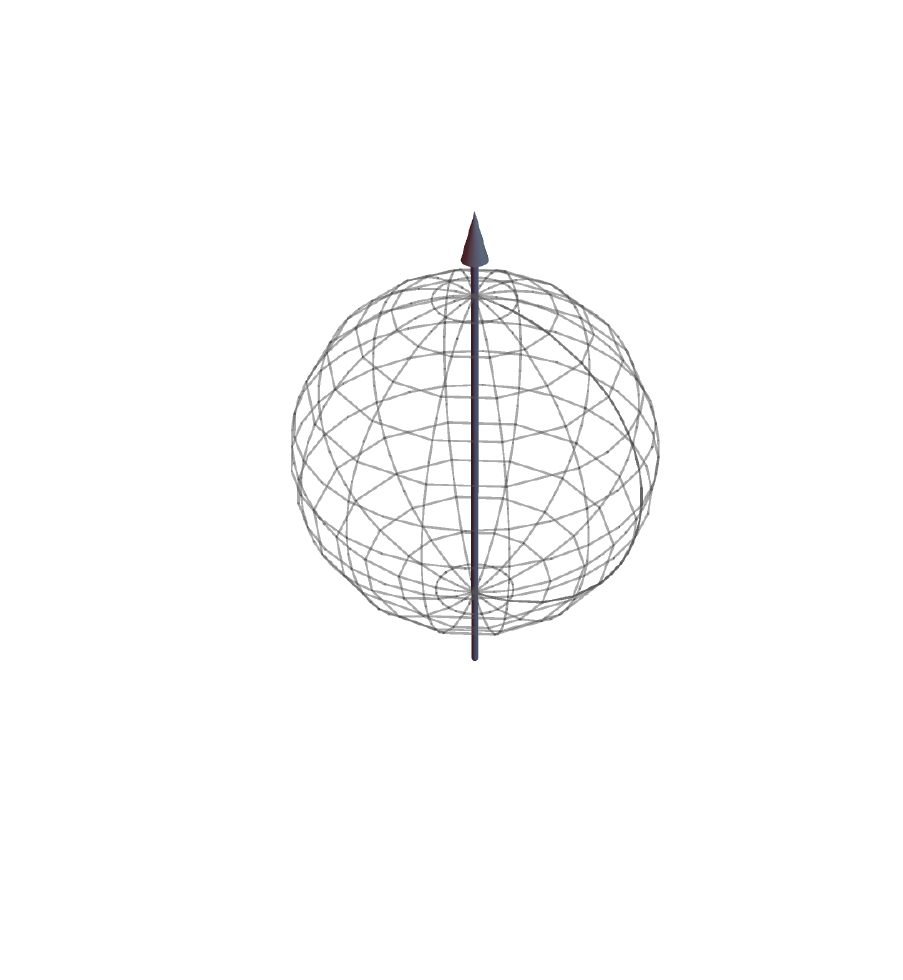}
\includegraphics[trim={2cm 2cm 2cm 2cm},width=0.19\columnwidth]{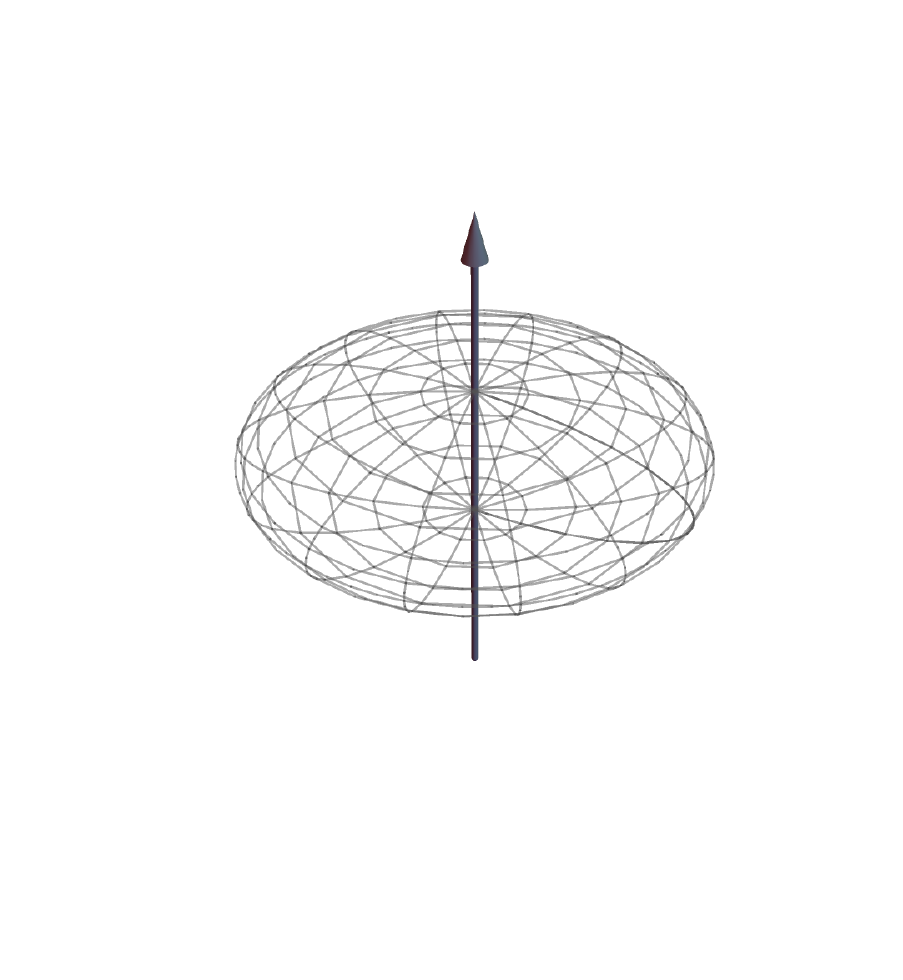}
\includegraphics[trim={2cm 2cm 2cm 2cm},width=0.19\columnwidth]{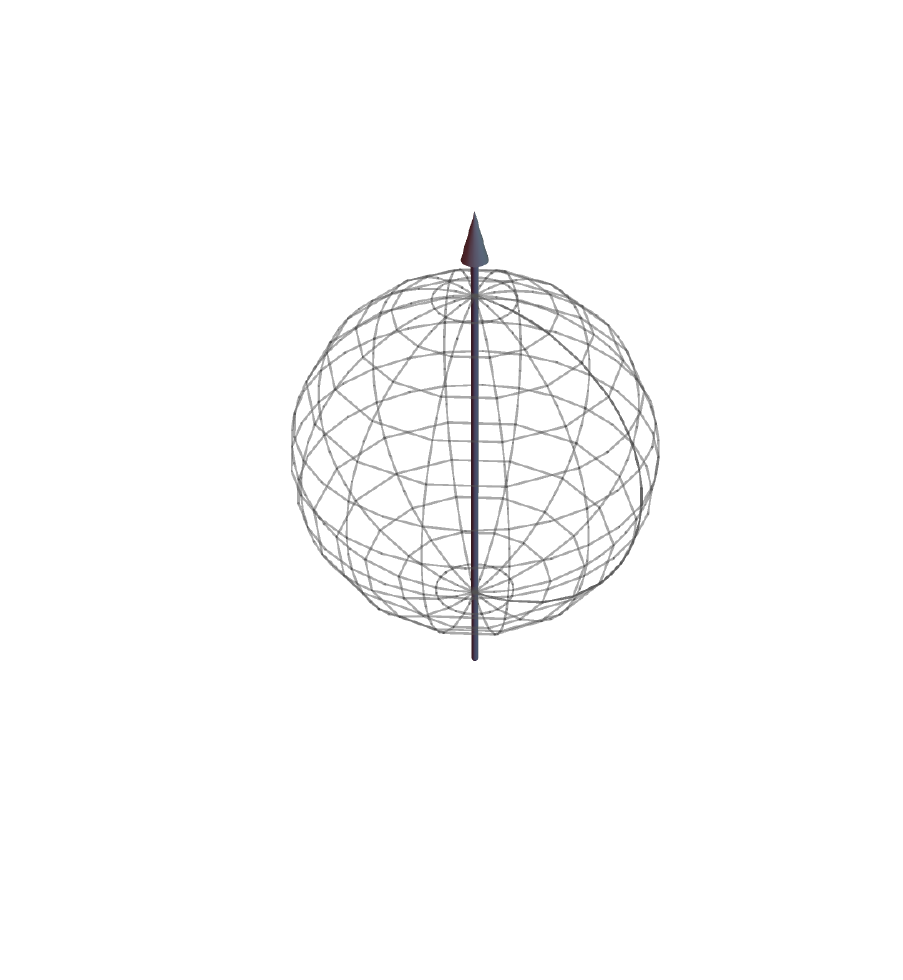}
\includegraphics[trim={2cm 2cm 2cm 2cm},width=0.19\columnwidth]{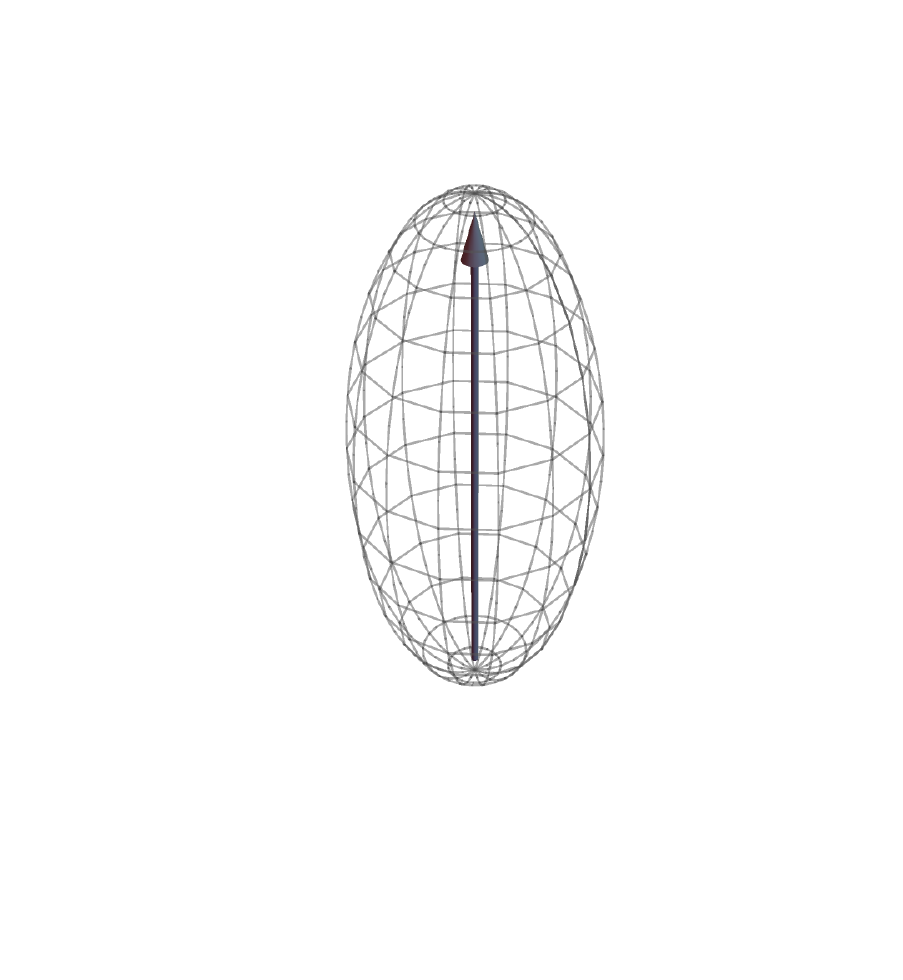}
\includegraphics[trim={2cm 2cm 2cm 2cm},width=0.19\columnwidth]{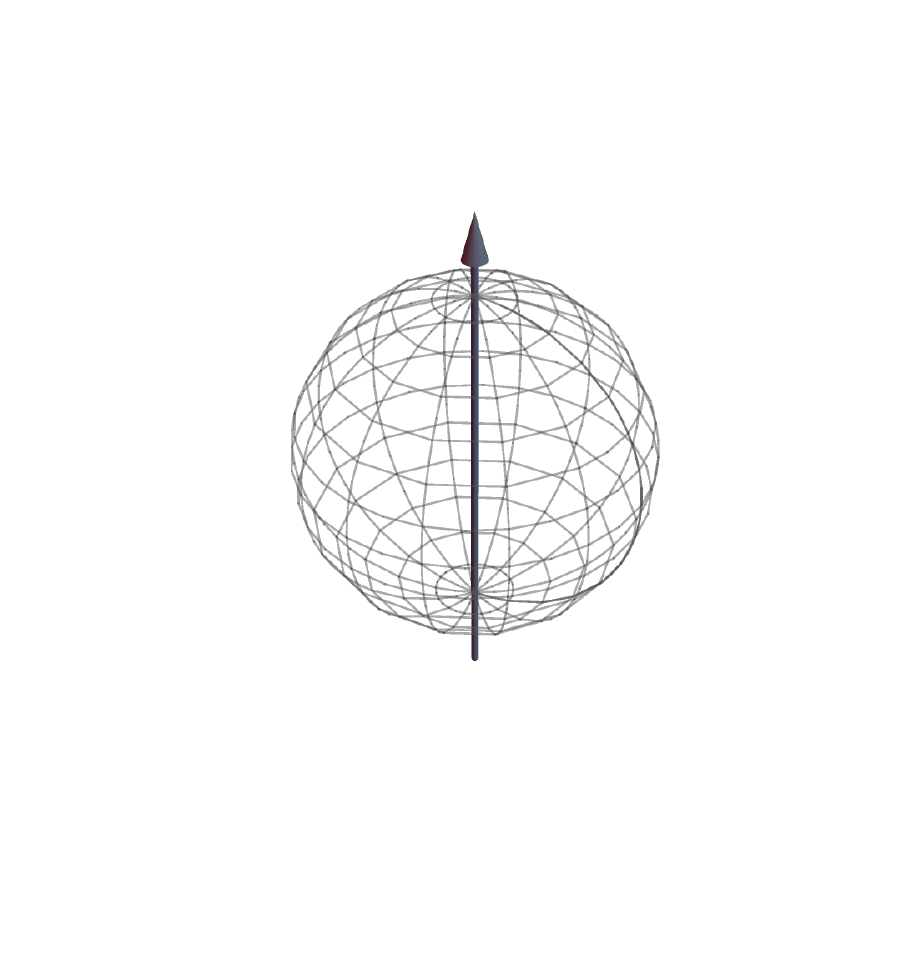}}\\
\subfloat[Full-trace scalar]
{\includegraphics[trim={2cm 2cm 2cm 2cm},width=0.19\columnwidth]{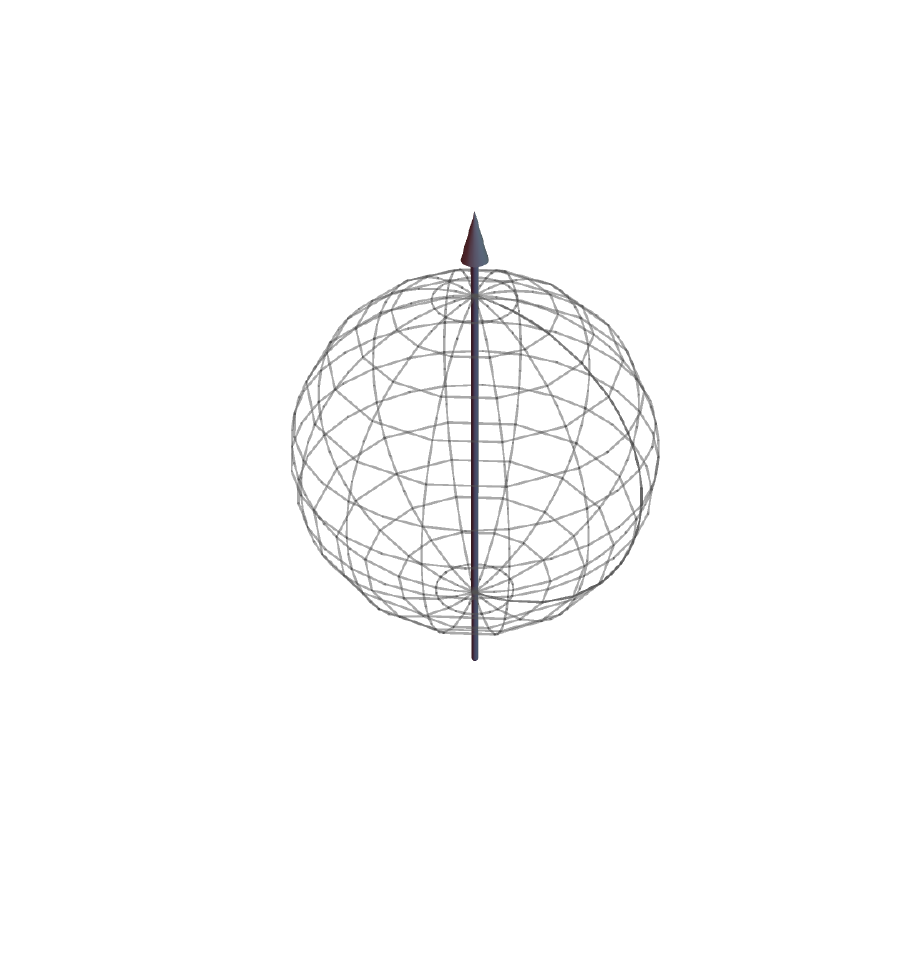}
\includegraphics[trim={2cm 2cm 2cm 2cm},width=0.19\columnwidth]{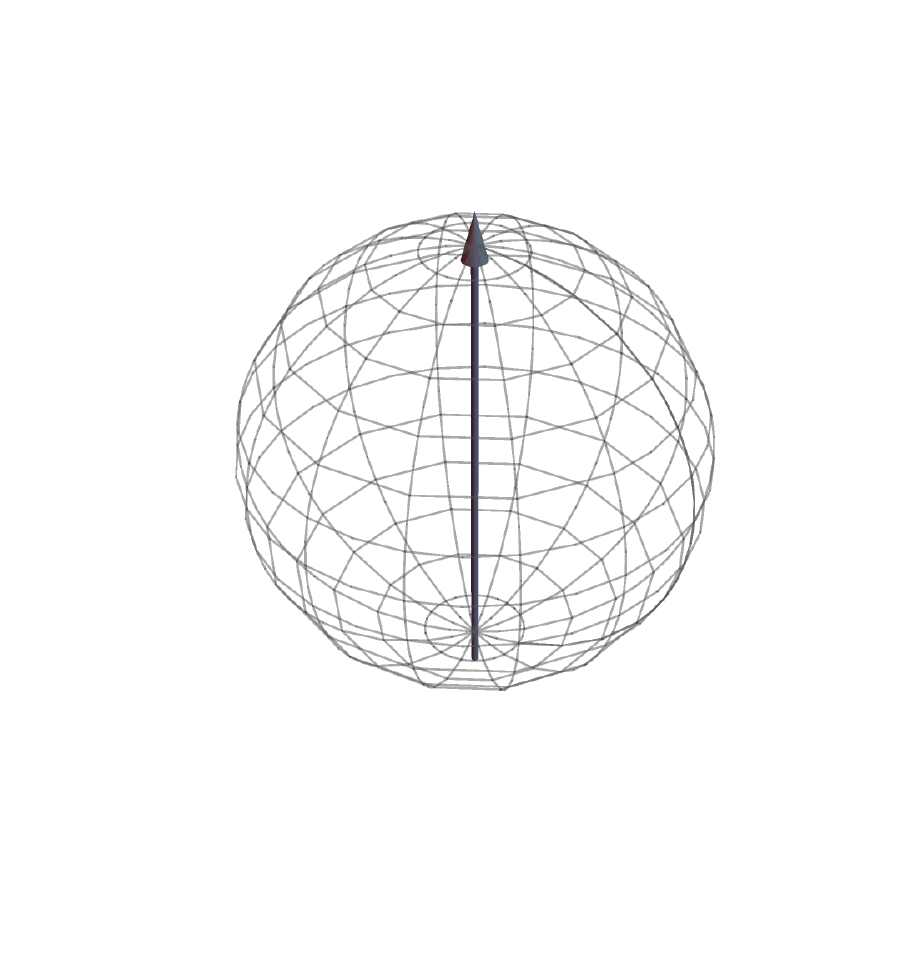}
\includegraphics[trim={2cm 2cm 2cm 2cm},width=0.19\columnwidth]{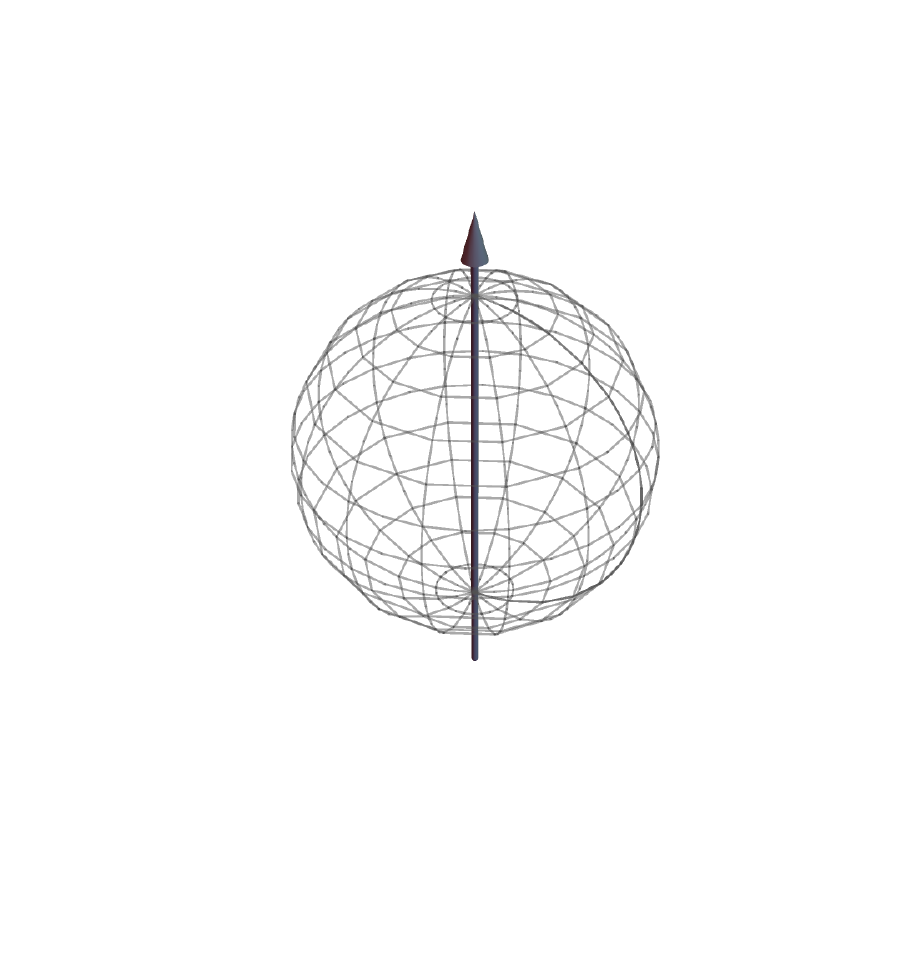}
\includegraphics[trim={2cm 2cm 2cm 2cm},width=0.19\columnwidth]{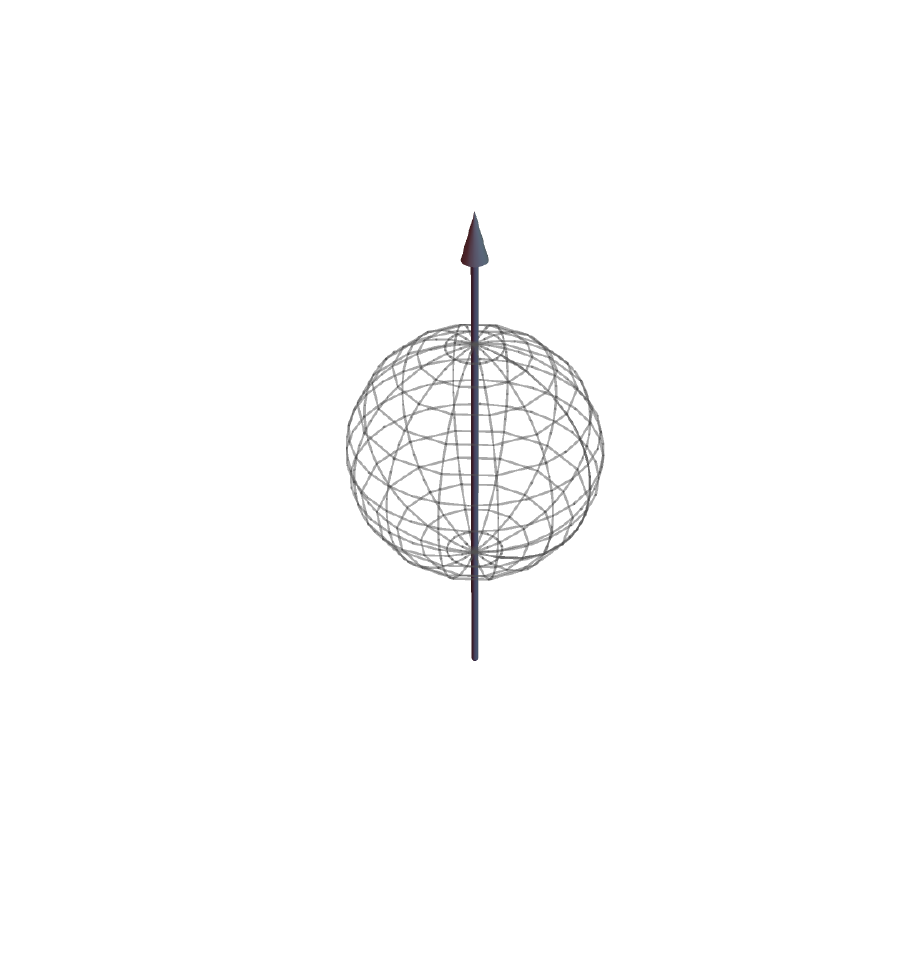}
\includegraphics[trim={2cm 2cm 2cm 2cm},width=0.19\columnwidth]{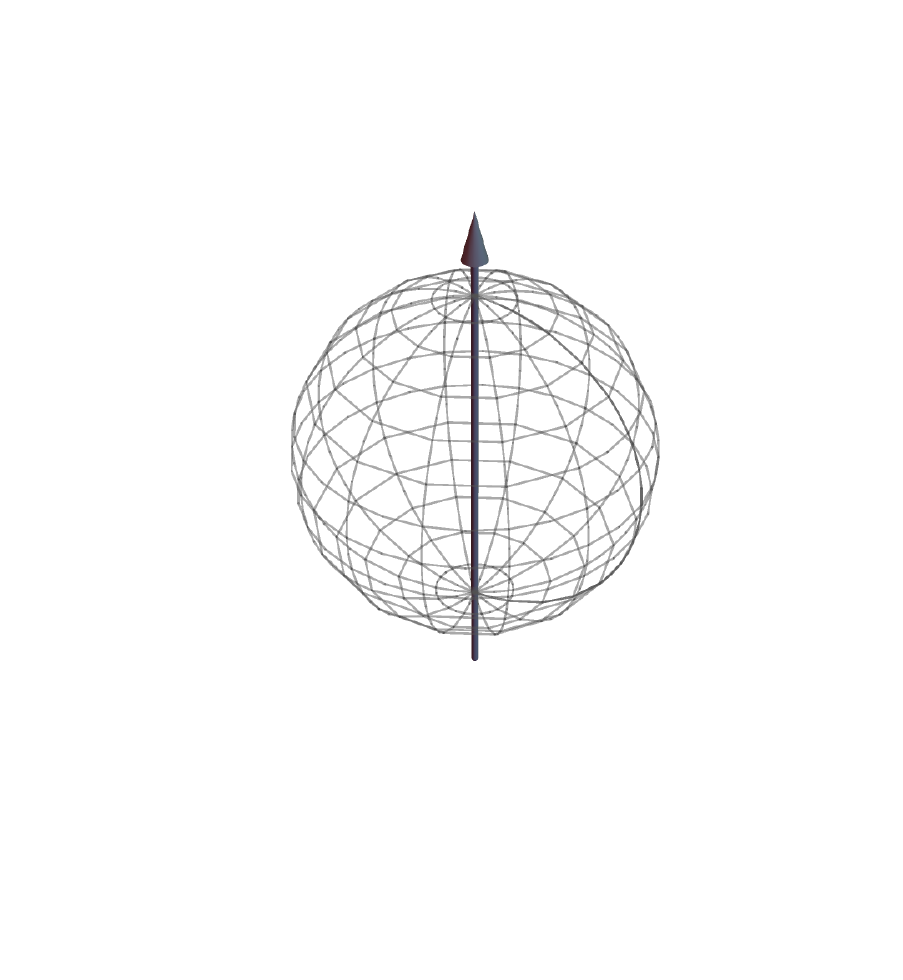}}
\caption{\protect\added{Scalar polarizations represented through their effect on a small sphere of freely-falling particles (mesh), as they propagate in the $z$-direction (arrow). Each row shows a different polarization at half-period intervals (columns), so that time evolves from left to right.
(a) The \emph{breathing} polarization stretches and squeezes space uniformly in the $x$-$y$ plane, and does not affect the $z$ direction.
(b) The \emph{longitudinal} polarization stretches and squeezes the $z$ direction exclusively.
(c) The \emph{traceless} scalar polarization is the linear combination of breathing and longitudinal given in Eq.~\eqref{eq:scalar_traceless} and alters all three directions without changing the sphere's volume.
(d) The \emph{full-trace} scalar polarization stretches and squeezes space isotropically in all three directions, thus being undetectable by instruments like LIGO and Virgo.
The space of scalar polarizations is spanned by any two of these four eigenstates.
}}
\label{fig:scalar_modes}
\end{figure}

The considerations presented above regarding wave frame orientation and antenna pattern symmetries apply just as well to the generalized polarization tensor of Eq.~\eqref{eq:hij_ngr}, except for the different properties that the beyond-GR \replaced{polarizatoins}{modes} exhibit under rotations around the direction of propagation.
Polarizations of different spin weight do not mix with each other under rotations.

A rotation by $\Delta \psi$ around the line of propagation transforms the two vector amplitudes by
\begin{subequations} \label{eq:htransf_v}
\beq
\hx \rightarrow \hx' = \hx \cos \Delta \psi - \hy \sin \Delta\psi \, ,
\eeq
\beq
\hy \rightarrow \hy' = \hx \cos \Delta \psi + \hy \sin \Delta\psi \, ,
\eeq
\end{subequations}
reflecting the fact that these are the components of a spin weight $|s|=1$ field (hence ``vector'').
Accordingly, any transformation in which the polarization angle entered as $2\psi$ for the tensor \replaced{polarizations}{modes} will look the same for vector \replaced{polarizations}{modes} but with the angle entering simply as $\psi$.
In particular, the two vector \replaced{polarizations}{modes} allow for the definition of right and left handed combinations in full analogy with \eq{circ}, 
\beq \label{eq:circ_vec}
e^{v,R/L}_{ij} \equiv \frac{1}{\sqrt{2}} \left(e^\xsym_{ij} \pm i e^\ysym_{ij} \right) ,
\eeq
except that they correspond to eigenstates of the helicity operator with eigenvalues $\pm 1$, instead of $\pm 2$.
These circular vector \replaced{polarizations}{modes} transform, in analogy with \eq{htransf_circ}, by
\begin{subequations} \label{eq:htransf_circ_vec}
\begin{align}
h_{v,R} &\rightarrow h_{v,R}' = h_{v,R} \exp(- i  \Delta \psi) \, ,\\
h_{v,L} &\rightarrow h_{v,L}' = h_{v,L} \exp(+ i  \Delta \psi)\, .
\end{align}
\end{subequations}
Just like we can define circular vector modes, we can also construct elliptically polarized vector states.
These take on the same fundamental role for vector GWs as detailed in Sec.~\ref{sec:ellip_modes} for their tensor counterparts; in this case, the mathematical formalism for polarization states is identical to that of electromagnetic waves, which also correspond to a field of spin weight $\left|s\right|=1$.

On the other hand, the two scalar \replaced{polarizations}{modes} are invariant under rotations around $z$,
\begin{subequations} \label{eq:htransf_s}
\beq
\hb \rightarrow \hb' = \hb\, ,
\eeq
\beq
\hlon \rightarrow \hlon' = \hlon\, ,
\eeq
\end{subequations}
revealing that these behave as spin-weight $s=0$ fields (hence ``scalar'').
Since these \replaced{polarizations}{modes} are already invariant under rotations, there is no meaningful notion of a circular (or elliptical) scalar polarization.
Furthermore, in the small-antenna limit, differential-arm GW detectors are only sensitive to the traceless linear combination of the two scalar polarizations.
In terms of the breathing and longitudinal \replaced{polarizations}{modes} above, this is
\beq \label{eq:scalar_traceless}
h_{\rm s} \equiv \hb - 2\hlon\, ,
\eeq
which is the only scalar \replaced{polarization}{mode} measurable by existing detectors.
\added[comment=FN]{The effect of this traceless scalar polarizations is to simultaneously stretch (squeeze) along the $x$ and $y$ directions while squeezing (stretching) along the $z$ direction;}\added{the complementary fully-trace scalar polarization, $\hb + \hlon$, stretches and squeezes space isotropically in all three directions, making it undetectable by current detectors.
All scalar polarizations are illustrated in Fig.~\ref{fig:scalar_modes}.}

\replaced{As an equivalent phrasing of the above}{Equivalently}, the geometric antenna patterns for the breathing and longitudinal \replaced{polarizations}{modes} are the same up to an overall constant (with our normalization, $F_{\bsym} = -F_{\lsym}$).
Therefore, the two terms are degenerate in Eq.~\eqref{eq:h_ngr} and their contributions cannot be disentangled in a model-independent way, i.e., without theory- and source-specific information about the detailed morphology of the $\hb(t)$ and $\hlon(t)$ functions.
For unmodeled analyses, it thus suffices to include only one scalar term in Eq.~\eqref{eq:h_ngr}---commonly that for the breathing \replaced{polarization}{mode}---so that the sum is over only five polarizations instead of six.
\added{One must be mindful of the chosen parametrization when deriving empirical results and comparing them to theory (see, e.g., \cite{Isi:2018miq}.)}

The rest of the mathematical formalism covered in Sec.~\ref{sec:ellip_modes} can easily be extended to accommodate nontensor \replaced{polarizations}{modes}.
In particular, a generalized definition of Stokes parameters was derived in \cite{Anile1974} to account for all helicities---this requires 36 Stokes parameters.
However, the practical utility of such fully-generalized Stokes parameters is unclear, since the polarizations of different helicites do not mix into each other under rotations around the direction of propagation.
Instead, it is possible to simply enhance the set of four tensor Stokes parameters by an additional four vector Stokes parameters (defined analogously), and two parameters for the intensity of each of the scalar \replaced{polarizations}{modes};
this adds up to 12 polarization parameters, instead of 36, at the expense of ignoring potential coherence across \replaced{polarizations}{modes} of different spin weight.

\section{Conclusion}
\label{sec:conclusion}

We have reviewed in detail the mathematical treatment of GW polarizations as it pertains practical applications for GW data analysis.
We began by showing how any GW signal can be decomposed into linear (Fig.~\ref{fig:rings}), circular (Fig.~\ref{fig:pol_diagram_circ}) or elliptical (Fig.~\ref{fig:pol_diagram_ellip}) \replaced{polarizations}{modes}, after choosing a physical polarization frame.
Arguing for the conceptual importance of elliptical (i.e., fully-polarized) modes, we outlined several of their key properties and reviewed a number of standard mathematical tools (Jones vectors, Poincar\'{e} sphere, Stokes parameters) useful in their description.
Since a large number of signal morphologies can be captured by superpositions of fully-polarized states, we emphasized their practical importance for GW data analyiss in unmodeled (or loosely modeled) applications, as well as in connection to the decompositions of the GW strain from planar sources (e.g., nonprecessing \acp{CBC}) into spin-weighted spherical harmonic.

We then clarified the conceptual distinctions between different notions of ``polarization angle'' ($\psi$, $\theta$, and $\Psi$ or $\Omega$) and showed how the different angles can often (but not always) be used interchangeably in practice.
In the process, we described in detail the different coordinate frames that appear in the practice of GW data analysis, including for making waveform predictions, for describing the wave propagation, and for deriving the measured signal over a network of detectors.
The current LIGO-Virgo conventions for all these frames are illustrated in Figs.~\ref{fig:diagram_waveframe}, \ref{fig:diagram_sourceframe} and \ref{fig:diagram_skyview}, which clarify the relations between all the relevant angles.
(Appendix \ref{app:spins} describes an additional coordinate frame used to specify generic spins in a binary, even though it is not directly relevant to GW polarizations.)

To lay out the connections between analyses that make use of different polarization parametrizations, we computed Jacobians for the corresponding coordinate transformations.
This allowed us to understand the implications of parametrizations for the polarization of a signal assumed by, e.g., \textsc{BayesWave} or ringdown analyses.
We found that parametrizing the GW signal in terms of the circular polarization amplitudes (Fig.~\ref{fig:jac_Aeps_Arl}), the linear polarization amplitudes (Fig.~\ref{fig:jac_Aphi}), or the cosine and sine quadratures of the linear polarizations (Fig.~\ref{fig:jac_Axy}) leads to implicitly favoring high signal intensities.
The parametrizations in terms of the linear amplitudes or their quadratures, as well as a parametrization in terms of an ellipse shape angle $\chi$ (Fig.~\ref{fig:jac_Aeps_Achi}), lead to favoring linear polarizations.
This preference is particularly pronounced for the parametrization in terms of the linear amplitudes, which also picks a preferred direction for the polarization ellipse, aligned with the plus and cross axes as determined by the implicit definition of the physical waveframe (choice of $\psi$).
We also showed how to relate the ellipticity to the inclination of a planar source, and how an isotropic prior in the source inclination is highly nonuniform in terms of ellipticity, favoring circular polarizations.

In the last section, we briefly touched on the generalization to metric theories of gravity with additional (nontensor) \replaced{polarizations}{modes}, for which the mathematical treatment is, for the most part, exactly analogous.

\begin{acknowledgments}
I would like to thank Will Farr, Katerina Chatziioannou and Leo Stein for insighful discussions, as well as Jose Mar\'ia Ezquiaga and Jolien Creighton for comments on the draft.
The Flatiron Institute is a division of the Simons Foundation.
This paper carries LIGO document number \dcc{}.
\end{acknowledgments}

\appendix

\section{Compact-binary spin frames}
\label{app:spins}

\begin{figure}
  \includegraphics[width=\columnwidth]{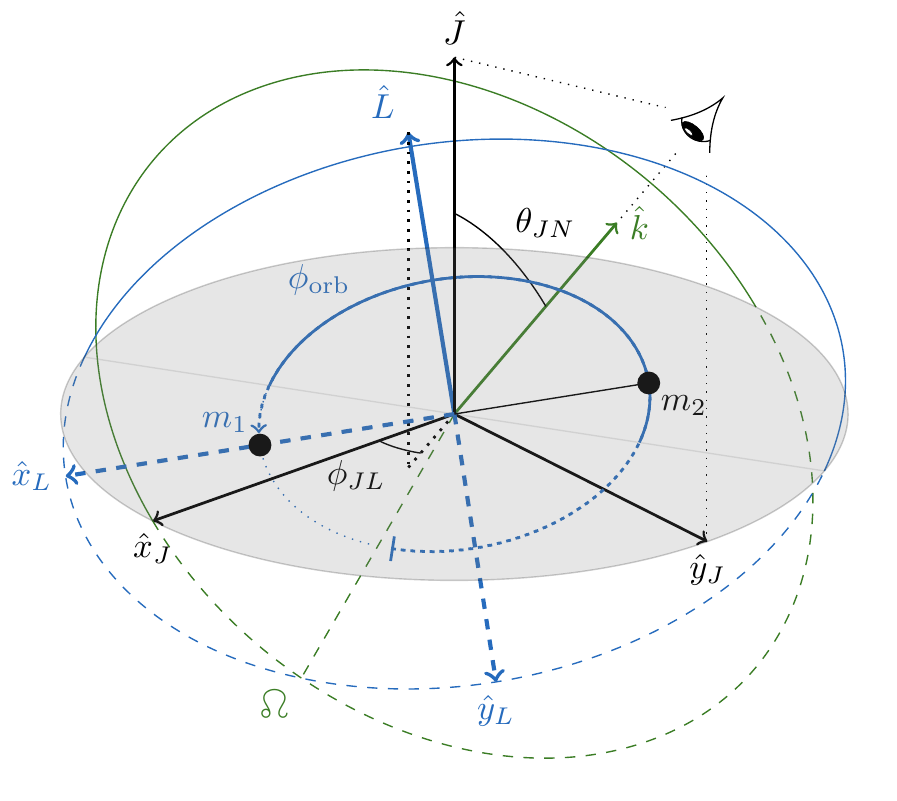}
  \caption{Coordinate frames used to specify the component spin vectors for a precessing \ac{CBC}. The Cartesian components of each individual spin vector $\vec{S}_{1/2}$ (not shown) are prescribed in the orbital frame given by $(\hat{x}_L, \hat{y}_L, \hat{L})$, where $\hat{x}_L$ points in the direction from $m_2$ to $m_1$ at a reference time (blue); the orientation of $\hat{x}_L$ is set by an orbital phase angle $\phi_{\rm orb}$ defined with respect to the line of nodes ($\ascnode$).
  The total angular momentum $\vec{J}$ defines an angle $\theta_{JN}$ with respect to the wavevector $\vec{k}$; in the $J$-based frame (black), $\vec{L}$ is oriented so as to define an angle $\phi_{JL}$ relative to $\hat{x}_J$, which is itself perpendicular to both $\vec{J}$ and $\vec{k}$. For comparison with Figs.~\ref{fig:diagram_waveframe}, \ref{fig:diagram_sourceframe} and \ref{fig:diagram_skyview} we show the plane of the sky (faint green circle), which is orthogonal to $\hat{k}$.
  Dashed traces mark elements below the $\{\hat{x}_J, \hat{y}_J\}$ plane.}
  \label{fig:spins}
  \end{figure}

Besides the polarization-related frames discussed in the main text, additional coordinates come into play when describing a precessing \ac{CBC}.
These are required to specify the orientations of the spins of the individual objects in the binary, since these are not aligned with the orbital angular momentum for the case of a precessing system.
Even though these coordinates are not directly relevant to the description of GW polarizations, we describe them here for completeness following the current LIGO-Virgo convention \cite{LALSuite:spins} (conventions occasionally change \cite{Pfeiffer:T1800226}).

The component spin vectors, $\vec{S}_{1/2}$, are prescribed at an arbitrary reference time (e.g., the moment when the signal reaches 20 Hz at the detector) in a Cartesian frame with $z$-axis along the orbital angular momentum, $\vec{L}$, with $x$-axis along the line pointing from the lighter object ($m_2$) to the heavier object ($m_1$), and with $y$-axis completing the right-handed triad; this $L$-based coordinate frame is shown in blue Fig.~\ref{fig:spins}.

In the above frame, the spin components are specified relative to the orbital plane.
For a precessing system, the orientation of the binary itself is set with respect to the observer through the angle $\theta_{JN}$ between the direction of propagation $\hat{k}$ and the total angular momentum of the binary, $\vec{J} \equiv \vec{L} + \vec{S}_1 + \vec{S}_2$ (Fig.~\ref{fig:spins} in black); this angle is similar to $\iota$ except for being defined relative to $\vec{J}$ instead of $\vec{L}$ (the two angles are the same for nonprecessing systems).
An additional angle, $\phi_{JL}$, establishes the orientation of $\vec{L}$ around $\vec{J}$, measured azimuthally with respect to the vector $\hat{x}_J$ perpendicular to the plane containing both $\vec{J}$ and $\hat{k}$, i.e., $\hat{x}_J = \hat{k} \times \hat{J} / |\hat{k} \times \hat{J}|$ in Fig.~\ref{fig:spins}.
The final degree of freedom is set by specifying the orbital phase $\phi_{\rm orb}$ at the reference time, defined as the angle spanned by the location of the primary body with respect to the line of nodes ($\ascnode$) within the orbital plane.

For a nonprecessing binary, $\vec{J}$ is parallel to $\vec{L}$, and so $\phi_{JL}$ is undefinded.
Meanwhile, $\theta_{JN}$ reduces to the angle $\iota$, which is defined to be the angle between $\hat{k}$ and $\hat{L}$ (Fig.~\ref{fig:diagram_sourceframe}).
The term ``inclination angle'' can refer either to $\theta_{JN}$ or $\iota$ depending on context.

\bibliography{refs}

\end{document}